\documentclass[
    a4paper,
    colorlinks=true,
    linkcolor=medblau,
    citecolor=medblau,
    urlcolor=medblau,
]{article}

\usepackage{arxiv}

\usepackage[utf8]{inputenc} 
\usepackage[T1]{fontenc}    
\usepackage{url}            
\usepackage{booktabs}       
\usepackage{amsfonts}       
\usepackage{amsmath}
\usepackage{nicefrac}       

\usepackage{microtype}      
\usepackage{lipsum}		   
\usepackage[numbers,sort&compress]{natbib}
\usepackage{doi}
\usepackage{siunitx}
\DeclareSIUnit\bar{bar}
\usepackage{enumitem}
\usepackage[version=4]{mhchem}
\usepackage{titlesec}
\usepackage[all]{nowidow}

\usepackage{graphicx}
\usepackage{xcolor} 

\definecolor{faublau}{RGB}{4,49,106}
\definecolor{faudunkelblau}{RGB}{4,30,66}
\definecolor{philgelb}{RGB}{253,183,53}
\definecolor{philorange}{RGB}{232,119,34}
\definecolor{rwrot}{RGB}{197,15,60}
\definecolor{rwdunkelrot}{RGB}{151,27,47}
\definecolor{medblau}{RGB}{24,180,241}
\definecolor{meddunkelblau}{RGB}{0,82,135}
\definecolor{natgrün}{RGB}{123,183,37}
\definecolor{natdunkelgrün}{RGB}{38,97,65}
\definecolor{tfmetallic}{RGB}{140,159,177}
\definecolor{tfdunkelmetallic}{RGB}{47,88,110}
\definecolor{schwarz}{RGB}{0,0,0}

\definecolor{faublau062}{RGB}{97,125,161}
\definecolor{faublau037}{RGB}{160,177,198}
\definecolor{faublau025}{RGB}{192,203,218}
\definecolor{faublau012}{RGB}{211,220,242}

\definecolor{faudunkelblau062}{RGB}{97,113,136}
\definecolor{faudunkelblau037}{RGB}{160,169,183}
\definecolor{faudunkelblau025}{RGB}{192,205,208}
\definecolor{faudunkelblau012}{RGB}{223,226,241}

\definecolor{philgelb062}{RGB}{254,206,118}
\definecolor{philgelb037}{RGB}{254,228,178}
\definecolor{philgelb025}{RGB}{254,237,204}
\definecolor{philgelb012}{RGB}{255,245,224}

\definecolor{philorange062}{RGB}{239,163,105}
\definecolor{philorange037}{RGB}{246,203,171}
\definecolor{philorange025}{RGB}{249,221,200}
\definecolor{philorange012}{RGB}{252,237,226}

\definecolor{rwrot062}{RGB}{221,115,124}
\definecolor{rwrot037}{RGB}{235,171,174}
\definecolor{rwrot025}{RGB}{241,200,201}
\definecolor{rwrot012}{RGB}{252,220,227}

\definecolor{rwdunkelrot062}{RGB}{190,113,125}
\definecolor{rwdunkelrot037}{RGB}{216,167,198}
\definecolor{rwdunkelrot025}{RGB}{230,198,205}
\definecolor{rwdunkelrot012}{RGB}{244,226,229}

\definecolor{medblau062}{RGB}{109,208,246}
\definecolor{medblau037}{RGB}{167,236,250}
\definecolor{medblau025}{RGB}{197,236,251}
\definecolor{medblau012}{RGB}{227,250,255}

\definecolor{meddunkelblau062}{RGB}{90,164,179}
\definecolor{meddunkelblau037}{RGB}{158,189,209}
\definecolor{meddunkelblau025}{RGB}{191,218,245}
\definecolor{meddunkelblau012}{RGB}{222,241,252}

\definecolor{natgrün062}{RGB}{172,210,117}
\definecolor{natgrün037}{RGB}{205,228,172}
\definecolor{natgrün025}{RGB}{222,237,200}
\definecolor{natgrün012}{RGB}{232,252,220}

\definecolor{natdunkelgrün062}{RGB}{118,155,135}
\definecolor{natdunkelgrün037}{RGB}{172,195,174}
\definecolor{natdunkelgrün025}{RGB}{201,215,207}
\definecolor{natdunkelgrün012}{RGB}{222,235,220}

\definecolor{tfmetallic062}{RGB}{182,194,206}
\definecolor{tfmetallic037}{RGB}{211,218,225}
\definecolor{tfmetallic025}{RGB}{235,240,247}
\definecolor{tfmetallic012}{RGB}{243,245,247}

\definecolor{tfdunkelmetallic062}{RGB}{124,150,163}
\definecolor{tfdunkelmetallic037}{RGB}{176,191,200}
\definecolor{tfdunkelmetallic025}{RGB}{228,233,236}
\definecolor{tfdunkelmetallic012}{RGB}{229,233,236}

\definecolor{schwarz062}{RGB}{94,94,94}
\definecolor{schwarz037}{RGB}{158,158,158}
\definecolor{schwarz025}{RGB}{191,191,191}
\definecolor{schwarz012}{RGB}{222,222,222}

\colorlet{col1}{faublau}           
\colorlet{col2}{tfdunkelmetallic}  
\colorlet{col3}{rwrot}             
\colorlet{col4}{natgrün}           
\colorlet{col5}{philgelb}          
\colorlet{col6}{philorange}        
\colorlet{col7}{rwdunkelrot}       
\colorlet{col8}{natdunkelgrün}     
\colorlet{col9}{medblau}           
\colorlet{col10}{meddunkelblau}    
\colorlet{col11}{tfmetallic}       
\colorlet{col12}{faudunkelblau}    
\colorlet{col13}{philgelb062}      
\colorlet{col14}{tfdunkelmetallic012} 
\colorlet{col15}{schwarz}          

\usepackage{transparent}
\usepackage{adjustbox}
\usepackage{subcaption}
\usepackage[labelfont=bf]{caption}
\usepackage{graphicx}
\usepackage{tikz}

\usepackage{tabularx}
\usepackage{multirow}
\usepackage{booktabs,makecell} 

\usepackage{pgfplots}
\pgfplotsset{compat=1.18}
\usepgfplotslibrary{statistics}
\usepackage{tikz}
\usetikzlibrary{tikzmark,patterns}
\usetikzlibrary{arrows,arrows.meta,shapes.arrows}

\usepackage[automake,nogroupskip,nonumberlist,nopostdot]{glossaries-extra}
\usepackage{glossary-mcols}

\newglossarystyle{mycoltreeheader}{%
  \setglossarystyle{mcoltree*}%
  \setlength{\columnsep}{8pt} %
  \newcolumntype{Y}{>{\raggedright\arraybackslash}p{1.15cm}}
  \newcolumntype{U}{>{\raggedright\arraybackslash}p{.7cm}}
}

\setabbreviationstyle[acronym]{long-short}
\setglossarystyle{mycoltreeheader}
\newglossary[slg]{symbolslist}{syi}{syg}{Nomenclature}
\makeglossaries

\newacronym{hpdc}{HPDC}{High-Pressure Die Casting}
\newacronym{cfd}{CFD}{Computational Fluid Dynamics}
\newacronym{vof}{VOF}{Volume of Fluid}
\newacronym{rans}{RANS}{Reynolds-Averaged Navier--Stokes}
\newacronym{esc}{ESC}{Externally Solidified Crystals}
\newacronym{dns}{DNS}{Direct Numerical Simulation}
\newacronym{ct}{CT}{Computed Tomography}
\newacronym{fvm}{FVM}{Finite Volume Method}
\newacronym{fem}{FEM}{Finite Element Method}
\newacronym{fdm}{FDM}{Finite Difference Method}
\newacronym{di}{DI}{Density Index}
\newacronym{cad}{CAD}{Computer Aided Design}
\newacronym{cfl}{CFL}{Courant-Friedrichs-Lewy number}
\newacronym{hpc}{HPC}{High-Performance Computing}
\newacronym{gci}{GCI}{Grid Convergence Index}
\newacronym{vr}{VR}{Venting Region}
\newacronym{mr}{MR}{Mid Region}
\newacronym{ir}{IR}{Ingate Region}
\newacronym{tifsa}{TIFSA}{Time-Integrated free surface Area}
\newacronym{tivf}{TIVF}{Time-Integrated Volumetric Flow}
\newacronym{tmvf}{TMVF}{Temporal Mean Volume Fraction}
\newacronym{les}{LES}{Large Eddy Simulation}
\newacronym{des}{DES}{Detached Eddy Simulation}
\newacronym[longplural={Evaluation Volumes}]{ev}{EV}{Evaluation Volume}
\newacronym[longplural={Evaluation Points}]{ep}{EP}{Evaluation Point}
\newacronym[longplural={Evaluation Surfaces}]{es}{ES}{Evaluation Surface}
\newacronym[longplural={Evaluation Regions}]{er}{ER}{Evaluation Region}

\newglossaryentry{sym:we}{
  type=symbolslist,
  name={$We$},
  sort={01_we},
  description={Weber number},
  user1 = $-$,
}
\newglossaryentry{sym:weg}{
  type=symbolslist,
  name={$We_g$},
  sort={01_we_g},
  description={Weber number at ingate},
  user1 = $-$,
}

\newglossaryentry{sym:rho}{
  type=symbolslist,
  name={$\rho$},
  sort={02_rho},
  description={Fluid density},
}
\newglossaryentry{sym:u}{
  type=symbolslist,
  name={$\mathbf{u}$},
  sort={02_u},
  description={Velocity field},
}
\newglossaryentry{sym:p}{
  type=symbolslist,
  name={$p$},
  sort={02_p},
  description={Pressure},
}
\newglossaryentry{sym:g}{
  type=symbolslist,
  name={$\mathbf{g}$},
  sort={02_g},
  description={Gravitational acceleration vector},
}

\newglossaryentry{sym:alpha}{
  type=symbolslist,
  name={$\alpha$},
  sort={03_alpha},
  description={Liquid volume fraction},
}
\newglossaryentry{sym:ur}{
  type=symbolslist,
  name={$\mathbf{u}_r$},
  sort={03_u_r},
  description={Artificial compression velocity},
}
\newglossaryentry{sym:as}{
  type=symbolslist,
  name={$A_{s}$},
  sort={03_A_s},
  description={Free surface area},
}

\newglossaryentry{sym:vin}{
  type=symbolslist,
  name={$v_{in}$},
  sort={04_v_in},
  description={Inlet velocity},
}
\newglossaryentry{sym:vg}{
  type=symbolslist,
  name={$v_{g}$},
  sort={04_v_g},
  description={Ingate velocity},
}
\newglossaryentry{sym:v_mean}{
  type=symbolslist,
  name={$\bar{v}$},
  sort={04_v_bar},
  description={Spatially averaged velocity},
}

\newglossaryentry{sym:tau}{
  type=symbolslist,
  name={$\tau_{ij}$},
  sort={05_tau_ij},
  description={Reynolds stress tensor},
}
\newglossaryentry{sym:uiuj}{
  type=symbolslist,
  name={$u_i', u_j'$},
  sort={05_uprime},
  description={Fluctuating velocity components},
}
\newglossaryentry{sym:k}{
  type=symbolslist,
  name={$k$},
  sort={05_k},
  description={Turbulent kinetic energy},
}
\newglossaryentry{sym:varepsilon}{
  type=symbolslist,
  name={$\varepsilon$},
  sort={05_epsilon},
  description={Turbulent dissipation rate},
}
\newglossaryentry{sym:Pk}{
  type=symbolslist,
  name={$P_k$},
  sort={05_Pk},
  description={Production rate of turbulent kinetic energy},
}
\newglossaryentry{sym:nu}{
  type=symbolslist,
  name={$\nu$},
  sort={05_nu},
  description={Molecular kinematic viscosity},
}
\newglossaryentry{sym:nut}{
  type=symbolslist,
  name={$\nu_t$},
  sort={05_nu_t},
  description={Turbulent (eddy) viscosity},
}
\newglossaryentry{sym:sigmak}{
  type=symbolslist,
  name={$\sigma_k$},
  sort={05_sigma_k},
  description={Turbulent Prandtl number for $k$},
}
\newglossaryentry{sym:sigmaeps}{
  type=symbolslist,
  name={$\sigma_\varepsilon$},
  sort={05_sigma_epsilon},
  description={Turbulent Prandtl number for $\varepsilon$},
}
\newglossaryentry{sym:Cmu}{
  type=symbolslist,
  name={$C_\mu$},
  sort={05_Cmu},
  description={Model constant for eddy viscosity},
}
\newglossaryentry{sym:C1eps}{
  type=symbolslist,
  name={$C_{1\varepsilon}$},
  sort={05_C1eps},
  description={Empirical model constant for $\varepsilon$ equation},
}
\newglossaryentry{sym:C2eps}{
  type=symbolslist,
  name={$C_{2\varepsilon}$},
  sort={05_C2eps},
  description={Empirical model constant for $\varepsilon$ equation},
}

\newglossaryentry{sym:h}{
  type=symbolslist,
  name={$h$},
  sort={06_h},
  description={Specific enthalpy},
}
\newglossaryentry{sym:T}{
  type=symbolslist,
  name={$T$},
  sort={06_T},
  description={Temperature},
}
\newglossaryentry{sym:lameff}{
  type=symbolslist,
  name={$\lambda_{\text{eff}}$},
  sort={06_lambda_eff},
  description={Effective thermal conductivity},
}
\newglossaryentry{sym:mueff}{
  type=symbolslist,
  name={$\mu_{\text{eff}}$},
  sort={06_mu_eff},
  description={Effective dynamic viscosity},
}
\newglossaryentry{sym:Phi}{
  type=symbolslist,
  name={$\Phi$},
  sort={06_Phi},
  description={Viscous dissipation},
}

\newglossaryentry{sym:phi}{
  type=symbolslist,
  name={$\phi$},
  sort={07_phi},
  description={Instantaneous flow variable},
  user1 = $-$,
}
\newglossaryentry{sym:phi_over}{
  type=symbolslist,
  name={$\overline{\phi}$},
  sort={07_phi_over},
  description={Flow variable, time-averaged},
}
\newglossaryentry{sym:phi_acc}{
  type=symbolslist,
  name={$\phi'$},
  sort={07_phi_prime},
  description={Fluctuating component $\phi$},
}
\newglossaryentry{sym:p_av}{
  type=symbolslist,
  name={$\overline{p}$},
  sort={07_p_over},
  description={Average pressure},
}

\newglossaryentry{sym:vair}{
  type=symbolslist,
  name={$V_{Air}$},
  sort={08_V_air},
  description={Entrapped air volume},
}
\newglossaryentry{sym:n}{
  type=symbolslist,
  name={$\mathbf{n}$},
  sort={08_n},
  description={Outward normal vector},
}

\newglossaryentry{sym:zeta_tifsa}{
  type=symbolslist,
  name={$\zeta_{\mathrm{TIFSA}}$},
  sort={09_zeta_TIFSA},
  description={TIFSA criterion},
}
\newglossaryentry{sym:zeta_tmvf}{
  type=symbolslist,
  name={$\zeta_{\mathrm{TMVF}}$},
  sort={09_zeta_TMVF},
  description={TMVF criterion},
}
\newglossaryentry{sym:zeta_tivf}{
  type=symbolslist,
  name={$\zeta_{\mathrm{TIVF}}$},
  sort={09_zeta_TIVF},
  description={TIVF criterion},
}

\newglossaryentry{sym:a_f}{
  type=symbolslist,
  name={$A_f$},
  sort={10_A_f},
  description={Surface area of face $f$},
}

\usepackage{flafter}
\usepackage{placeins,etoolbox} 
\usepackage{comment}

\titlespacing{\section}{0pt}{6pt}{3pt} 
\titlespacing{\subsection}{0pt}{4pt}{1pt} 
\usepackage{hyperref}

\title{Development and Experimental Validation of Novel Evaluation Criteria for Turbulent Two-Phase VOF Simulations in High-Pressure Die Casting}

\addauthor{Mehran Shahzadeh}{1}{mehran.shahzadeh@fau.de}{}
\addauthor{Fabian Teichmann}{1}{fabian.teichmann@fau.de}{0000-0001-9353-9168}
\addauthor{Sebastian Müller}{1}{seb.mueller@fau.de}{0000-0001-8693-6341}

\addaffil{1}{Friedrich-Alexander-Universität Erlangen-Nürnberg, Chair of Casting Technology, Fürth, 90762, Germany}

\hypersetup{
pdftitle={Development and Experimental Validation of Novel Evaluation Criteria for Turbulent Two-Phase VOF Simulations in High-Pressure Die Casting},
pdfauthor={Mehran ~Shazedeh, Fabian ~Teichmann, Sebastian ~Müller},
pdfkeywords={High-Pressure Die Casting, Volume of Fluid, CompressibleInterFoam, Turbulence Modeling, Reynolds-Averaged Navier-Stokes, Air Entrapment, Free Surface, Mesh Sensitivity, OpenFOAM}
}

\definecolor{col1}{HTML}{020109}
\definecolor{col6}{HTML}{1E1149}
\definecolor{col2}{HTML}{4E117B}
\definecolor{col7}{HTML}{882881}
\definecolor{col3}{HTML}{B93879}
\definecolor{col8}{HTML}{E95462}
\definecolor{col4}{HTML}{FC8B63}
\definecolor{col9}{HTML}{FEC78B}
\definecolor{col5}{HTML}{FEC78B}
\definecolor{col10}{HTML}{C83D73}

\begin{document}
\maketitle
\begin{abstract}
	Air entrapment during mold filling critically affects porosity and overall casting quality in \gls{hpdc}. This study assesses the feasibility of applying the \gls{vof} method within \texttt{OpenFOAM} to simulate compressible, turbulent mold filling in a thin-walled geometry. Three-dimensional simulations with the \texttt{compressibleInterFoam} solver were carried out under ambient initial cavity conditions, using both laminar flow and the $k$\,\textit{$\epsilon$} turbulence model. The free surface dynamics were examined across a range of inlet velocities to evaluate their influence on interface morphology, cavity pressurization, and gas entrapment. To quantify these effects, three evaluation criteria were introduced: the \gls{tifsa} as a measure of oxidation risk, the \gls{tmvf} as an indicator of filling continuity and air entrapment, and the \gls{tivf} as a proxy for surface loading. Results show that turbulence modeling accelerates pressurization and limits the persistence of entrapped gas, with velocity governing the balance between smooth filling, turbulent breakup, and exposure duration. Comparison with experimental casting trials, including CT based porosity analysis and photogrammetric surface evaluation, validated that the model captures key defect mechanisms and provides quantitative guidance for process optimization.
\end{abstract}



\printglossary[type=symbolslist,title={Nomenclature}]

\section{Introduction}
\label{sec:intro}

\gls{hpdc} is known to enable the rapid production of near-net-shape products with complex geometries and stringent mechanical requirements. By means of this method, the need for additional postprocessing steps is significantly reduced, thereby lowering overall manufacturing costs~\cite{cornacchia2019}. Demand for this production method has been strongly driven by the automotive industry, where high output rates of a wide variety of components are required. Consequently, almost \SI{60}{\percent} of light metal castings are currently produced using this process~\cite{campatelli2012}. In practice, molten metal is poured into a casting chamber and then injected into a mold at high velocity through piston acceleration. The mold filling time is typically on the order of milliseconds to a few seconds, and the molten metal commonly reaches velocities between \SI{30}{\metre\per\second} and \SI{100}{\metre\per\second}~\cite{markezic2019,bonollo2015,podprocka2015}. However, the complexity of the process and the intricate nature of the resulting parts lead to relatively high reject rates. While other manufacturing lines can achieve defect rates as low as parts per million, this method generally yields defect rates in the percentage range~\cite{campatelli2012}. Given these challenges, an enhanced understanding of how different filling velocities affect melt flow behavior is required, with particular focus on air entrapment and surface formation. Through the integration of high fidelity \gls{cfd} simulations, it is shown that advanced computational techniques can substantially improve the accuracy of casting predictions, thereby enabling more reliable outcomes and optimized process parameters for defect reduction in \gls{hpdc}.

\subsection{HPDC defects and challenges}
\label{ssec:defects}
Defects in \gls{hpdc} arise from interactions of gate design, melt injection velocity, applied pressure, and thermal conditions of alloy and mold, which, if not controlled, lead to shrinkage, cold shuts, soldering, cracking, gas entrapment, and porosity~\cite{markezic2019,bonollo2015,podprocka2015}. Among these, porosity is the most critical, as it reduces mechanical performance, causes blistering during heat treatment, and limits weldability~\cite{li2016}. Li et al.~\cite{li2016} classified porosity into gas shrinkage pores, gas pores, net shrinkage, and island shrinkage, with cracks often initiating at interconnected pores in weak cross sections~\cite{li2016,jiao2021}. Gas is readily trapped by vortex flows, limiting diffusion and promoting pore formation~\cite{cao2019}. Experiments showed that reducing cavity pressure from \SIrange{100}{1013}{\milli\bar} markedly decreases pore size and volume, improving heat treatment response and fatigue life~\cite{cao2019,cao2020,szalva2020}. High injection velocities favor turbulence and recirculation and thus entrapment; turbulence can also fragment inclusions and equiaxed crystals and refine microstructure, improving properties~\cite{lordan2021}. Jiao et al.~\cite{jiao2021} showed that shot speed controls \gls{esc} and that optimized flow improves uniformity and reduces porosity~\cite{jiao2021}. In AlSi10MnMg, \gls{esc} fraction increases from surface to center with a globular to dendritic transition; higher shot speeds suppress dendritic \gls{esc} networks and reduce shrinkage porosity, while lower speeds promote dendritic structures and degrade properties~\cite{chen2022}. Another major defect source is oxide films (bifilms)~\cite{campbell2011}, formed when turbulent surface folding brings oxidized surfaces together and traps gas; these unbonded double layers behave like internal cracks and nucleate porosity~\cite{tiryakioglu2020}. Shock tube studies of oxidizing liquid metal droplets link high Weber number (\gls{sym:we}) interfacial fracture to bifilm creation mechanics in casting: Galinstan droplets show standard breakup modes with transitions near \gls{sym:we}~$~\approx~\numlist{15;35;80}{}$, but rupture occurs earlier in nondimensional time and by sharp edged, fracture like bag failure that yields coarse, non-spherical fragments due to the elastic oxide skin~\cite{hopfes2021a}. Field metal exhibits similar morphology and timings of initial deformation, with slightly larger bag inflation and a marginally later breakup onset at low \gls{sym:we}, and more numerous smaller fragments, consistent with oxidation kinetics effects~\cite{hopfes2021b}. These results imply that in \gls{hpdc} the combination of surface turbulence, oxide membrane inflation, and oxide yielding fosters folding and cracking of the surface skin, trapping gas between double layers and seeding bifilms that later open as crack like defects or porosity, reinforcing the need to control \gls{sym:we} at ingates, suppress free surface turbulence, and use vacuum assistance~\cite{hopfes2021a,hopfes2021b,campbell2011,tiryakioglu2020}.

\subsection{Multiphase Simulation in Casting}
Significant advancements have been made in \gls{vof} techniques over the past decades. The SOLA-VOF method, introduced by Hirt and Nichols~\cite{hirt1981}, has been extensively adopted for casting filling simulations. By combining the momentum and continuity equations with the \gls{vof} approach, accurate and stable free surface simulations can be achieved. Under high pressure conditions, Schneiderbauer et~al.~\cite{schneiderbauer2012} demonstrated the effectiveness of \gls{vof} in capturing complex flow phenomena in casting processes. Otsuka~\cite{otsuka2014} investigated gas entrapment in \gls{hpdc} using \gls{vof} simulations, showing that coarse mesh resolutions can result in poor predictions of wave and splash behavior, thereby emphasizing the necessity for fine mesh refinement to obtain reliable results. In industrial practice, \gls{cfd} software packages such as ANSYS Fluent, FLOW-3D, and ProCAST implement various numerical discretization techniques, including the \gls{fvm}, \gls{fem}, and \gls{fdm}, to simulate complex casting processes. These include filling, solidification, and stress analysis, often in combination with \gls{vof} for interface tracking. Such tools are extensively used to optimize process parameters, reduce defect rates, and enhance the quality of \gls{hpdc} products~\cite{chen2022}. Building on these foundations, multiphase modeling strategies have been applied directly to \gls{hpdc}. Tavakoli et al. \cite{tavakoli2006} applied a \gls{vof} formulation to capture interface evolution during mold filling, demonstrating how recirculation and jet impingement drive gas entrapment. Cao et al. \cite{cao2017} extended such approaches by integrating gas transport modeling with \gls{vof} simulations to predict the size and distribution of entrapped pores in zinc alloy die casting, highlighting the role of cavity pressurization in realistic porosity prediction. Complementary to these multiphase studies, Wang et al. \cite{wang2022} combined \gls{cfd} simulations and orthogonal experiments to map the influence of filling velocity and feeding conditions in thin-walled \gls{hpdc} components, confirming that porosity distribution is highly sensitive to process parameters. In parallel, open-source \gls{cfd} frameworks such as \texttt{OpenFOAM} have been investigated for their applicability to casting processes. Kohlstädt et al. \cite{kohlstdt2021} demonstrated the use of \texttt{OpenFOAM} to determine critical velocities in the shot sleeve of \gls{hpdc} machines, showing that open-source \gls{cfd} can provide quantitative design guidance under industrially relevant conditions. More recently, Tang et al. \cite{tang2024} validated advanced turbulence closures within \texttt{OpenFOAM} for multiphase flows in continuous casting, underlining the maturity of turbulence modeling and its applicability to gas-liquid interactions relevant for \gls{hpdc}. The \gls{vof} approach was therefore selected in the present work to accurately capture the transient free surface dynamics and turbulent metal-air interactions during mold filling. This capability is essential for predicting critical defect mechanisms, including air entrapment, oxide film formation, and the onset of porosity. By resolving these interfacial phenomena, the numerical model provides a more reliable basis for optimizing shot speeds, gating design, and process parameters to improve casting quality.

A central methodological choice in this study is the use of water as a surrogate fluid for molten aluminum. Direct numerical simulation of liquid aluminum is computationally demanding, as its thermophysical properties vary strongly with temperature and alloy composition. By contrast, water offers a robust alternative for capturing the dominant hydrodynamic mechanisms of mold filling, such as free surface development, jet breakup, and air entrainment, while avoiding uncertainties associated with melt property specification. The use of water analogues is well established in casting research: Yuan et al.~\cite{yuan2008} validated numerical die casting models against water analogue experiments during the slow shot phase, demonstrating close agreement between predicted and observed flow fields. Javurek and Wincor~\cite{javurek2020} applied similar strategies in continuous casting, where water models provided reliable validation data for turbulent multiphase flow. In low-pressure casting, Viswanath et al.~\cite{viswanath2017} combined simulations and water analogue experiments to track free surface evolution, while Renukananda and Ravi~\cite{renukananda2015} compared liquid metal and water flows in multi-gate systems to assess gating design.

Beyond individual advances, three recurring themes emerge from prior research: (i) the use of surrogate experiments with water to isolate hydrodynamic effects in casting processes~\cite{yuan2008,javurek2020,viswanath2017,renukananda2015}, (ii) the refinement of \gls{cfd} strategies for \gls{hpdc}, ranging from laminar approximations to \gls{rans} and \gls{les} turbulence closures, and (iii) the recognition that compressibility strongly influences porosity prediction through its impact on cavity pressurization and free surface evolution. Despite this progress, systematic investigations directly comparing turbulence closures under compressible multiphase conditions in thin-walled geometries remain scarce. This work addresses that gap by systematically evaluating how turbulence modeling shapes free surface morphology, air pocket survival, and cavity pressurization in compressible \gls{vof} simulations. By linking these mechanisms to defect formation, the study provides new insights into the role of turbulence representation in porosity prediction and offers guidance for process optimization in \gls{hpdc}.

\subsection{Summary and research goals}
\label{ssec:sumGoal}
The introduction highlighted the key challenges of \gls{hpdc}, namely defect formation due to turbulent mold filling, air entrapment, and oxide related porosity. While advances in \gls{cfd} provide new opportunities to analyze such phenomena, several open questions remain regarding the role of compressibility, turbulence, and evaluation metrics in predicting defect prone regions. Building on this motivation, the study is guided by the following research questions:

\begin{itemize}
  \item Can a compressible multiphase \gls{cfd} framework capture cavity pressurization and pore contraction, and how does this depend on inlet velocity and filling speed?
  \item To what extent do laminar and $k$-$\varepsilon$ models differ in predicting free surface dynamics and localized air entrapment?
  \item How does the timing of porosity assessment affect predictions, given that rising cavity pressure compresses or dissolves small pores? Which flow parameters (velocity, pressure) provide robust criteria for defining a meaningful evaluation window?
  \item To what degree do simulation based predictions reproduce \gls{ct} derived porosity distributions?  
\end{itemize}

By addressing these questions, the study aims to establish a validated \gls{cfd} based framework for linking filling dynamics, air entrapment, and porosity formation in \gls{hpdc}. Beyond providing mechanistic insights, the introduced criteria (\gls{tifsa}, \gls{tmvf}, \gls{tivf}) also serve as quantitative quality measures that allow entire simulations or selected time windows to be evaluated consistently. This constitutes a methodological contribution of the present work, as it provides a structured set of diagnostic indicators that can be applied both to process optimization and to the systematic evaluation of \gls{cfd} simulations in industrial \gls{hpdc} practice.

\section{Governing equations}
\label{sec:gov}
The simulations are based on the compressible Navier-Stokes equations coupled with a \gls{vof} formulation to resolve liquid-gas interfaces \citep{hirt1981}. Turbulence is represented within the \gls{rans} framework, where any instantaneous flow variable \gls{sym:phi} is decomposed into a mean and a fluctuating part,
\begin{equation}
\phi = \overline{\phi} + \phi',
\end{equation}
with \gls{sym:phi_over} the time-averaged component and \gls{sym:phi_acc} the fluctuation. Substitution into the Navier-Stokes equations leads to additional Reynolds stress terms in the averaged momentum equations,
\begin{equation}
\tau_{ij} = -\rho\, \overline{u_i' u_j'} ,
\end{equation}
where \gls{sym:tau} are the Reynolds stresses, \gls{sym:rho} is the fluid density, and \gls{sym:uiuj} are the fluctuating velocity components. To close the system, the standard $k$-$\varepsilon$ turbulence model is employed \citep{launder1974}. It introduces two additional transport equations: one for the turbulent kinetic energy \gls{sym:k} and one for the turbulent dissipation rate \gls{sym:varepsilon}. The transport equation for \gls{sym:k} reads
\begin{equation}
\frac{\partial k}{\partial t} + \overline{u}_j \frac{\partial k}{\partial x_j} 
= P_k - \varepsilon 
+ \frac{\partial}{\partial x_j} \left[ \left( \nu + \frac{\nu_t}{\sigma_k} \right) \frac{\partial k}{\partial x_j} \right],
\end{equation}
where \gls{sym:Pk} is the production rate of turbulent kinetic energy, \gls{sym:nu} the molecular kinematic viscosity, \gls{sym:nut} the turbulent (eddy) viscosity, and \gls{sym:sigmak} the turbulent Prandtl number for \gls{sym:k}. The dissipation rate equation is given by
\begin{equation}
\frac{\partial \varepsilon}{\partial t} + \overline{u}_j \frac{\partial \varepsilon}{\partial x_j} 
= C_{1\varepsilon} \frac{\varepsilon}{k} P_k - C_{2\varepsilon} \frac{\varepsilon^2}{k} 
+ \frac{\partial}{\partial x_j} \left[ \left( \nu + \frac{\nu_t}{\sigma_\varepsilon} \right) \frac{\partial \varepsilon}{\partial x_j} \right],
\end{equation}
where \gls{sym:C1eps}, \gls{sym:C2eps}, and \gls{sym:sigmaeps} are empirical model constants calibrated for turbulent flows. The eddy viscosity is then modeled as
\begin{equation}
\nu_t = C_\mu \frac{k^2}{\varepsilon},
\end{equation}
with \gls{sym:Cmu} another closure constant. The standard values of these empirical coefficients are: \gls{sym:sigmaeps}$=1.3$, \gls{sym:sigmak}$=1$, \gls{sym:C1eps}$=1.44$, \gls{sym:C2eps}$=1.92$, \gls{sym:Cmu}$=0.09$.The liquid-gas interface is tracked with the \gls{vof} method by solving a transport equation for the liquid volume fraction \gls{sym:alpha},
\begin{equation}
\frac{\partial \alpha}{\partial t} + \nabla \cdot (\alpha \mathbf{u}) 
+ \nabla \cdot \big[\alpha (1-\alpha) \, \mathbf{u}_r \big] = 0,
\end{equation}
where \gls{sym:u} is the velocity field and \gls{sym:ur} an artificial compression velocity that preserves a sharp interface \citep{okagaki2021}. Cells with $\alpha = 1$ are fully liquid, $\alpha = 0$ are fully gas, and intermediate values indicate the presence of an interface. The compressible \gls{rans} equations with $k$-$\varepsilon$ closure, combined with the \gls{vof} interface capturing scheme, form the mathematical framework of this study \citep{launder1974,hirt1981}. This formulation resolves the coupled effects of turbulence, compressibility, and free surface dynamics, which are decisive for predicting air entrapment and porosity formation in \gls{hpdc}. The volume fraction field \gls{sym:alpha} serves as a continuous indicator of phase distribution and is visualized in Figure~\ref{fig:alpha_rendered_grid}, which illustrates how the \gls{vof} method captures the evolving liquid-gas interface during mold filling at comparable filling stages for three inlet velocities.

\section{Methods}
\label{sec:methods}
To investigate mold filling behavior and defect formation mechanisms in \gls{hpdc}, a combined experimental and numerical approach was applied. Casting trials were carried out under representative industrial conditions and analyzed ex-situ for surface flow patterns and porosity using photogrammetry and X-ray \gls{ct}. In parallel, \gls{cfd} simulations were performed in \texttt{OpenFOAM} to capture transient flow phenomena and free surface dynamics. To enable systematic comparison between simulations and experiments, three complementary evaluation criteria were introduced.

\subsection{Experimental methods}
\label{ssec:exp_methods}
The experimental program focused on a thin-walled plate geometry that reflects typical \gls{hpdc} casting conditions. Casting trials of Silafont\textsuperscript{\textregistered}~36 plates were carried out under industrially relevant process parameters to provide a realistic basis for validation. The resulting components were then examined using photogrammetric analysis to characterize flow-induced surface features and by X-ray \gls{ct} to quantify internal porosity.

\subsubsection{HPDC experiments}
\label{sssec:hpdc}
\begin{figure}[hbt]
\small
\centering
  \begin{subfigure}[b]{0.45\textwidth}
    \centering
    {\def\svgwidth{160pt} 
    \input{figure1_a.tex}}
    \caption{}
    \label{fig:CADmodel}
  \end{subfigure}
  \hfill
  \begin{subfigure}[b]{0.45\textwidth}
  \centering
  {\def\svgwidth{135pt} 
\begingroup%
  \makeatletter%
  \providecommand\color[2][]{%
    \errmessage{(Inkscape) Color is used for the text in Inkscape, but the package 'color.sty' is not loaded}%
    \renewcommand\color[2][]{}%
  }%
  \providecommand\transparent[1]{%
    \errmessage{(Inkscape) Transparency is used (non-zero) for the text in Inkscape, but the package 'transparent.sty' is not loaded}%
    \renewcommand\transparent[1]{}%
  }%
  \providecommand\rotatebox[2]{#2}%
  \newcommand*\fsize{\dimexpr\f@size pt\relax}%
  \newcommand*\lineheight[1]{\fontsize{\fsize}{#1\fsize}\selectfont}%
  \ifx\svgwidth\undefined%
    \setlength{\unitlength}{155.99999856bp}%
    \ifx\svgscale\undefined%
      \relax%
    \else%
      \setlength{\unitlength}{\unitlength * \real{\svgscale}}%
    \fi%
  \else%
    \setlength{\unitlength}{\svgwidth}%
  \fi%
  \global\let\svgwidth\undefined%
  \global\let\svgscale\undefined%
  \makeatother%
  \begin{picture}(1,1.98966649)%
    \lineheight{1}%
    \setlength\tabcolsep{0pt}%
    \put(0,0){\includegraphics[width=\unitlength,page=1]{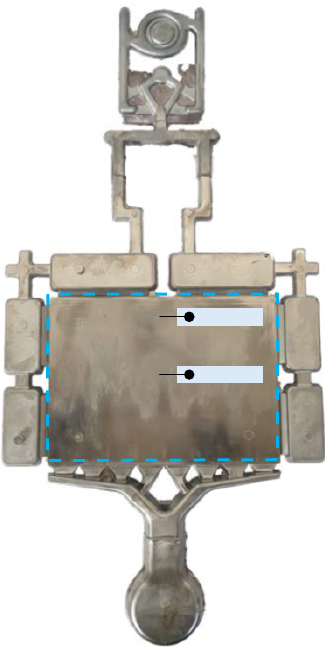}}%
    \put(0.33925425,0.99180133){\color[rgb]{0,0,0}\makebox(0,0)[t]{\lineheight{4.85710382}\smash{\begin{tabular}[t]{c}X-CT VR\end{tabular}}}}%
    \put(0.34651559,0.81404704){\color[rgb]{0,0,0}\makebox(0,0)[t]{\lineheight{4.85710382}\smash{\begin{tabular}[t]{c}X-CT MR\end{tabular}}}}%
    \put(0,0){\includegraphics[width=\unitlength,page=2]{figure1_bk.pdf}}%
    \put(0.34651587,0.63629289){\color[rgb]{0,0,0}\makebox(0,0)[t]{\lineheight{4.85710382}\smash{\begin{tabular}[t]{c}X-CT IR\end{tabular}}}}%
    \put(0.05465147,0.36944515){\color[rgb]{0,0,0}\makebox(0,0)[lt]{\lineheight{1.19999826}\smash{\begin{tabular}[t]{l}$ES_{TIVF}$\end{tabular}}}}%
    \put(0,0){\includegraphics[width=\unitlength,page=3]{figure1_bk.pdf}}%
  \end{picture}%
\endgroup%
}
  \caption{}
  \label{fig:Casting}
\end{subfigure}
\caption{Casted plate of \SI{3}{\milli\metre} thickness: (a) \gls{cad} model with \gls{ev} for fluid flow $EV_{FF}$, \glspl{ep} for inlet velocity $EP_{iv}$ and inlet pressure $EP_p$ as well as (b) photograph with \gls{es} for the TIVF criterion $ES_{TIVF}$. Both views also highlight the \gls{ir}, \gls{mr}, and \gls{vr} used in simulation and experiment.}
\label{fig:platte}
\end{figure}

In this study a flat plate geometry is employed, which is representative for \gls{hpdc} casting conditions. As illustrated in Figure~\ref{fig:CADmodel}, the CAD model is viewed from the fixed die half with highlighted regions of interest for the fluid flow analysis. The actual casting part, depicted in Figure~\ref{fig:Casting} viewed from the ejector side, shows the corresponding areas investigated experimentally by \gls{ct} and by the conducted simulations (Fig.~\ref{fig:exp_flow}). To evaluate the simulations, various \glspl{ev}, \glspl{es}, \glspl{ep} as well as the three evaluation regions \gls{ir}, \gls{mr} and \gls{vr} were defined within the \gls{cad} geometry as the figure indicates. 
The components were produced from Silafont\textsuperscript{\textregistered}~36 (AlSi10MnMg), an Al-Si-Mn-Mg alloy widely used in structural \gls{hpdc} applications due to its combination of castability, strength, and ductility. Its nominal composition limits are listed in Table~\ref{tab:chem_comp}. 
\begin{table}[b]
\centering
\small
\caption{Nominal chemical composition limits of Silafont\textsuperscript{\textregistered}\ 36 (AlSi10MnMg).}
\label{tab:chem_comp}
\begin{tabularx}{\textwidth}{l *{9}{>{\centering\arraybackslash}X} c}
\toprule
\textbf{Element [\%]} & \textbf{Si} & \textbf{Fe} & \textbf{Cu} & \textbf{Mn} & \textbf{Mg} & \textbf{Zn} & \textbf{Ti} & \textbf{Sr} & \textbf{P} & \textbf{Other} \\
\midrule
min. & 9.5 & -- & -- & 0.5 & 0.1 & -- & 0.04 & 0.010 & -- & -- \\
max. & 11.5 & 0.15 & 0.03 & 0.8 & 0.5 & 0.07 & 0.15 & 0.025 & 0.001 & 0.10 \\
\bottomrule
\end{tabularx}
\end{table}
Within this range, a magnesium content of \SI{0.283}{\percent} was chosen for the present study, allowing targeted adjustment of mechanical properties. The casting trials were performed on a \textit{B\"{u}hler Evolution B53D} die casting machine with a locking force of \SI{5250}{\kilo\newton}. The process parameters reflected typical industrial \gls{hpdc} practice. The casting had a total weight of \SI{1.07}{\kilogram}, with the plate component itself contributing \SI{0.25}{\kilogram}. A cycle time of approximately \SI{50}{\second} was maintained, with the plunger operating in two stages. During the first phase, a velocity of \SI{0.5}{\metre\per\second} was applied to gently push the melt through the shot sleeve, followed by a second phase acceleration to \SI{2.25}{\metre\per\second} for mold filling. An intensification pressure of \SI{700}{\bar} was then applied to ensure complete cavity filling and compensate for solidification shrinkage. The melt was injected at \SI{720}{\degreeCelsius}, with the die maintained at a chamber temperature of \SI{250}{\degreeCelsius}. The melt quality, assessed by the density index, was \SI{0.7}{\percent}, indicating low hydrogen content and high quality feeding conditions. Following solidification, a T6 heat treatment was applied to enhance strength and ductility. This procedure consisted of solution annealing at \SI{490}{\degreeCelsius} for \SI{2}{\hour}, including a controlled heating phase of \SI{1}{\hour}, immediate water quenching at room temperature, and subsequent artificial aging at \SI{160}{\degreeCelsius} for \SI{2}{\hour}. Together, the defined alloy chemistry, process parameters, and heat treatment establish the metallurgical and thermal framework for analyzing porosity formation and filling behavior in the \gls{hpdc} plates.

\subsubsection{Photogrammetric analysis of flow patterns}
\label{sssec:photo}
The cast plates were evaluated for internal defects and surface flow patterns to enable a direct comparison between numerically predicted flow fields and experimentally observed results, thereby validating the numerical \gls{cfd} framework applied in this work. The three defined evaluation regions (\gls{er}), shown in Fig.~\ref{fig:platte} (\gls{ir}, \gls{mr}, \gls{vr}), were selected to capture areas where pronounced variations in flow behavior and defect susceptibility were expected. Flow patterns were analyzed using photogrammetric measurements, which experimentally characterize flow-induced surface features on the castings. These features arise from the interaction of the molten metal with release agent residues and oxide layers during mold filling. High-resolution images were acquired using a \textit{Sony Alpha 7R\,III} camera equipped with a 42\,megapixel sensor. A total of ten \SI{3}{\milli\metre}-thick \gls{hpdc} plates were produced under the process conditions described in the previous sections. For each plate, the left and right halves relative to the central symmetry plane were evaluated separately by mirroring one side, resulting in a dataset of twenty images. The images capture surface features in detail, preserving the imprints left by the advancing melt front during cavity filling. For quantitative analysis, each image was converted into a grayscale intensity map, where higher intensity values correspond to brighter, more reflective surface regions. Such regions are interpreted as areas where the lubricant layer was thinned or removed, oxide films were modified, or the surface microtexture was altered due to locally increased shear and turbulent impingement. After normalization and averaging of the grayscale intensity data from all twenty images, a representative intensity distribution was obtained, highlighting zones that consistently experienced elevated mechanical or thermal loading during mold filling. 

\subsubsection{X-ray computed tomography}
\label{sssec:xray}
To complement the surface analysis, inner defects were characterized using X-ray computed tomography. In total, 30 bar-shaped specimens were milled out of the casted plates, as depicted in Figure~\ref{fig:platte}. All scans were performed using a \textit{General Electric Phoenix V|tome|x S 240} CT scanner with an acceleration voltage of $U_B = \SI{200}{\kilo\volt}$, a beam current of $I_B = \SI{150}{\micro\ampere}$, and an exposure time of $t_{exp} = \SI{220}{\second}$. The resulting voxel edge length was $vx = \SI{33}{\micro\meter}$. Each 3D \gls{ct} volume was reconstructed from 1441 2D greyscale images and analyzed for porosity using \textit{Volume Graphics VGStudioMax 3.3} and the pore detection algorithm \textit{VGDefx}.

\subsection{Numerical methods}
The numerical workflow followed established practices in \gls{cfd} for casting processes and consisted of three principal stages: geometry preparation, spatial discretization, and numerical solution with subsequent postprocessing. In the first stage, the computational domain was created from the CAD model of a \SI{3}{\milli\metre} flat plate mold, designed to capture the essential hydrodynamic features of the HPDC process without the added complexity of three-dimensional tooling details (Fig.~\ref{fig:CADmodel}). The second stage involved spatial discretization, where the domain was divided into finite control volumes using native \texttt{OpenFOAM} utilities. The mesh was refined in regions of expected high gradients, such as the ingate and the advancing flow front, to ensure accurate resolution of interface dynamics. Finally, the governing equations were solved using the \texttt{compressibleInterFoam} solver in \texttt{OpenFOAM} v8, which is well suited for compressible multiphase flows \citep{ivanov2021}. The solver employs the volume-of-fluid (\gls{vof}) approach to capture the liquid-gas interface \citep{hirt1981,okagaki2021} and enables the tracking of entrained air pockets as well as the evolving flow front during mold filling (Fig.~\ref{fig:figureH}-\ref{fig:figureJ}). The gas phase was initialized at ambient conditions ($p=1013\ \text{mbar}$) and treated as compressible, such that the subsequent rise in cavity pressure emerged naturally from the coupled filling calculation. Results were visualized in \texttt{ParaView}, facilitating quantitative and qualitative assessment of velocity fields, pressure distributions, and free surface behavior.

The governing equations (Section \ref{sec:gov}) were implemented in the \texttt{compress\-ibleInterFoam} solver, available within the \texttt{OpenFOAM} framework, which is designed for the simulation of compressible, multiphase flows and is therefore well suited for modeling \gls{hpdc} processes \citep{ivanov2021}. In contrast to incompressible solvers, \texttt{compressibleInterFoam} incorporates compressibility effects, which can significantly influence flow behavior during high speed mold filling and subsequent solidification. By resolving the interactions between molten metal and the entrained gas phase within the mold, the solver facilitates detailed analyses of phenomena such as air entrapment, turbulence, and pressure fluctuations, which are critical for predicting the formation of defects including porosity and oxide inclusions. In \texttt{compressibleInterFoam}, the compressible Navier-Stokes equations are solved in conjunction with the \gls{vof} method \citep{hirt1981} to capture the evolution of the metal-gas interface. The formulation is based on three main conservation laws: mass, momentum, and energy. Mass conservation is enforced through the continuity equation,
\begin{equation}
\frac{\partial \rho}{\partial t} + \nabla \cdot (\rho \, \mathbf{u}) = 0 ,
\end{equation}
where \gls{sym:rho} denotes the fluid density and \gls{sym:u} is the velocity vector. The conservation of momentum is described by
\begin{equation}
\frac{\partial (\rho \, \mathbf{u})}{\partial t} + \nabla \cdot (\rho \, \mathbf{u} \, \mathbf{u}) 
= -\nabla p + \nabla \cdot \left[ \mu_{\text{eff}} \left( \nabla \mathbf{u} + \nabla \mathbf{u}^T \right) \right] + \rho \, \mathbf{g} ,
\end{equation}
where \gls{sym:p} represents the pressure, \gls{sym:mueff} is the effective dynamic viscosity (including turbulent contributions), and \gls{sym:g} is the gravitational acceleration vector. Thermal effects are incorporated via the energy equation,
\begin{equation}
\frac{\partial (\rho h)}{\partial t} + \nabla \cdot (\rho \, \mathbf{u} \, h) 
= \frac{\partial p}{\partial t} + \nabla \cdot (\lambda_{\text{eff}} \nabla T) + \Phi ,
\end{equation}
where \gls{sym:h} denotes the specific enthalpy, \gls{sym:lameff} is the effective thermal conductivity, \gls{sym:T} is the temperature, and \gls{sym:Phi} accounts for viscous dissipation. Together, these equations form the basis of the solver, which incorporates \gls{rans} based turbulence models, such as the $k$-$\varepsilon$ closure \citep{launder1974}, to simulate turbulent multiphase flows under compressible conditions. The \gls{vof} method is employed to capture transient metal-gas interfaces with high fidelity, enabling detailed analysis of air entrapment, oxide film formation, and pressure fluctuation effects in \gls{hpdc} \citep{hirt1981,okagaki2021}. The choice of \texttt{compressibleInterFoam} in the present study was motivated by the need to accurately capture both compressibility effects and detailed free surface dynamics in \gls{hpdc}. Unlike incompressible solvers, this approach allows for the representation of pressure wave propagation, density variations, and gas phase compression, all of which influence defect formation. By combining the compressible Navier-Stokes formulation with \gls{vof} based interface capturing and turbulence modeling \citep{launder1974,hirt1981,okagaki2021}, the solver provides a robust framework for simulating the complex, transient interactions that govern defect development in industrial \gls{hpdc} operations.

As outlined in the introduction, water was employed as the working fluid to represent molten aluminum. This substitution, widely adopted in both experimental and numerical casting studies~\cite{yuan2008,javurek2020,viswanath2017,renukananda2015}, enables the capture of dominant hydrodynamic effects while reducing computational cost and uncertainty in thermophysical data. Table~\ref{tab:fluid_comparison} compares the key properties of water and aluminum. Their dynamic viscosities are of the same order of magnitude, ensuring comparable Reynolds number regimes under representative velocities and cavity dimensions. Although differences exist in density and surface tension, these primarily affect absolute pressure levels and interface stability rather than the qualitative mechanisms of splashing, rupture, and void formation. Accordingly, water provides a controlled and reliable medium for studying filling kinematics, free surface activity, and entrainment mechanisms in \gls{vof} based simulations.

\begin{table}[htb]
\newcolumntype{K}{>{\centering\arraybackslash}X}
\centering
\small
\caption{Thermophysical properties of water at \SI{20}{\degreeCelsius} and molten Aluminum at \SI{720}{\degreeCelsius} \cite{incropera2011fundamentals, mills2002thermophysical}.}
\label{tab:fluid_comparison}
\begin{tabularx}{\textwidth}{l K K K K}
\toprule
\textbf{Property} & \textbf{Symbol} & \textbf{Unit} & \textbf{Water} & \textbf{Aluminum} \\
\midrule
Density & \gls{sym:rho} & \SI{}{\kilogram\per\meter\cubed} & 998.2 & 2375 \\
Dynamic viscosity & $\mu$ & \SI{}{\pascal\second} & $1.002 \times 10^{-3}$ & $1.2 \times 10^{-3}$ \\
Kinematic viscosity & $\nu$ & \SI{}{\meter\squared\per\second} & $1.004 \times 10^{-6}$ & $5.05 \times 10^{-7}$ \\
Surface tension & $\sigma$ & \SI{}{\newton\per\meter} & 0.0728 & 0.90 \\
Specific heat capacity & $c_p$ & \SI{}{\joule\per\kilogram\per\kelvin} & 4182 & 1080 \\
Thermal conductivity & $k$ & \SI{}{\watt\per\meter\per\kelvin} & 0.598 & 80 \\
Speed of sound & $c$ & \SI{}{\meter\per\second} & 1482 & 4320 \\
\bottomrule
\end{tabularx}
\end{table}

\subsection{Computational setup and performance considerations}
The simulations were executed on the \gls{hpc} cluster \emph{Meggie} at Friedrich-Alexander-Universität Erlangen-Nürnberg using OpenFOAM. To enhance computational efficiency, the domain consisting of 2.6 million cells was decomposed into 64 subdomains with the \texttt{scotch} method \cite{wang2012}, which minimizes communication overhead by reducing interfacial areas between partitions. This strategy improved scalability and reduced wall clock time for large scale runs. The case was executed on 64 processors distributed across eight nodes, with eight physical tasks per node. Each compute node was equipped with Intel Xeon E5-2630v4 Broadwell processors, comprising two chips with ten physical cores each, resulting in 20 cores per node. The processors operated at a base clock speed of 2.2~GHz with a shared cache of 25~MB, and each node provided 64~GB of RAM. Parallelization employed a distributed memory paradigm (MPI) with a hybrid communication model suited for medium to high interprocess communication demands.
Boundary conditions included prescribed constant inlet velocities in the range of 0.05 to 2.5~m/s, no-slip conditions at rigid mold walls, and a sealed outlet to enforce mass conservation within the cavity. The upper range of inlet velocities is representative of the industrial casting conditions considered in this study, as the real component was manufactured at an inlet velocity of \gls{sym:vin}$=\SI{2.25}{\metre\per\second}$. For this reason, this operating point was examined in detail to assess numerical performance characteristics such as Courant number behavior, residual convergence, and wall-clock time. Under these conditions, the runtime with the $k$-$\varepsilon$ turbulence model was 46,497~s, compared to 36,860~s for the corresponding laminar case. The observed difference reflects the additional transport equations required by turbulence modeling and highlights its performance impact in \gls{hpdc} flow predictions. After time integration, the distributed solution fields were reconstructed into a unified dataset. Postprocessing placed particular emphasis on the liquid-gas interface, as this directly governs the occurrence of entrapped gas and the evolution of free surface instabilities. Unlike conventional fluid-domain visualizations, the analysis focused on the advancing free surface, thereby enabling a direct link between numerical predictions and experimentally observable flow features.

\subsection{Evaluation metrics for mold filling}
To systematically evaluate mold filling in \gls{hpdc}, we introduce three complementary field based metrics: \gls{tifsa}, \gls{tmvf}, and \gls{tivf}. While all are derived from the transient multiphase \gls{vof} solution, each targets a different physical mechanism relevant to defect formation, as described in the following.

\subsubsection{Time-Integrated free surface Area (TIFSA)}
\gls{tifsa} quantifies the cumulative liquid-gas interface area present during the entire filling phase, condensing this information into a single scalar measure that also captures oxidation related risks linked to free surface exposure (Fig.~\ref{fig:figureK}, Figure~\ref{fig:figureL}). Physically, it reflects the degree of oxidation risk in aluminum alloys, which arises because oxide films form essentially instantaneously once a fresh surface is exposed to air. Even very short exposure events therefore contribute to bifilm creation, although the resulting oxide may remain extremely thin. Prolonged or repeated exposure, on the other hand, allows oxide thickening and increases the likelihood that films are folded back into the melt, where they can act as crack initiators or pore nucleation sites. By integrating both surface area and exposure duration, \gls{tifsa} captures the combined effect of interface generation and persistence, providing a process specific fingerprint of surface exposure. This criterion is particularly relevant in \gls{hpdc}, where the free surface is continually created and transported by splashing, jet impingement, or large scale flow separation. High \gls{tifsa} values therefore indicate filling conditions that favor extensive oxide formation and entrainment. Reducing the complex, transient interface evolution to a single quantitative value makes it possible to compare different operating conditions, alloys, or geometries in a consistent manner. In this sense, \gls{tifsa} provides both a graphical interpretation (through the evolving free surface) and a global numerical indicator that can be used to characterize and rank processes with respect to oxidation risk. Formally, if \gls{sym:as}$(t)$ denotes the instantaneous free surface area reconstructed from the \gls{vof} field (e.g., the $\alpha=0.5$ isosurface) and $t_{end}$ the last timestep available, the global criterion \gls{sym:zeta_tifsa} is defined as:
\begin{equation}
  \zeta_{\mathrm{TIFSA}} =
  \int_{0}^{t_{\mathrm{end}}} A_{\mathrm{s}}(t)\,\mathrm{d}t,
  \label{eq:TIFSA}
\end{equation}
and its discrete evaluation is given by
\begin{equation}
  \zeta_{\mathrm{TIFSA}} \approx
  \sum_{k=1}^{N_t} A_{\mathrm{s}}(t_k)\,\Delta t_k .
\end{equation}

\subsubsection{Temporal Mean Volume Fraction (TMVF)}
\gls{tmvf} is a pointwise, time-averaged indicator that condenses the local history of liquid occupation during filling into a single scalar value at each fixed position. This compact formulation provides a robust footprint of filling continuity and exposure history across the entire domain. The interpretation of \gls{tmvf} is most informative in its spatial distribution. Broad plateaus with values close to 1 indicate regions that became liquid-filled early and remained continuously wetted. These zones suggest robust filling and a low likelihood of persistent air. In contrast, pockets with low \gls{tmvf} mark areas of late filling or intermittent wetting, where voids may persist. Narrow bands with sharp \gls{tmvf} gradients, or mottled regions with intermediate values, reveal repeated front passage, local recirculation, or surface re-exposure. Such features are signatures of vigorous mixing, even though \gls{tmvf} itself does not resolve turbulence. In practice, these gradient bands align with splash and impingement regions and correlate with defect prone areas observed experimentally. \gls{tmvf} depends on the chosen evaluation horizon $t_{\mathrm{end}}$, as it condenses filling history up to that time. As a time-averaged measure, it may underrepresent very brief gas incursions unless such events recur. It is worth noting that, due to its stationary and averaged nature, \gls{tmvf} cannot capture the transient migration of entrained air pockets or the detailed movement of the filling front. Formally, \gls{sym:zeta_tmvf} is defined from the local liquid volume fraction $\alpha(\mathbf{x},t)$ as
\begin{equation}
  \zeta_{\mathrm{TMVF}}(x,y)=\frac{1}{t_{\mathrm{end}}}\int_{0}^{t_{\mathrm{end}}}\alpha(x,y,t)\,\mathrm{d}t,
  \label{eq:TMVF}
\end{equation}
with the discrete form
\begin{equation}
  \zeta_{\mathrm{TMVF}}(x,y)\approx\frac{1}{t_{\mathrm{end}}}\sum_{k=1}^{N_t}\alpha(x,y,t_k)\,\Delta t_k.
\end{equation}

\subsubsection{Time-Integrated Volumetric Flow (TIVF)}
\label{ssec:tivf}

The \gls{tivf} metric is introduced to characterize the cumulative flow exposure of surface-adjacent regions during mold filling and thereby quantify the intensity of surface overrun. Elevated \gls{tivf} values mark zones repeatedly subjected to liquid impingement, enhanced local shear, or frequent free surface passage, all of which are relevant in \gls{hpdc} due to their potential influence on lubricant removal, localized thermal loading, and surface alteration. Moreover, \gls{tivf} serves as the numerical counterpart to the experimentally derived grayscale fields presented later in this work, where persistent contrast patterns in averaged surface images are interpreted as signatures of preferential flow exposure. In its formal volumetric definition, \gls{sym:zeta_tivf} represents the time-integrated volumetric flux across a surface patch consisting of a finite number of faces. For a face \(f\) with area \gls{sym:a_f}, outward normal vector \gls{sym:n}$_f$, and local velocity \gls{sym:u}$_f(t)$, the continuous expression reads

\begin{equation}
\zeta_{\mathrm{TIVF}} =
\int_{0}^{t_{\mathrm{end}}} \int_{A}
\bigl[\mathbf{u}(t)\cdot\mathbf{n}\bigr]\,\mathrm{d}A\,\mathrm{d}t ,
\end{equation}

which yields a volumetric measure. In this formulation, \gls{sym:u}$(t)$ denotes the local fluid velocity vector at time \(t\), \gls{sym:n} the outward-pointing unit normal vector of the considered surface patch \(A\), and \(t_{\mathrm{end}}\) the end time of the filling stage. The face areas \gls{sym:a_f} depend on the local mesh resolution and therefore vary spatially. Because the present analysis focuses on relative spatial distributions of surface exposure rather than absolute volumetric magnitudes, and because nonuniform mesh sizes would introduce an additional layer of variability into the interpretation, an area-independent formulation was adopted. The evaluated \gls{tivf} field thus corresponds to the temporal integral of the normal velocity component,

\begin{equation}
Z_{\mathrm{TIVF}}(x,y) =
\int_{0}^{t_{\mathrm{end}}}
\bigl[\mathbf{u}(x,y,t)\cdot \mathbf{n}(x,y)\bigr]\,\mathrm{d}t ,
\label{eq:localTIVF}
\end{equation}

which has units of length. Here, \((x,y)\) denotes the coordinates of a surface-adjacent evaluation point, and \gls{sym:n}\((x,y)\) the corresponding outward surface normal. This formulation captures the temporal accumulation of normal flow activity while intentionally avoiding the direct influence of varying face areas \gls{sym:a_f}. The resulting field provides a mesh-robust and directly interpretable measure of cumulative surface flow exposure. Its spatial patterns are well suited for comparative analysis and for correlating simulated flow activity with experimental grayscale imprints, where relative variations rather than absolute volumetric values are of primary relevance.

\section{Results}
\label{sec:res}
The following section presents the numerical and experimental findings in the order defined 
by the workflow summarized in the following list. The results are organized to first assess numerical robustness and solver performance, then analyze the role of turbulence modeling in shaping free surface dynamics, and finally validate the computational predictions against \gls{ct} based porosity data 
and water analogue experiments:
\begin{enumerate}
    \item \textbf{Exploration of numerical solver}: Mesh dependence and resolution effects (Fig.~\ref{fig:figureE}); Courant number control and residual monitoring (Figs.~\ref{fig:figureB} and~\ref{fig:figureC}); inlet velocity sweep and resulting cavity pressure response (Fig.~\ref{fig:figureG}); ingate and region-averaged velocity histories used to identify characteristic filling phases (Fig.~\ref{fig:figureO} and~\ref{fig:figureON}).

    \item \textbf{Simulation setup and runs}: Initialization at ambient cavity pressure; prescribed inlet velocities at the gating inlet (Fig.~\ref{fig:figureG}); reference point and evaluation regions defined in the geometry (Fig.~\ref{fig:CADmodel}); termination at $0.7\times10^{8}\,\si{\pascal}$; ten laminar and ten $k$-$\varepsilon$ runs.

    \item \textbf{Free surface and turbulence}: Laminar vs.\ $k$-$\varepsilon$ free-surface morphology during early and intermediate filling (Figs.~\ref{fig:figureH} and~\ref{fig:figureI}); late-stage divergence of interface topology (Fig.~\ref{fig:figureJ}); entrapped air--pressure coupling in \gls{mr} and \gls{vr} (Fig.~\ref{fig:figureM}); quantification of free-surface exposure via \gls{tifsa} (Figs.~\ref{fig:figureK} and~\ref{fig:figureL}).

    \item \textbf{Validation}: Porosity distribution from \gls{ct} (Fig.~\ref{fig:porosity_bar}); simulated \gls{tmvf} field (Fig.~\ref{fig:tmvf}); experimental validation of flow patterns using grayscale intensity and simulated \gls{tivf} (Fig.~\ref{fig:exp_flow}).
\end{enumerate}

\subsection{Exploration of the numerical solver}
\label{sec:explore}
As a first step, the performance of the compressible multiphase solver is examined with respect 
to mesh resolution, timestep control, and stability. This ensures that the computed fields 
reliably capture the relevant flow physics without being dominated by numerical artifacts. The 
observations from this stage provide the basis for selecting suitable parameters for the 
subsequent analyses of filling dynamics and defect formation. 

\subsubsection{Mesh sensitivity and resolution effects}
\label{ssse:mesh}
\begin{figure}[hb]
\footnotesize
\centering
\begin{tabular}{m{1.5mm}*{4}{c}}
\small
\raisebox{2\normalbaselineskip}[0pt][0pt]{\rotatebox[origin=c]{90}{\textbf{Mesh}}} &
\subfloat[]{\includegraphics[width=.21\linewidth]{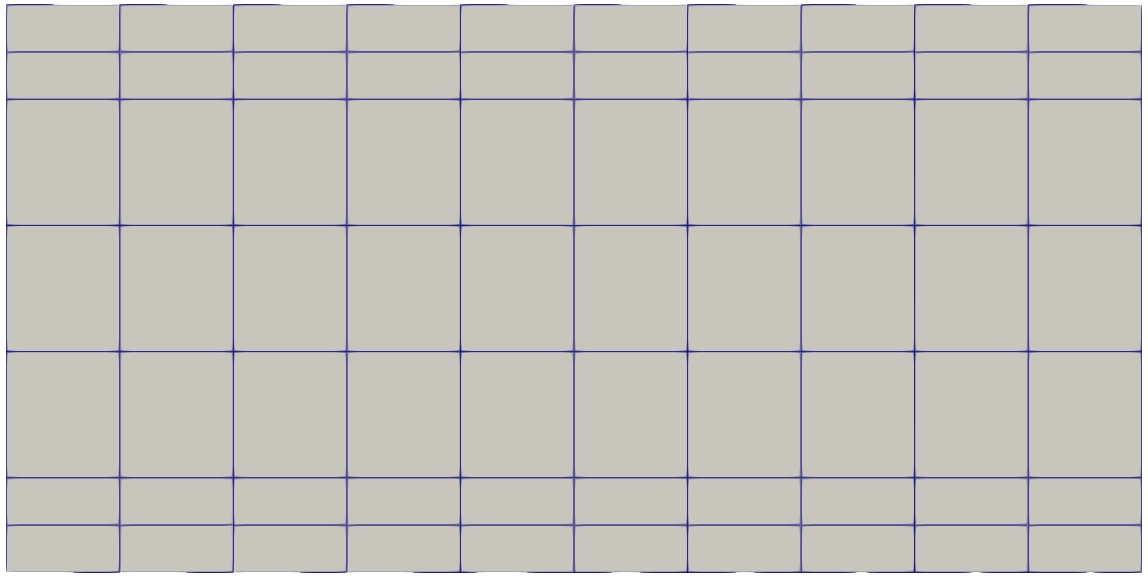}} & 
\subfloat[]{\includegraphics[width=.21\linewidth]{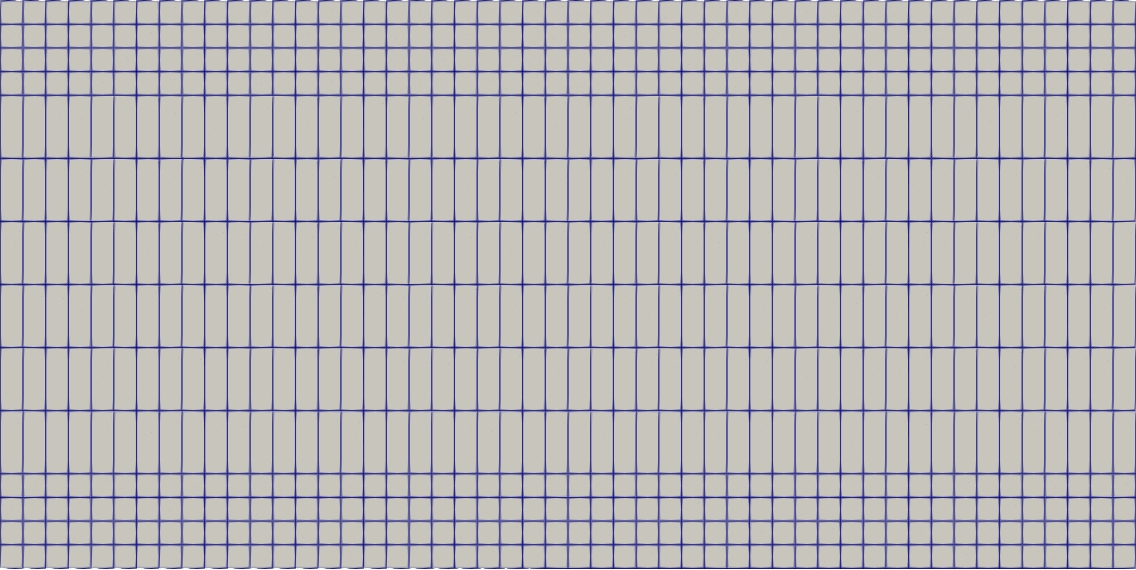}} & 
\subfloat[]{\includegraphics[width=.21\linewidth]{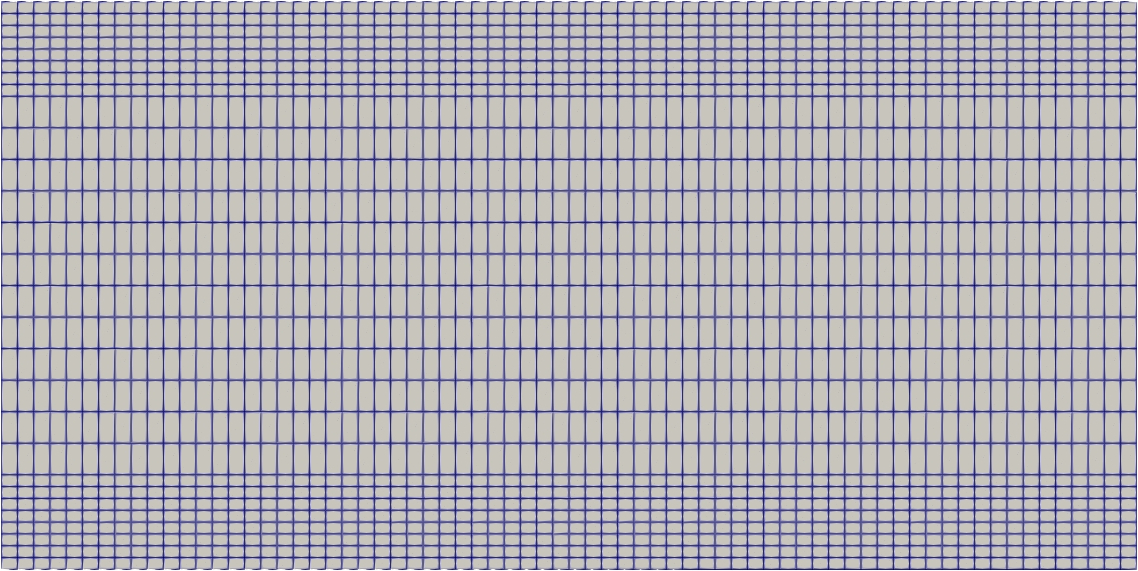}} & 
\subfloat[]{\includegraphics[width=.21\linewidth]{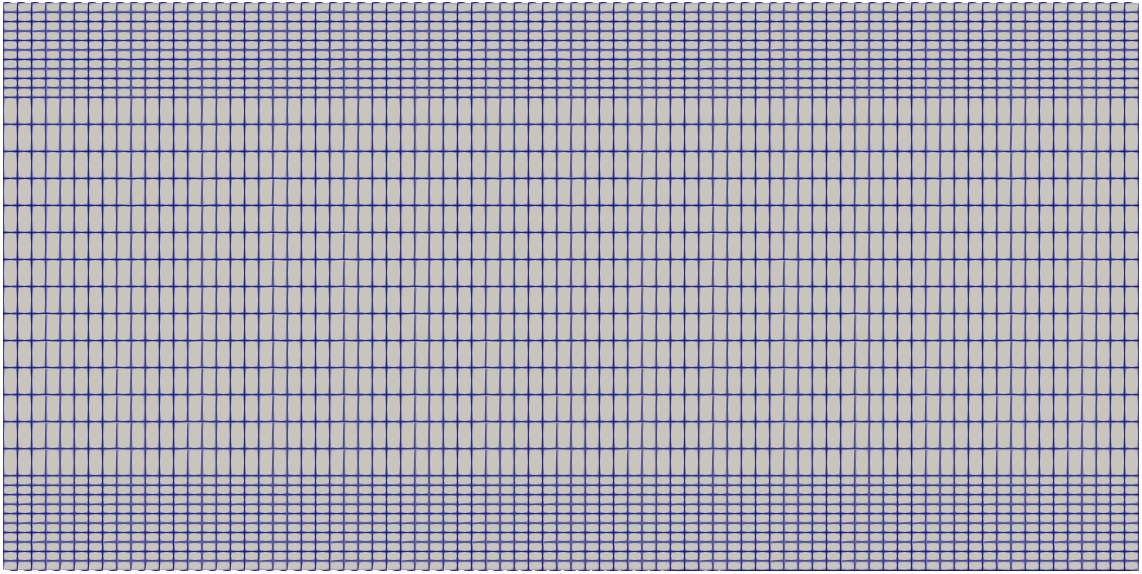}} \\ 
\raisebox{4\normalbaselineskip}[0pt][0pt]{\rotatebox[origin=c]{90}{\textbf{Detail 1}}}  & 
\subfloat[]{\includegraphics[width=.21\linewidth]{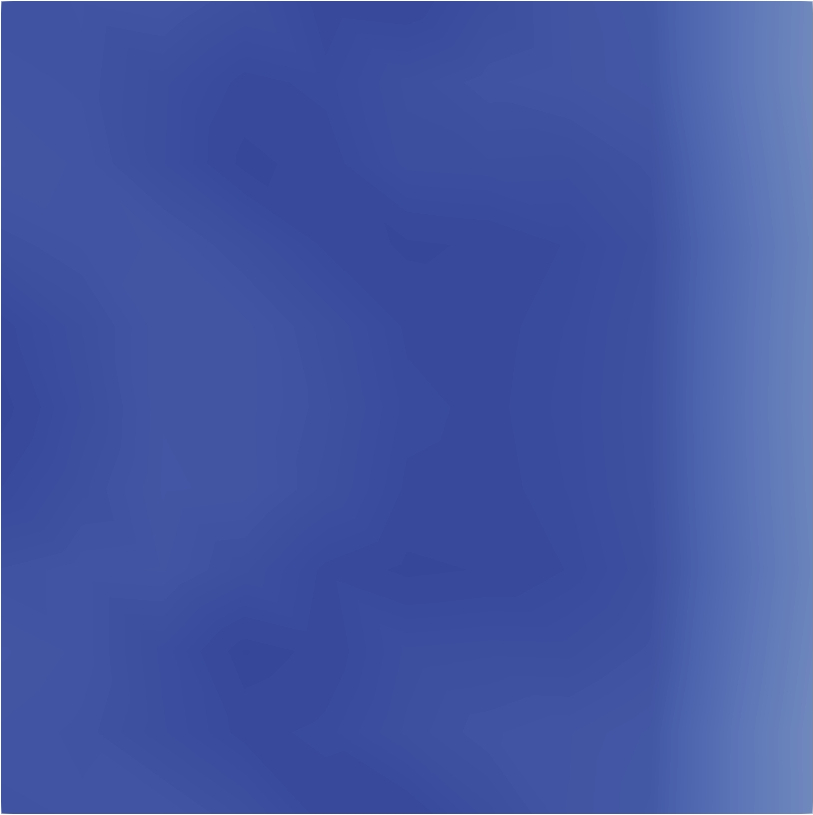}} & 
\subfloat[]{\includegraphics[width=.21\linewidth]{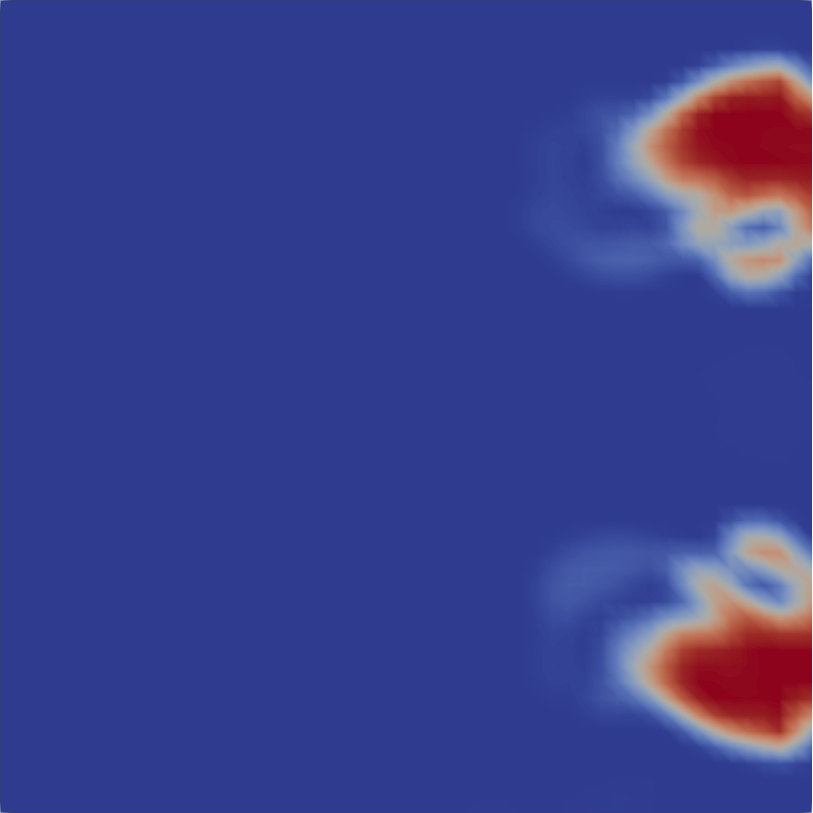}} & 
\subfloat[]{\includegraphics[width=.21\linewidth]{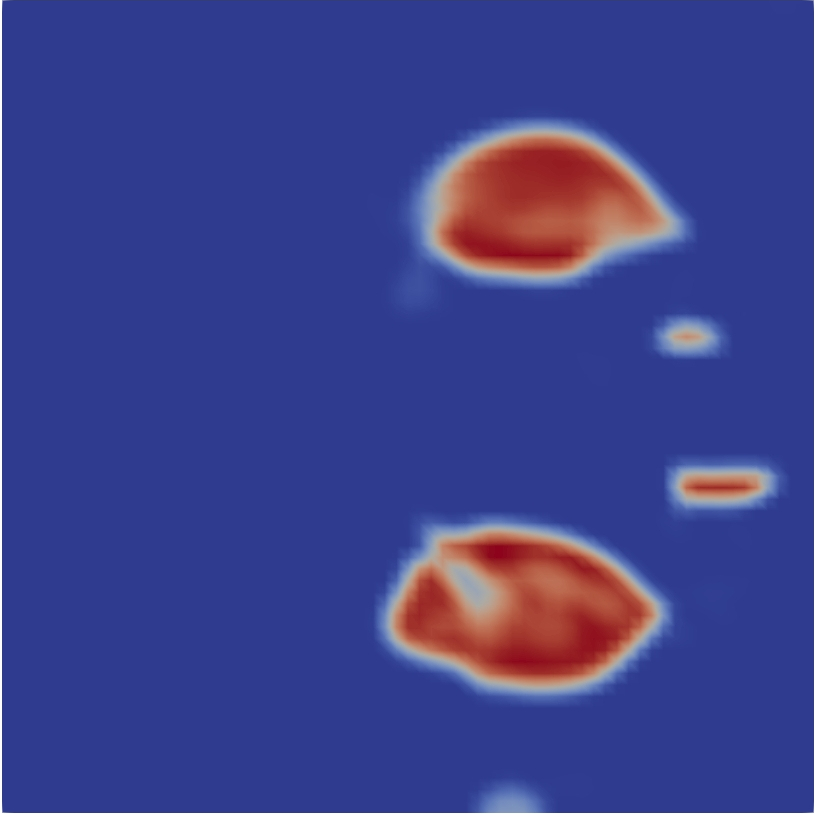}} & 
\subfloat[]{\includegraphics[width=.21\linewidth]{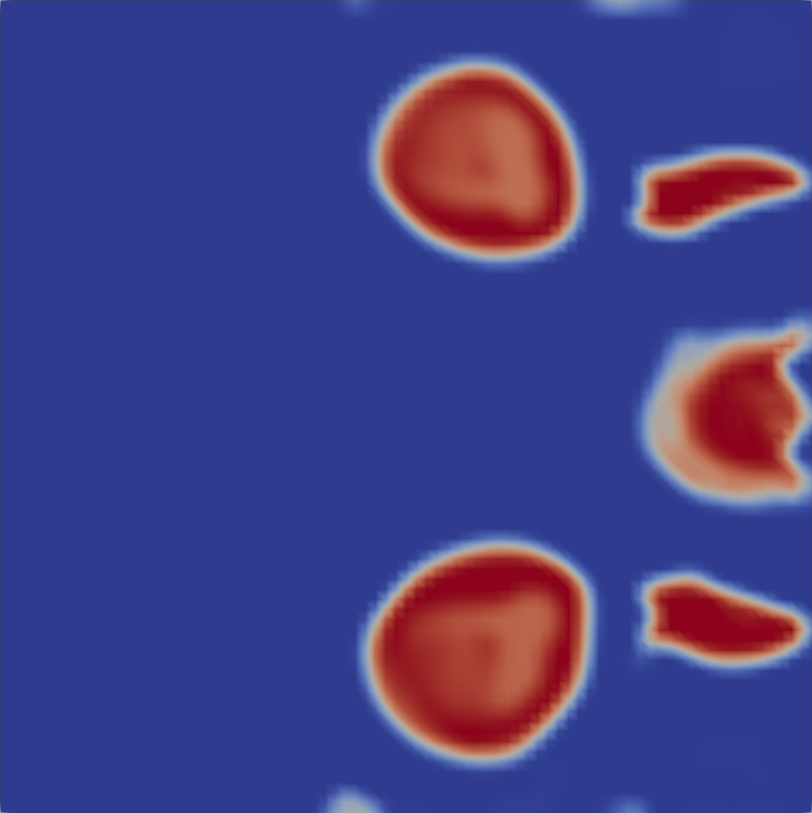}} \\ 
\raisebox{2\normalbaselineskip}[0pt][0pt]{\rotatebox[origin=c]{90}{\textbf{Detail 2}}} & 
\subfloat[]{\includegraphics[width=.21\linewidth]{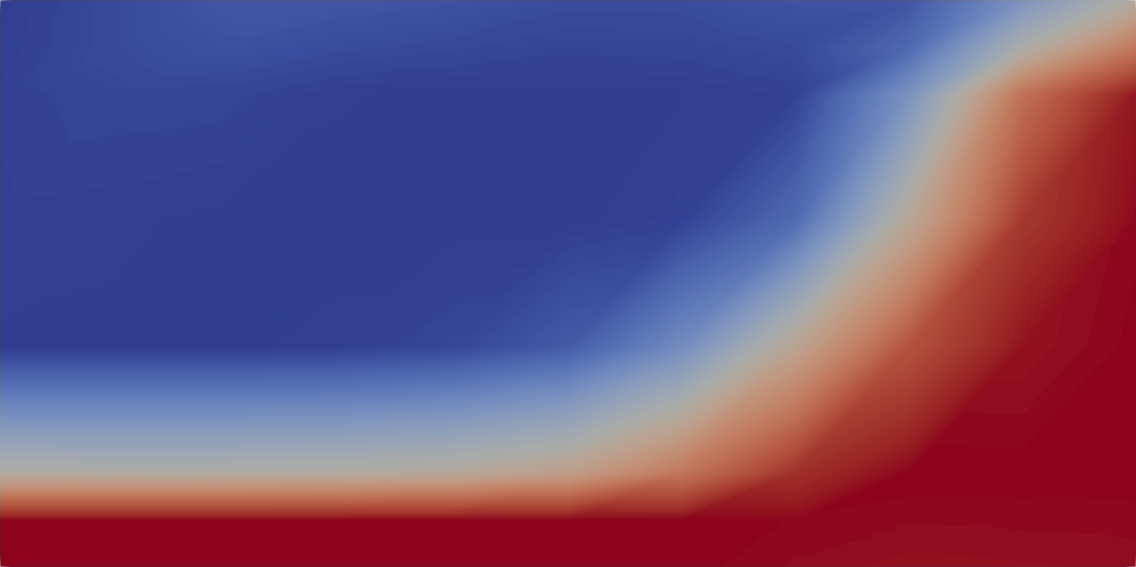}} & 
\subfloat[]{\includegraphics[width=.21\linewidth]{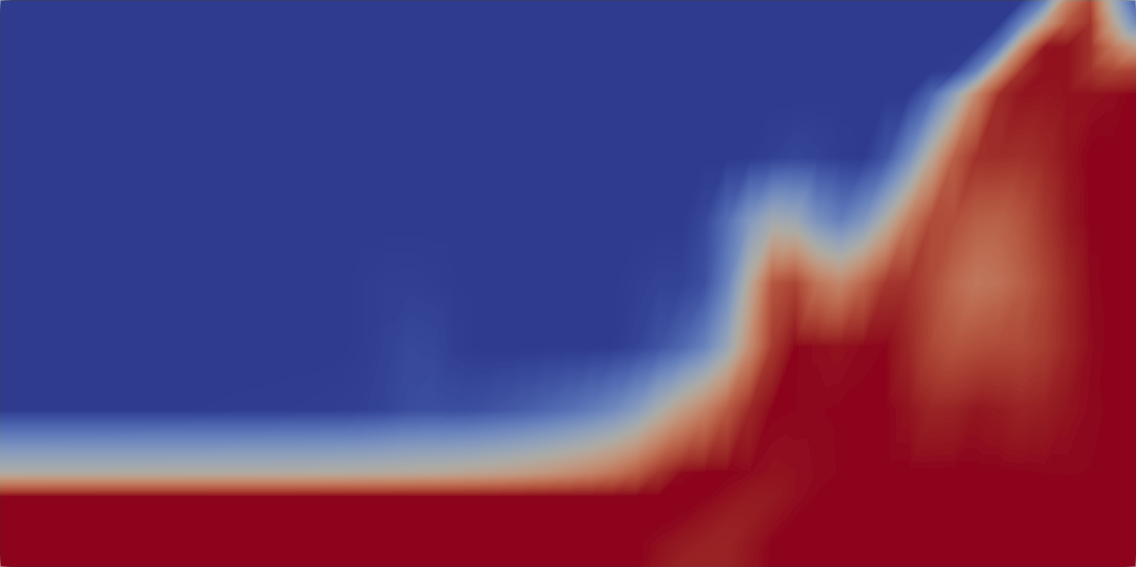}} & 
\subfloat[]{\includegraphics[width=.21\linewidth]{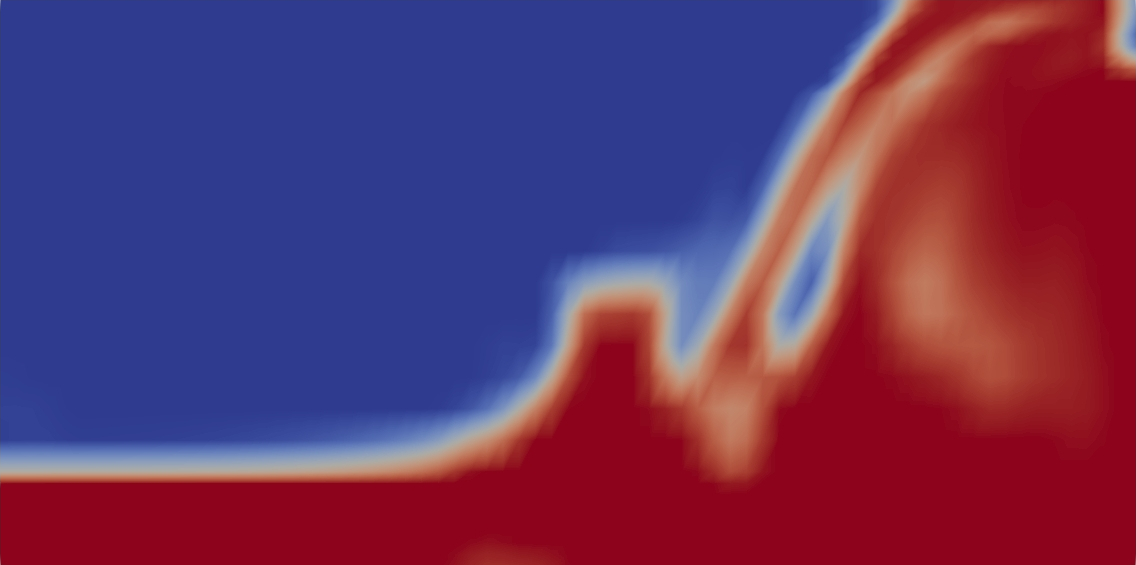}} & 
\subfloat[]{\includegraphics[width=.21\linewidth]{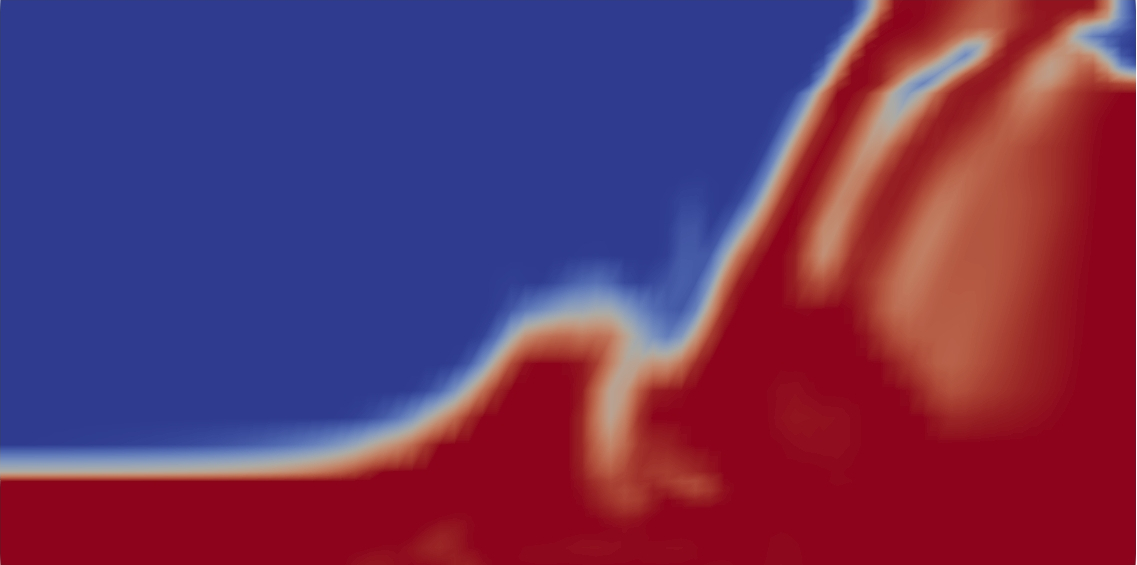}} \\ 
& & & & \\
& \textbf{\#1:} $700/0.7/0.2, 0.79$ & \textbf{\#2:} $35000/230/0.05, 0.89$& \textbf{\#3:} $137200/2404/0.04, 0.60$& \textbf{\#4:} $217600/6473/0.05, 0.99$\\
& & & & \\
\multicolumn{5}{c}{Cells / Runtime [s] / \glstext{cfl} (mean, max) } \\
\end{tabular}
\caption{Mesh (a-d) and multiphase fluid flow for four test cases (\#1-4), in side view (i-l) and top view (e-h) illustrating mesh dependency with increasing mesh resolution}
\label{fig:figureE}
\end{figure}

A mesh sensitivity study was conducted to assess the influence of spatial resolution on the predicted free surface dynamics. The simulations were performed using a \gls{vof}-based multiphase flow model for a  cross section of \SI{10}{\milli\meter} at a constant Weber number of \gls{sym:we}$ = 142.9$. All boundary conditions, material properties, and solver settings were kept identical in order to isolate discretization effects. A formal \gls{gci} analysis was not pursued, as the quantities of interest are predominantly qualitative and interface dominated. Figure~\ref{fig:figureE} shows side and top views of the simulated free surface evolution for four meshes of increasing resolution. The results exhibit a pronounced dependence on mesh resolution. Increasing the mesh density by more than an order of magnitude from Case~1 to Case~2 leads to a visible improvement in interface definition, but remains insufficient to reproduce essential free surface features. Despite stable Courant number levels, the interface remains overly diffused and characteristic interfacial structures are not captured, indicating that numerical stability alone does not guarantee physical fidelity. A central criterion for mesh adequacy is the prediction of flow separation, defined as the detachment of the advancing fluid front from the mold wall. This mechanism increases free surface area and is directly linked to air entrapment and oxide formation in \gls{hpdc}. Flow separation is not observed on the two coarsest meshes, even though the computational cost already increases by more than two orders of magnitude between these cases. This demonstrates that the absence of flow separation is a resolution driven limitation rather than a numerical or temporal one. Both finer meshes (Cases~3 and~4) resolve the onset and evolution of flow separation. Compared to Case~2, Case~3 increases the mesh resolution by approximately a factor of four and captures the relevant free surface topology in close agreement with the finest mesh. Further refinement to Case~4 roughly doubles the number of cells and increases the computational cost by nearly a factor of three, while yielding only minor local differences without affecting the dominant flow behavior. Case~3 is therefore considered mesh-converged for the quantities of interest and is selected for subsequent simulations.
\subsubsection{Solver convergence and residual control}
\label{sssec:solvConv}
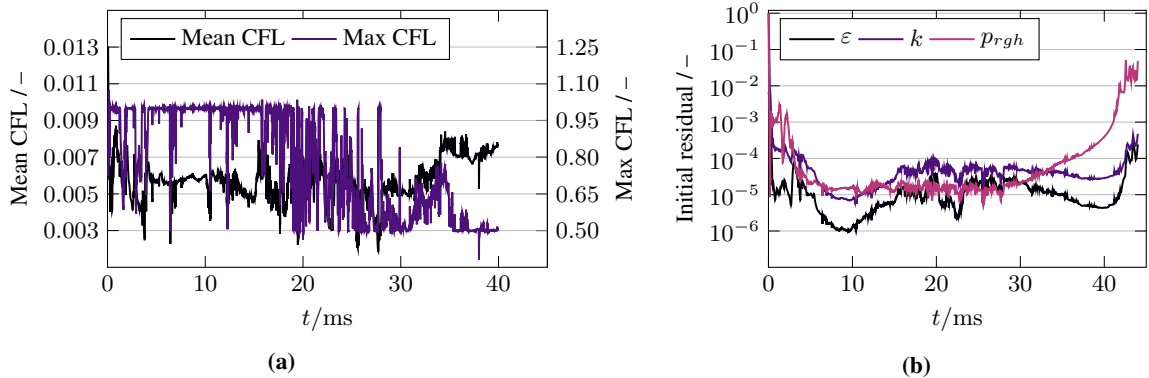
\begin{figure}[hb]
\small
\centering
\begin{subfigure}[t]{.45\linewidth}
\newcommand{\MyMarkerSize}{1.5pt}
\newcommand{\MyLineWidth}{.75pt}
\newcommand{\MyWidth}{1\textwidth}
\newcommand{\MyHeight}{5cm}

\begin{tikzpicture}
\centering
  \begin{axis}[
    width=\MyWidth,
    height=\MyHeight,
    xlabel={$t / \SI{}{\milli\second}$},
    ylabel={Mean CFL / --},
    xmin = 0,
    xmax = 45,     
    ymin = .001,
    ymax = .015, 
    ymajorgrids=true,
    tick align=inside,
    tick style={black},
    font=\small,
    clip=false,
    yticklabel style={
      /pgf/number format/fixed,
      /pgf/number format/precision=3,
      /pgf/number format/fixed zerofill
    },
    scaled y ticks=false,
    ytick={0.003,0.005,0.007,0.009,0.011,0.013},
  ]
    \addplot[
      line width=\MyLineWidth, 
      sharp plot, 
      mark size=\MyMarkerSize,
      color=col1
    ] table [x=time, y=mean_courant, col sep=comma]{simulation_data_FT.dat};
    \label{plot_one}
    \addlegendimage{Mean CFL}
  \end{axis}

  \begin{axis}[
    width=\MyWidth,
    height=\MyHeight,
    xmin = 0,
    xmax = 45,      
    ymin = 0.35,
    ymax = 1.40, 
    axis y line*=right,
    axis x line=none,
    font=\small,
    ylabel={Max CFL / --},
    yticklabel style={
      /pgf/number format/fixed,
      /pgf/number format/precision=2,
      /pgf/number format/fixed zerofill
    },
    scaled y ticks=false,
    ytick={0.5,0.65,0.8,0.95,1.1,1.25},
    legend pos= north west,
    legend columns=3,
    legend cell align=left,
  ]
  \addlegendimage{/pgfplots/refstyle=plot_one}\addlegendentry{Mean CFL}
    \addplot[
      line width=\MyLineWidth, 
      sharp plot, 
      mark size=\MyMarkerSize,
      color=col2
    ] table [x=time, y=max_courant, col sep=comma]{simulation_data_FT.dat};
    \addlegendentry{Max CFL}
  \end{axis}
\end{tikzpicture}
    \vspace{-1.3em}
    \caption{}
    \label{fig:figureB} 
\end{subfigure}
\hspace{4em}
\begin{subfigure}[t]{.4\linewidth}
    \newcommand{\MyMarkerSize}{1.5pt}
\newcommand{\MyLineWidth}{.75pt}
\newcommand{\MyWidth}{1\textwidth}
\newcommand{\MyHeight}{5cm}

\begin{tikzpicture}
\centering
  \begin{axis}[
    width=\MyWidth,
    height=\MyHeight,
    xlabel={$t / \SI{}{\milli\second}$},
    ylabel={Initial residual / --},
    xmin = 0,
    xmax = 45,     
    ymin = 10e-8,
    ymax = 1.2,
    ymode = log,
    log basis y={10},
    ymajorgrids=true,
    legend pos=north west,
    legend cell align=left,
    legend columns=3,
    tick align=inside,
    tick style={black},
    font=\small,
    clip=false,
    yticklabel style={
      /pgf/number format/fixed,
      /pgf/number format/precision=1,
    },
    scaled y ticks=false,
    ytick={10e-1,10e-2,10e-3,10e-4,10e-5,10e-6,10e-7},
  ]
    \addplot[
      line width=\MyLineWidth, 
      sharp plot, 
      mark size=\MyMarkerSize,
      color=col1,
    ] table [x=time, y=initial_residual_epsilon, col sep=comma]{time_residual_ds50.dat};
    \addlegendentry{$\varepsilon$}
  
   \addplot[
      line width=\MyLineWidth, 
      sharp plot, 
      mark size=\MyMarkerSize,
      color=col2,
    ] table [x=time, y=initial_residual_k, col sep=comma]{time_residual_ds50.dat};
    \addlegendentry{$k$}

    \addplot[
      line width=\MyLineWidth, 
      sharp plot, 
      mark size=\MyMarkerSize,
      color=col3,
    ] table [x=time, y=initial_residual_p_rgh, col sep=comma]{time_residual_ds50.dat};
    \addlegendentry{$p_{rgh}$}   
  \end{axis}
\end{tikzpicture}
    \caption{}
    \label{fig:figureC}
\end{subfigure}
\caption{Numerical stability metrics for the test plate (see Fig.~\ref{fig:CADmodel}): (a) mean and maximum \gls{cfl} number; (b) residual histories of the governing equations.}
\label{fig:metrics}
\end{figure}

The numerical stability and reliability of the simulations were assessed for the reference case corresponding to the industrial process conditions discussed previously, i.e. an inlet velocity of \gls{sym:vin}$=\SI{2.25}{\metre\per\second}$. This case was selected for detailed analysis as it represents the most demanding operating point in terms of flow velocity, interface dynamics, and numerical stiffness. As shown in Fig.~\ref{fig:figureB}, the maximum \gls{cfl} number across the computational domain remained consistently below unity throughout the entire filling process, while the mean \gls{cfl} values were approximately one order of magnitude lower. This behavior satisfies the stability requirements for transient \gls{vof} simulations and indicates robust control of the time integration. Excessive Courant numbers are known to deteriorate interface reconstruction and induce spurious oscillations; their absence here confirms the effectiveness of the adaptive timestepping strategy. The timestep size was adjusted dynamically based on the local velocity field and mesh resolution, ensuring that regions of high velocity in refined areas did not violate the stability criterion. This approach enabled stable and accurate simulations without overly conservative global timestep restrictions, thereby balancing numerical efficiency and solution fidelity. Convergence behavior was further evaluated by monitoring the residuals of the governing equations, shown in Fig.~\ref{fig:figureC}. All monitored residuals remained within the prescribed tolerance limits. A pronounced decrease in residuals is observed during the early stages of mold filling, reflecting rapid solver adaptation to the evolving free surface topology. In the later stages, a moderate increase in residual levels occurs as the flow approaches cavity closure, driven by steep pressure gradients and compressibility effects at the advancing fluid front. These fluctuations remain bounded and do not indicate numerical instability or loss of physical consistency.
\subsubsection{Flow regime sensitivity to inlet velocity}
\label{sssec:flowReg}
\begin{figure}[b]
\small
\centering
\newcommand{\MyMarkerSize}{2pt}
\newcommand{\MyLineWidth}{1pt}
\newcommand{\MyWidth}{1\textwidth}
\newcommand{\MyHight}{5cm}

\begin{tikzpicture}
\centering
\begin{axis}[
    width=\MyWidth,
    height=\MyHight,
    xlabel={t / \SI[per-mode=symbol]{}{\milli\second}},
    ylabel={Pressure $p$ / \SI[per-mode=symbol]{}{\pascal}},
    ymin=0,
    xmin=0.04,
    grid=both,
    legend pos = north east,
    legend style={legend columns=5},
    legend style={draw=black, fill=white, font=\small},
    legend cell align=left,
    xmode=log,
    yticklabel style={
    /pgf/number format/fixed,
    /pgf/number format/precision=1,
    /pgf/number format/fixed zerofill
    },
]

\addplot[
    color=col1, ultra thick, sharp plot,
] table [x=Time, y=p, col sep=comma] {combined_5.dat};
\addlegendentry{$\SI[per-mode=symbol]{0.05}{\meter\per\second}$}

\addplot[
    color=col2, ultra thick, sharp plot,
] table [x=Time, y=p, col sep=comma] {combined_10.dat};
\addlegendentry{$\SI[per-mode=symbol]{0.10}{\meter\per\second}$}

\addplot[
    color=col3, ultra thick, sharp plot,
] table [x=Time, y=p, col sep=comma] {combined_20.dat};
\addlegendentry{$\SI[per-mode=symbol]{0.20}{\meter\per\second}$}

\addplot[
    color=col4, ultra thick, sharp plot,
] table [x=Time, y=p, col sep=comma] {combined_50.dat};
\addlegendentry{$\SI[per-mode=symbol]{0.50}{\meter\per\second}$}

\addplot[
    color=col5, ultra thick, sharp plot,
] table [x=Time, y=p, col sep=comma] {combined_100.dat};
\addlegendentry{$\SI[per-mode=symbol]{1.00}{\meter\per\second}$}

\addplot[
    color=col6, ultra thick, sharp plot,
] table [x=Time, y=p, col sep=comma] {combined_150.dat};
\addlegendentry{$\SI[per-mode=symbol]{1.50}{\meter\per\second}$}

\addplot[
    color=col7, ultra thick, sharp plot,
] table [x=Time, y=p, col sep=comma] {combined_175.dat};
\addlegendentry{$\SI[per-mode=symbol]{1.75}{\meter\per\second}$}

\addplot[
    color=col8, ultra thick, sharp plot,
] table [x=Time, y=p, col sep=comma] {combined_200.dat};
\addlegendentry{$\SI[per-mode=symbol]{2.00}{\meter\per\second}$}

\addplot[
    color=col9,
    mark=*,
    only marks,
    mark options={scale=0.8, solid, draw=black, line width=0.5pt},
    dashed,
    sharp plot, mark size=\MyMarkerSize,
    ultra thick,
] table [x=Time, y=p, col sep=comma] {combined_225.dat};
\addlegendentry{$\SI[per-mode=symbol]{2.25}{\meter\per\second}$}

\addplot[
    color=col10, ultra thick, sharp plot,
] table [x=Time, y=p, col sep=comma] {combined_250.dat};
\addlegendentry{$\SI[per-mode=symbol]{2.50}{\meter\per\second}$}

\end{axis}
\end{tikzpicture}
\caption{Evolution of cavity pressure $p$ for different inlet velocities \gls{sym:vin} with turbulence modeling.}
\label{fig:figureG}
\end{figure}
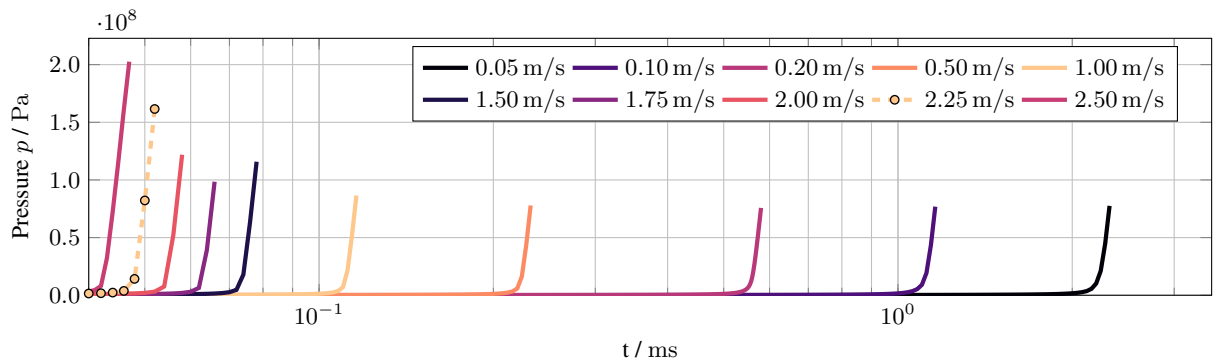

\begin{figure}[tb]
\small
    \centering
    \begin{subfigure}[b]{.45\linewidth}
        \centering
        \newcommand{\MyMarkerSize}{2.5pt}
\newcommand{\MyLineWidth}{1pt}

\begin{tikzpicture}
\begin{axis}[
    width=\linewidth,
    height=5cm,
    xlabel={Time / \SI{}{\milli\second}},
    ylabel={\gls{sym:vg}~/~\SI{}{\meter\per\second}},
    grid=major,
    ymin = 0,
    xmin = 0,
    xmax = 44,
    every mark/.append style={mark size=5pt},  
    legend style={anchor=north,legend columns=2},
    legend style={draw=black, fill=white, font=\small},
    legend pos=north east,
    legend cell align=left,
    scaled y ticks=false,
    yticklabel style={
    /pgf/number format/fixed,
    /pgf/number format/precision=0,
    /pgf/number format/fixed zerofill
    },
]
\addplot[
    color=col1,
    dashed,
    line width=1pt, 
    mark=*, 
    mark options = {solid, draw=black, line width=0.5pt},
    sharp plot, mark size=\MyMarkerSize,
] table[x=Time,y=Point 1,col sep=comma]{u_magnitude_combined.dat};
\addlegendentry{\gls{ep} 1}

\addplot[
    color=col2,
    line width=1pt, 
    mark=square*, 
    dashed,
    mark options = {solid, draw=black, line width=0.5pt},
    sharp plot, mark size=\MyMarkerSize,
] table[x=Time,y=Point 2,col sep=comma]{u_magnitude_combined.dat};
\addlegendentry{\gls{ep} 2}

\addplot[
    line width=1pt, 
    color=col3,
    mark=triangle*, 
    dashed,
    mark options = {solid, draw=black, line width=0.5pt},
    sharp plot, mark size=\MyMarkerSize,
] table[x=Time,y=Point 3,col sep=comma]{u_magnitude_combined.dat};
\addlegendentry{\gls{ep} 3}

\addplot[
    line width=1pt, 
    color=col4,
    mark=diamond*,
    dashed,
    mark options = {solid, draw=black, line width=0.5pt},
    sharp plot, mark size=\MyMarkerSize,
] table[x=Time,y=Point 4,col sep=comma]{u_magnitude_combined.dat};
\addlegendentry{\gls{ep} 4}

\end{axis}
\end{tikzpicture}
        \caption{}
        \label{fig:figureO}
    \end{subfigure}
    \hfill
    \begin{subfigure}[b]{.45\linewidth}
        \centering
        \newcommand{\MyMarkerSize}{2.5pt}
\newcommand{\MyLineWidth}{1pt}

\begin{tikzpicture}
  \begin{axis}[
    width=\linewidth,
    height=5cm,
    xlabel={Time / \SI{}{\milli\second}},
    ylabel={\gls{sym:v_mean} / \SI{}{\meter\per\second}},
    ymin=0, 
    xmin = 0,
    xmax = 44,
    grid=both,
    legend style={anchor=north,legend columns=2},
    legend style={draw=black, fill=white, font=\small},
    legend pos=north east,
    legend cell align=left,
    yticklabel style={
    /pgf/number format/fixed,
    /pgf/number format/precision=1,
    /pgf/number format/fixed zerofill
    },
  ]
    
    \addplot[
      line width=1pt, 
      color=col1,
      mark=*,
      only marks,
      mark options={solid, draw=black, line width=0.5pt},
      dashed,
      sharp plot, mark size=\MyMarkerSize,
    ] table [x=time, y=umagtop, col sep=comma] {Umags_UmagTop.dat};
      \addlegendentry{\gls{vr}}

    \addplot[
    line width=1pt, 
      color=col2,
      mark=square*,
      only marks,
      mark options={solid, draw=black, line width=0.5pt},
      dashed,
      sharp plot, mark size=\MyMarkerSize,
    ] table [x=time, y=umagmid, col sep=comma] {Umags_UmagTop.dat};
    \addlegendentry{\gls{mr}}
    
  \end{axis}

\end{tikzpicture}
        \caption{}
        \label{fig:figureON}
    \end{subfigure}
   \caption{Velocities as a function of time: (a) ingate velocities evaluated at the centroids of the \glspl{ep} along the left side of the gating system, and (b) mean velocities $\bar{v}$ evaluated over the \glspl{ev} in the \gls{mr} and \gls{vr}; compare Fig.~\ref{fig:CADmodel}.}
\end{figure}
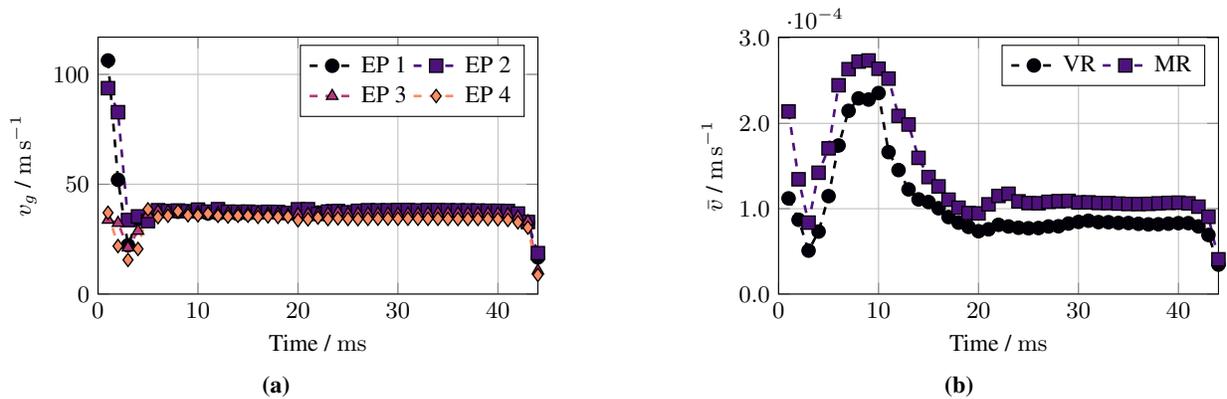

\begin{figure}
\centering 
\subfloat[\centering]{\includegraphics[width=.20\linewidth,trim={0 6cm 0 8cm},clip]{"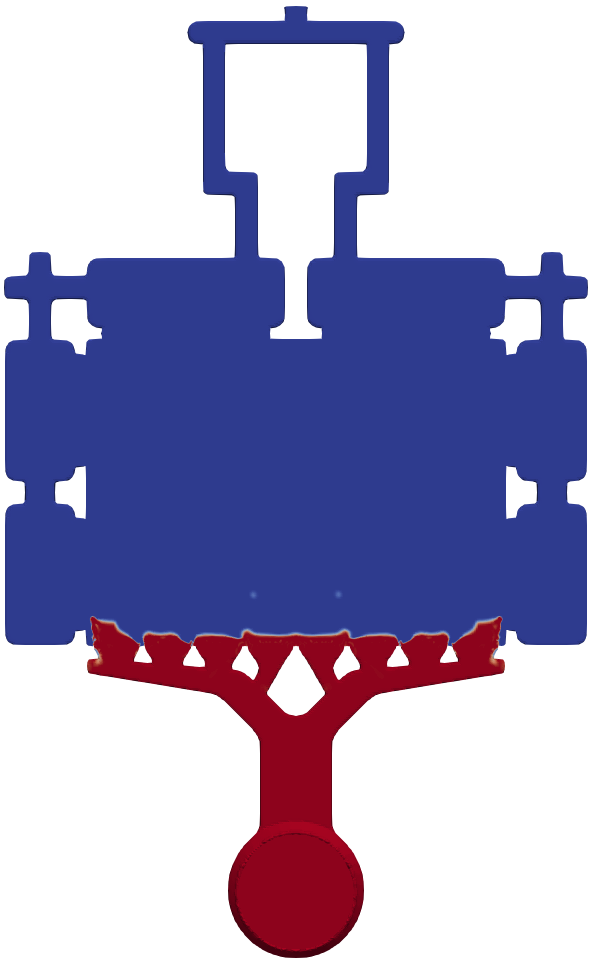"}}\hfill
\subfloat[\centering]{\includegraphics[width=.20\linewidth,trim={0 6cm 0 8cm},clip]{"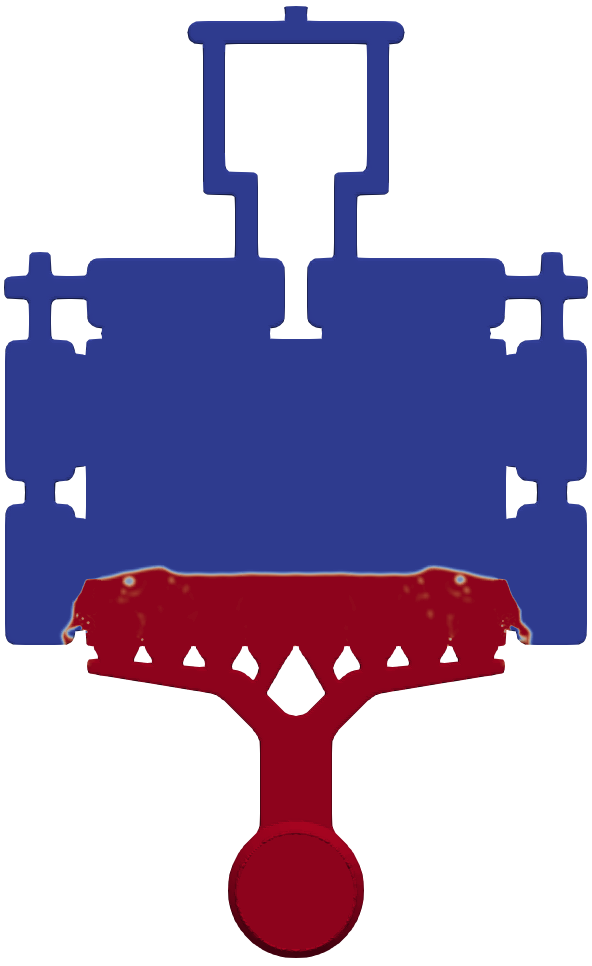"}}\hfill
\subfloat[\centering]{\includegraphics[width=.20\linewidth,trim={0 6cm 0 8cm},clip]{"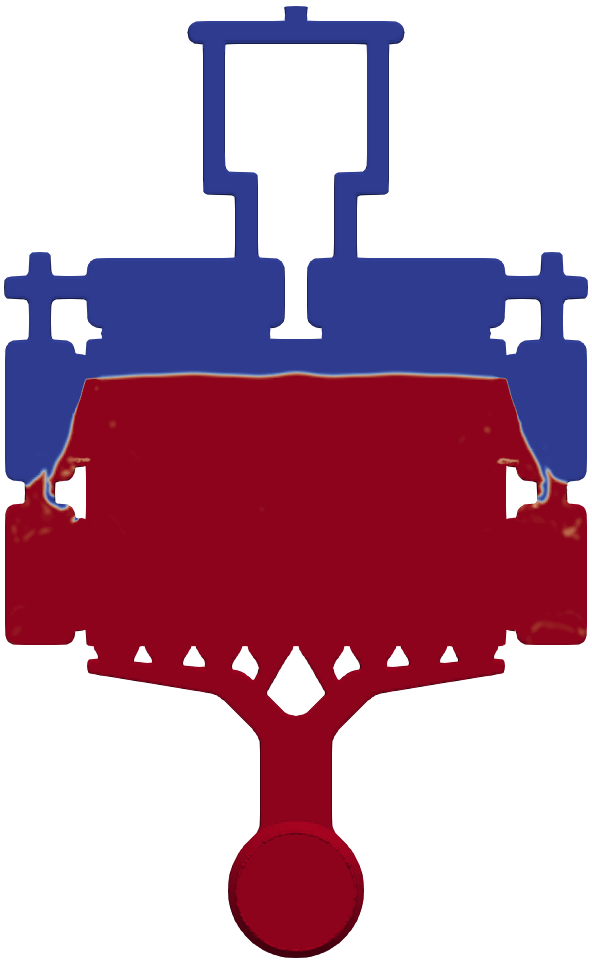"}}\hfill
\subfloat[\centering]{\includegraphics[width=.20\linewidth,trim={0 6cm 0 8cm},clip]{"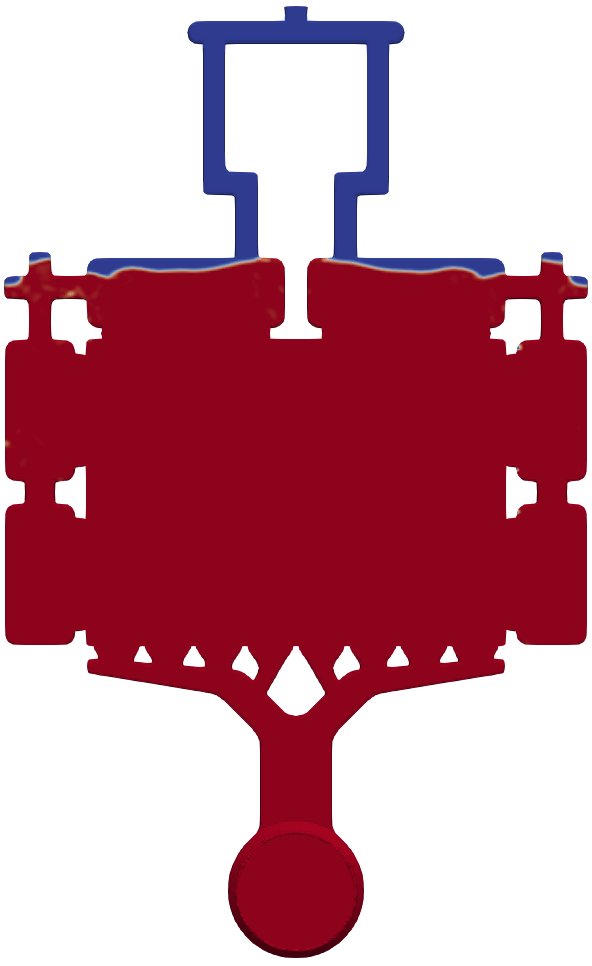"}}\\[0.8em]
\subfloat[\centering]{\includegraphics[width=.20\linewidth,trim={0 6cm 0 8cm},clip]{"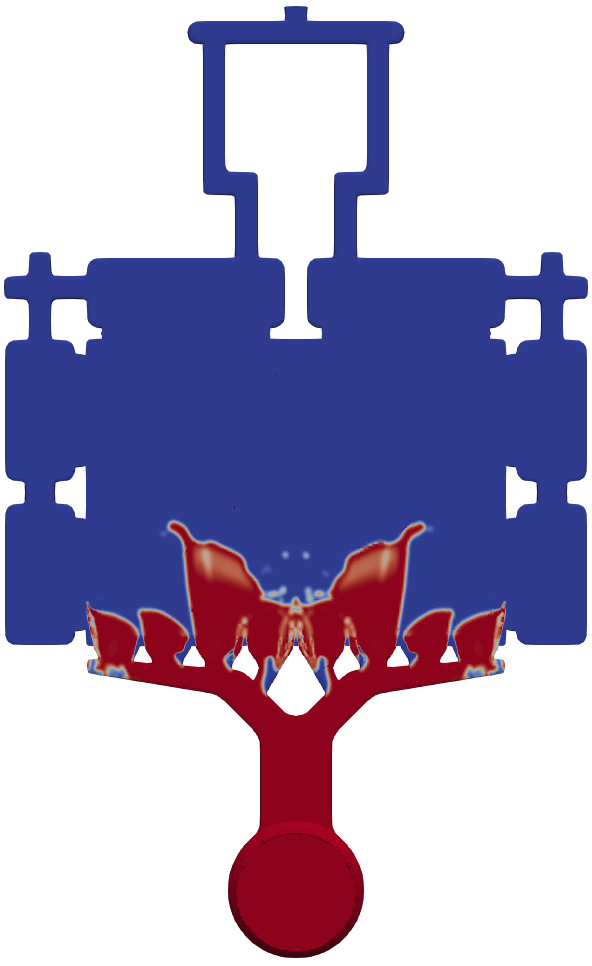"}}\hfill
\subfloat[\centering]{\includegraphics[width=.20\linewidth,trim={0 6cm 0 8cm},clip]{"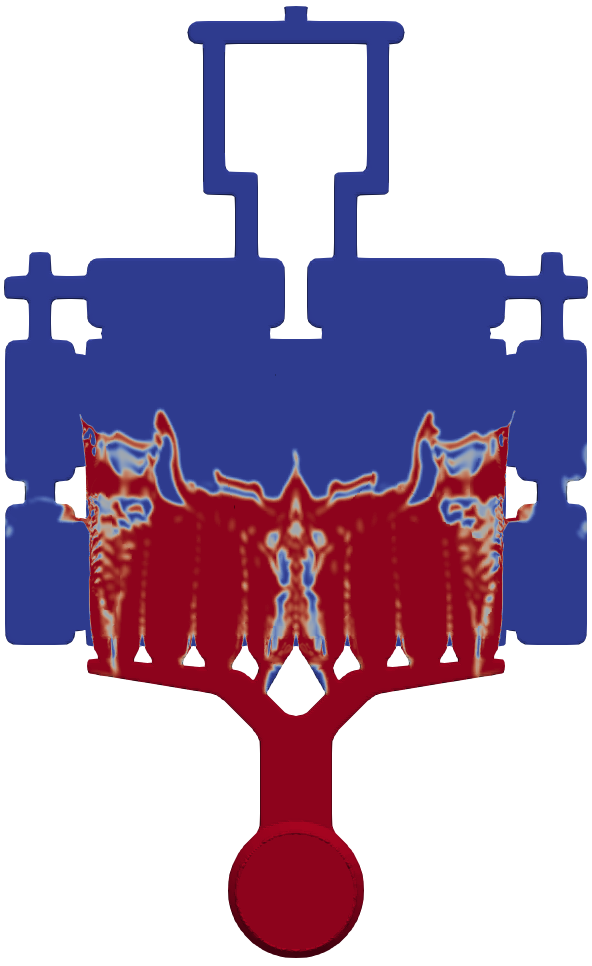"}}\hfill
\subfloat[\centering]{\includegraphics[width=.20\linewidth,trim={0 6cm 0 8cm},clip]{"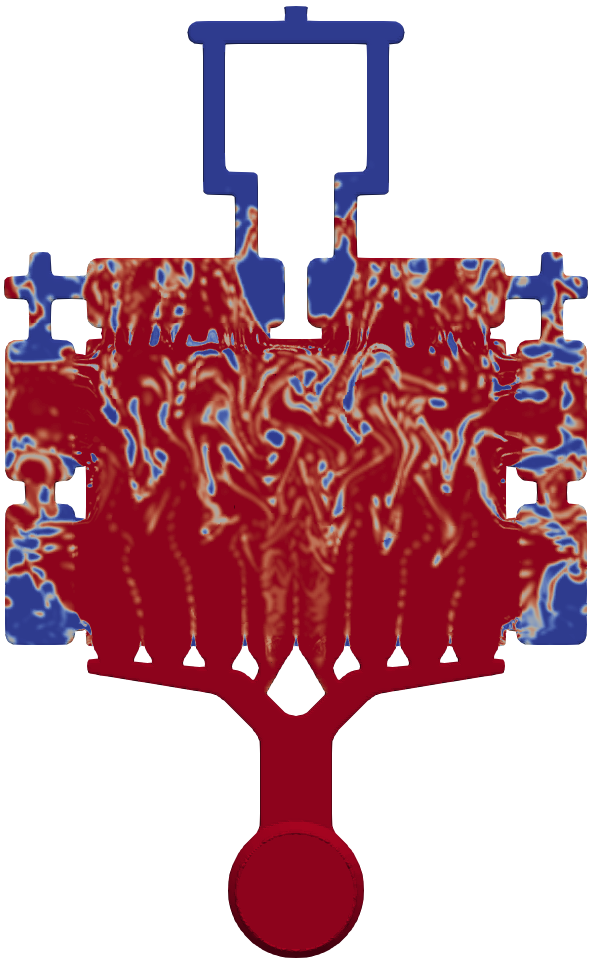"}}\hfill
\subfloat[\centering]{\includegraphics[width=.20\linewidth,trim={0 6cm 0 8cm},clip]{"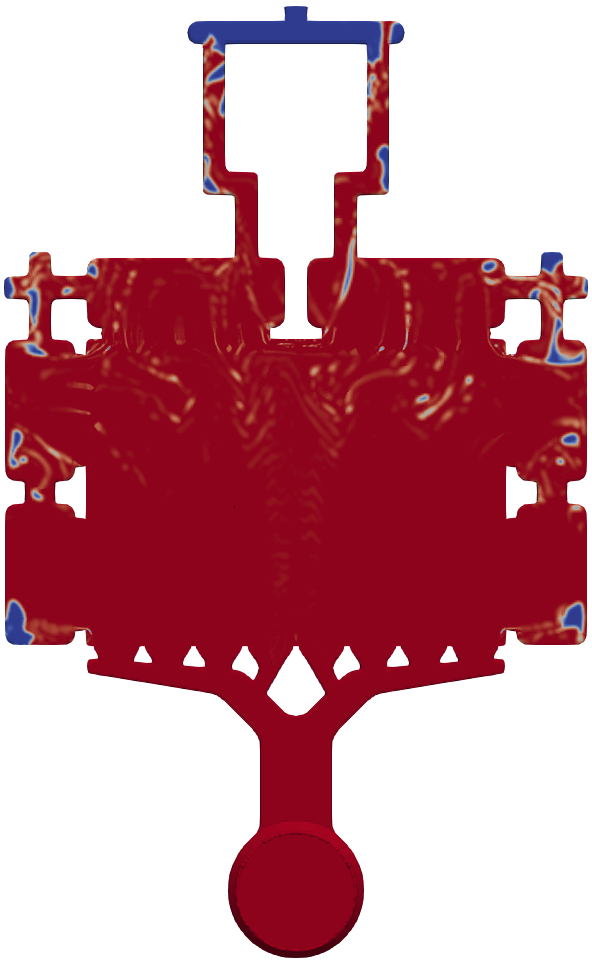"}}\\[0.8em]
\subfloat[\centering]{\includegraphics[width=.20\linewidth,trim={0 6cm 0 8cm},clip]{"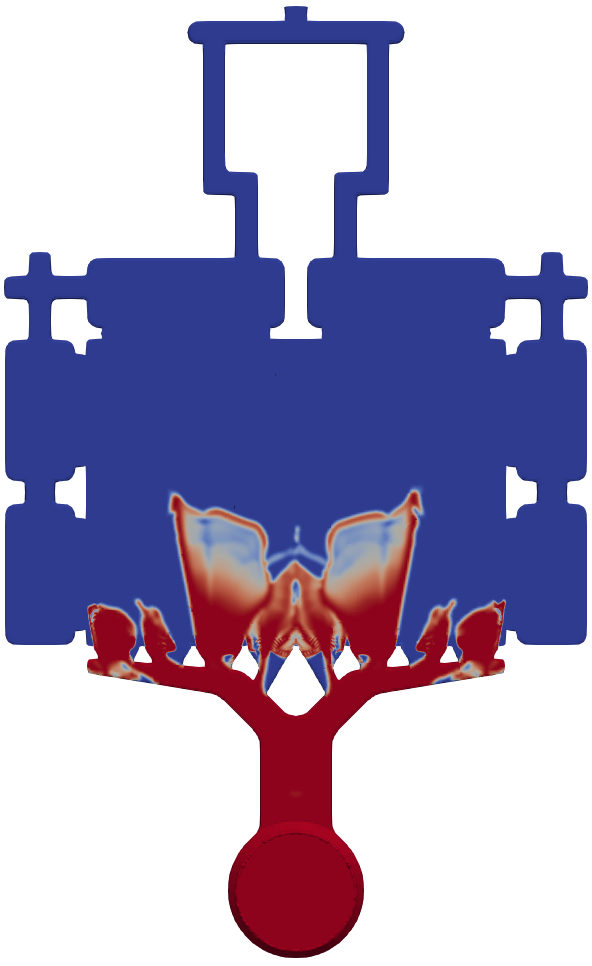"}}\hfill
\subfloat[\centering]{\includegraphics[width=.20\linewidth,trim={0 6cm 0 8cm},clip]{"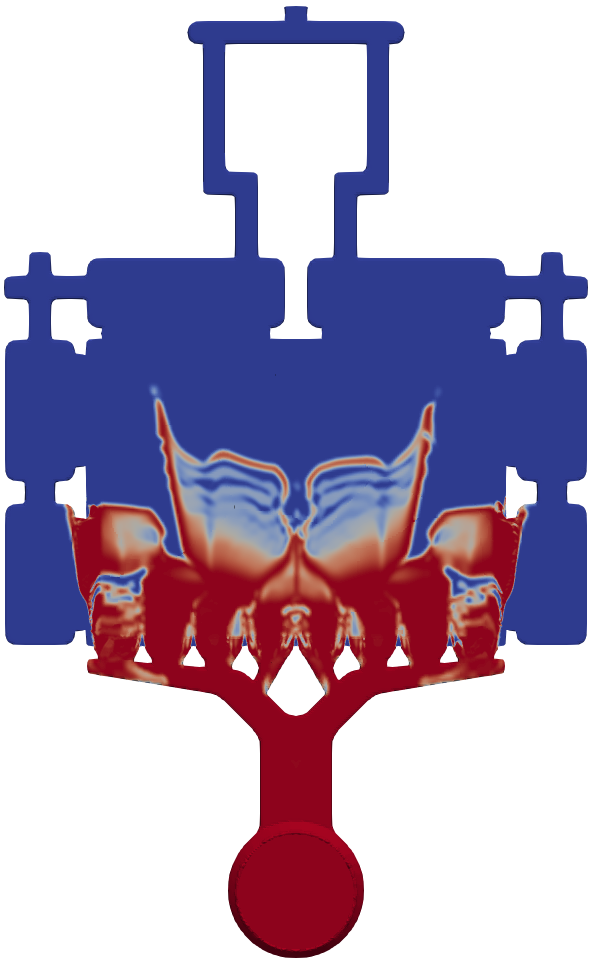"}}\hfill
\subfloat[\centering]{\includegraphics[width=.20\linewidth,trim={0 6cm 0 8cm},clip]{"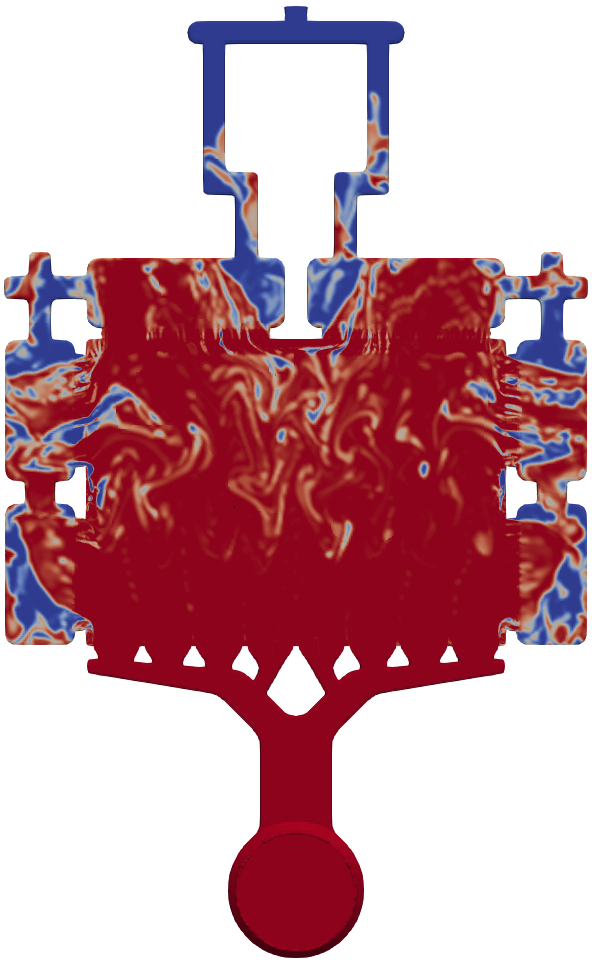"}}\hfill
\subfloat[\centering]{\includegraphics[width=.20\linewidth,trim={0 6cm 0 8cm},clip]{"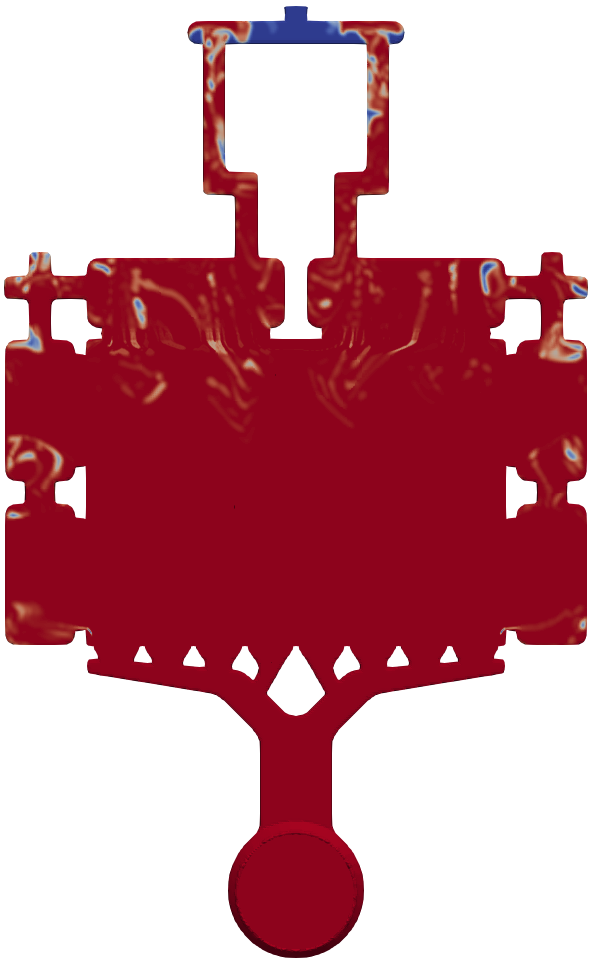"}}
\caption{Rendered \gls{sym:alpha} fields at three inlet velocities over four filling stages: (a--d) \gls{sym:vin}$=\SI{0.05}{\metre\per\second}$; (e--h) \gls{sym:vin}$=\SI{0.50}{\metre\per\second}$; (i--l) \gls{sym:vin}$=\SI{2.50}{\metre\per\second}$.}
\label{fig:alpha_rendered_grid}
\end{figure}

As an additional verification step, the influence of inlet velocity on compressible multiphase flow was examined to ensure that the solver produced physically consistent behavior across a broad range of filling conditions. Ten inlet velocities were simulated, spanning from \gls{sym:vin}$=\SI{0.05}{\metre\per\second}$ to \gls{sym:vin}$=\SI{2.5}{\metre\per\second}$. These values were prescribed at the cross section of the gating system (location indicated in Figure~\ref{fig:CADmodel}) and correspond approximately to piston velocities, accounting for the geometric reduction between the shot sleeve and the gating inlet. The upper end of the investigated inlet velocity range (\SIrange{2.0}{2.5}{\metre\per\second}) represents industrially relevant operating conditions and translates into realistic ingate velocities of approximately \SI{30}{\metre\per\second}, as demonstrated for the case with \gls{sym:vin}$=\SI{2.25}{\metre\per\second}$. Lower inlet velocities were deliberately included to illustrate how free surface structures and pressurization dynamics evolve across different flow regimes. The corresponding cavity pressure histories for all ten inlet velocities are shown in Figure~\ref{fig:figureG}. 
As expected, increasing inlet velocity leads to earlier and more pronounced pressure peaks. At higher values of \gls{sym:vin}, peak pressures occur within the first few milliseconds and reach significantly larger magnitudes due to rapid compression of entrapped air. In contrast, very low inlet velocities (e.g., \gls{sym:vin}$=\SI{0.05}{\metre\per\second}$) result in delayed and weaker pressure rises, reflecting gradual air displacement and reduced compressibility effects. Further insight into the filling dynamics for the reference case with an inlet velocity of \gls{sym:vin}$=\SI{2.25}{\metre\per\second}$ is provided by the ingate velocity \gls{sym:vg} data shown in Figure~\ref{fig:figureO}. 
The four monitored ingate centroids (Points~1--4) exhibit strong initial differences during the jetting phase. Point~1, located farthest from the central axis, peaks above \SI{100}{\metre\per\second}, while the others reach values between \SIrange[range-phrase={~and~},range-units=single]{35}{90}{\metre\per\second}. This variability reflects unequal flow path lengths, cross-sectional geometry, and the formation of local separation zones in the gating system. After this transient overshoot, all four profiles converge towards a quasi-steady plateau of approximately \SI{37}{\metre\per\second}, which is sustained throughout the main filling stage. Small oscillations around this level are attributed to turbulence and pressure fluctuations within the gating channels. A synchronized decline in ingate velocities begins at $t \approx \SI{41}{\milli\second}$, coinciding with the sharp drop in flow velocity in the \gls{vr} and \gls{mr} (Fig.~\ref{fig:figureON}) and with the late pressure surge observed in the \gls{vr} (Fig.~\ref{fig:figureM}). This timing indicates that counterpressure buildup within the cavity and backflow through the ingates are the dominant mechanisms governing the final collapse of entrapped air. The displacement of entrapped air is further illustrated in Figure~\ref{fig:figureON}, which compares the mean velocities \gls{sym:v_mean} in the \gls{vr} and \gls{mr}. It is noted that the absolute values of \gls{sym:v_mean} are comparatively low, as the evaluation regions include both fluid-filled and not-yet-filled cells during the early filling stage, resulting in a large fraction of zero-velocity contributions in the spatial averaging. Consequently, the temporal evolution of \gls{sym:v_mean} is governed primarily by the progression of the melt front through the respective regions. Between $t \approx \SI{28}{\milli\second}$ and $t \approx \SI{41}{\milli\second}$, a plateau-like behavior is observed, indicating that the evaluation regions are largely filled while through-flow is still maintained. The subsequent pronounced decline in \gls{sym:v_mean} marks the transition to a flow-stagnant state after complete filling, where velocities approach zero throughout the region. This transition is of particular relevance for porosity assessment: during this phase, the cavity is filled and convective transport diminishes, while the internal pressure has not yet increased significantly (see Fig.~\ref{fig:figureG}), allowing entrapped air to remain mobile and potentially be expelled. Persistent air in the \gls{vr} beyond this point indicates reduced venting efficiency or unfavorable flow recirculation, making this region particularly susceptible to porosity defects. These results suggest that targeted improvements in venting design or local flow guidance in the \gls{vr} could significantly reduce defect formation. Figure~\ref{fig:alpha_rendered_grid} shows rendered \gls{sym:alpha}-fields for three inlet velocities \gls{sym:vin}, aligned such that comparable global \gls{sym:alpha} values are reached across the sequences. This alignment enables a direct comparison of interface morphology at equivalent filling stages. At the lowest inlet velocity (\gls{sym:vin}$=\SI{0.05}{\metre\per\second}$, a--d), corresponding to a Weber number of \gls{sym:we}$\approx18$, the liquid front advances smoothly with only limited breakup. Surface tension still exerts a stabilizing influence, resulting in comparatively tranquil filling. However, the slow advance delays cavity pressurization and increases the risk of premature solidification. At an intermediate inlet velocity (\gls{sym:vin}$=\SI{0.50}{\metre\per\second}$, e--h), with \gls{sym:we}$\approx1.8\times10^{3}$, inertial forces dominate over surface tension. The advancing front becomes unstable, forming folds and recirculation cells that lead to pronounced interface fragmentation and conditions favorable for gas entrapment. At the highest inlet velocity (\gls{sym:vin}$=\SI{2.50}{\metre\per\second}$, i--l), corresponding to \gls{sym:we}$\approx4.6\times10^{4}$, inertia overwhelmingly exceeds surface tension effects. The filling process is characterized by strong splashing and highly turbulent motion, producing a strongly corrugated interface. While this promotes rapid cavity pressurization and efficient venting, it also generates extensive free surface area, thereby increasing the potential for oxidation. In the late filling stages, visual differences between the three cases diminish, as all exhibit highly turbulent and fragmented fronts that appear qualitatively similar. Distinguishing between these regimes therefore requires quantitative evaluation of free surface exposure, as provided by the \gls{tifsa} analysis in Figs.~\ref{fig:figureK}–\ref{fig:figureL}. It should be emphasized that Weber numbers defined at the ingate represent an upper bound; within the cavity, local velocities decay and characteristic length scales evolve, such that the effective Weber number \gls{sym:we} is lower than the nominal ingate Weber number \gls{sym:weg}.

\subsection{Influence of turbulence modeling on the free surface evolution}
\begin{figure}[b]
\centering 
\subfloat[\centering]{\includegraphics[width=.23\linewidth,trim={15cm 2cm 15cm 4cm},clip]{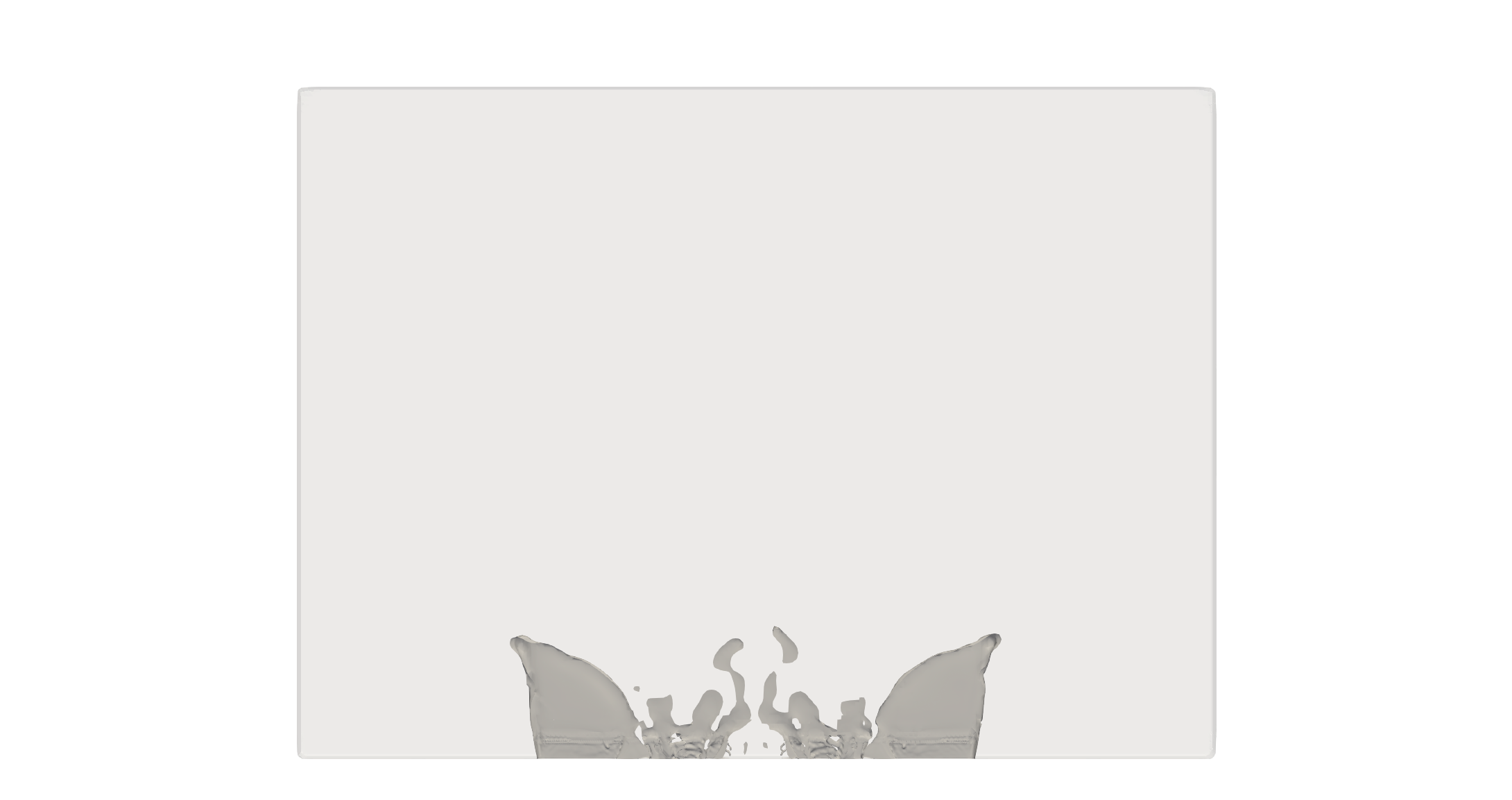}} \hfill
\subfloat[\centering]{\includegraphics[width=.23\linewidth,trim={15cm 2cm 15cm 4cm},clip]{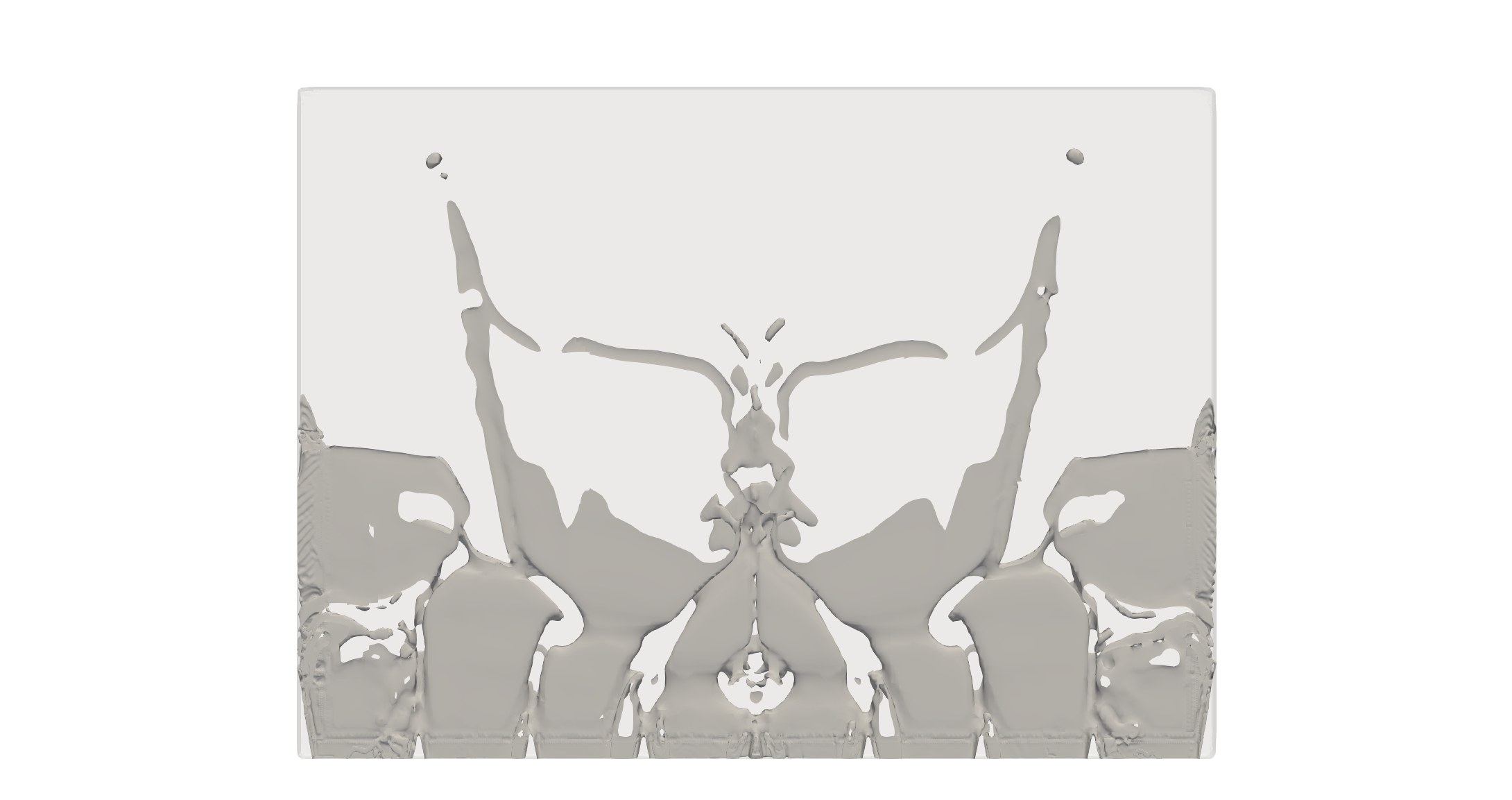}} \hfill
\subfloat[\centering]{\includegraphics[width=.23\linewidth,trim={15cm 2cm 15cm 4cm},clip]{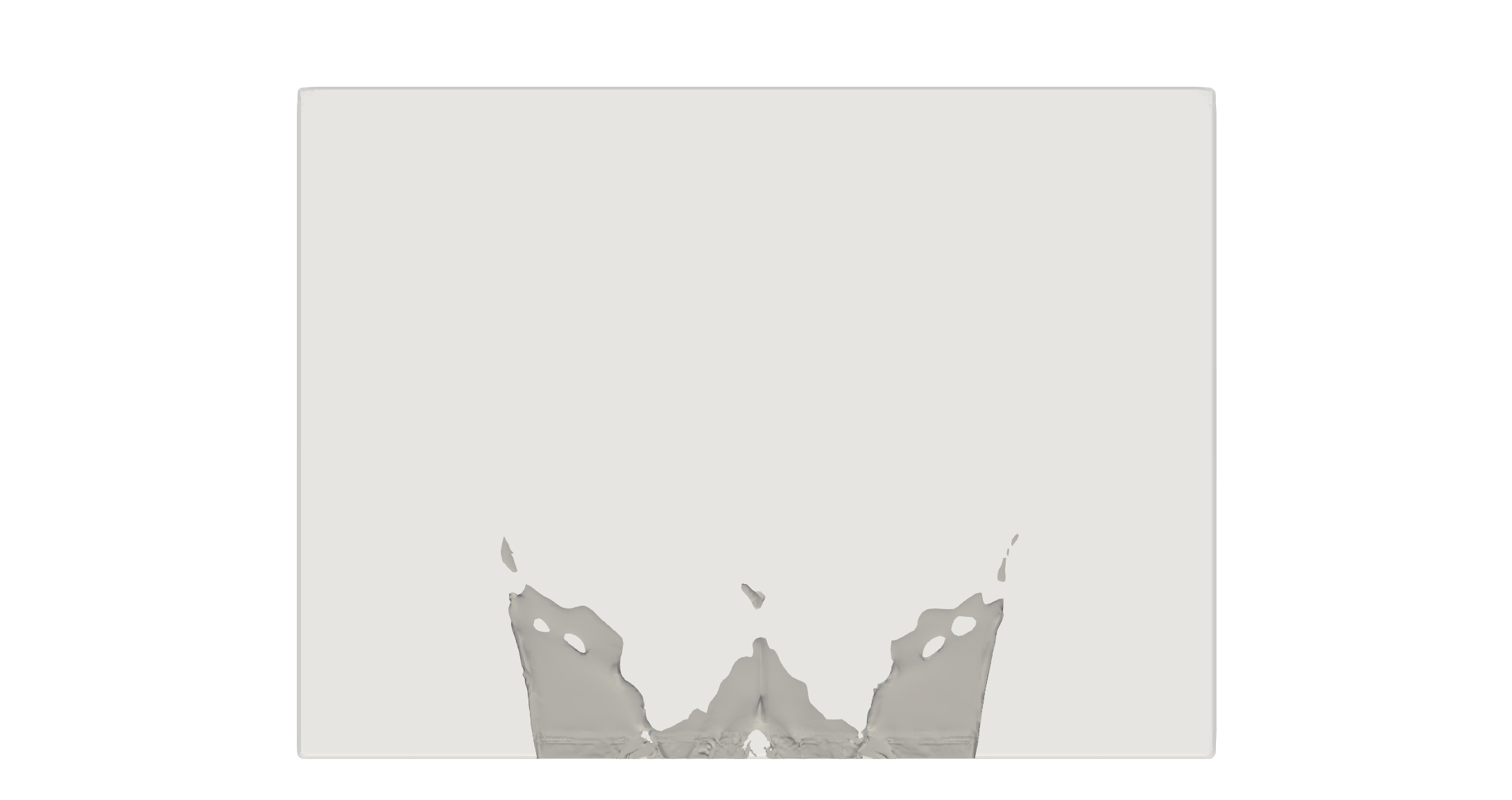}} \hfill
\subfloat[\centering]{\includegraphics[width=.23\linewidth,trim={15cm 2cm 15cm 4cm},clip]{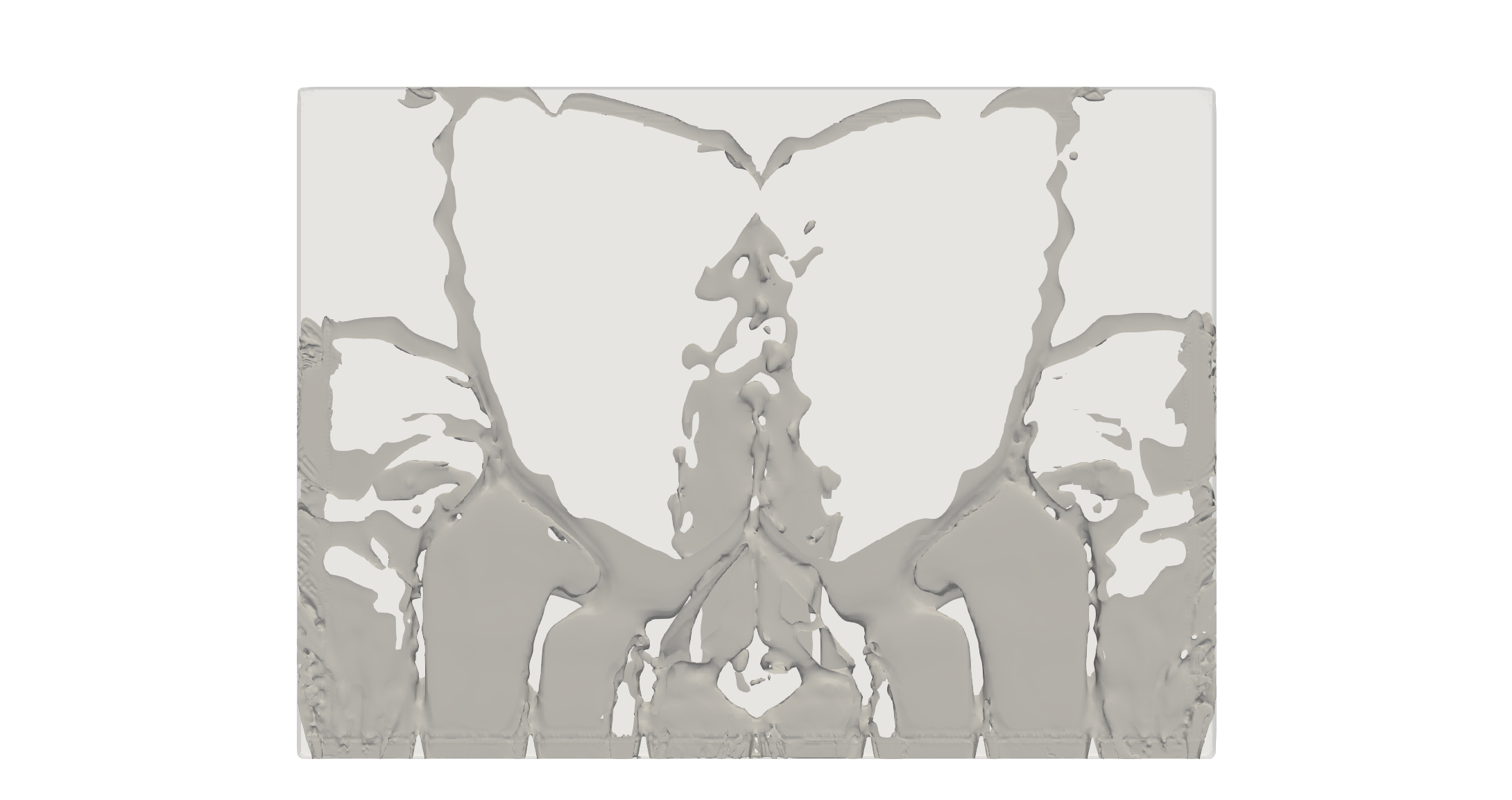}} \hfill
\caption{Early filling stages of the \SI{3}{\milli\metre} plate showing the free surface at $t=\SI{2}{\milli\second}$ and $t=\SI{5}{\milli\second}$ without turbulence modeling (a–b) and with $k$-$\varepsilon$ turbulence modeling (c–d).}
\label{fig:figureH}
\end{figure}

\begin{figure}[thb]
\centering 
\subfloat[\centering]{\includegraphics[width=.28\linewidth,trim={15cm 2cm 15cm 4cm},clip]{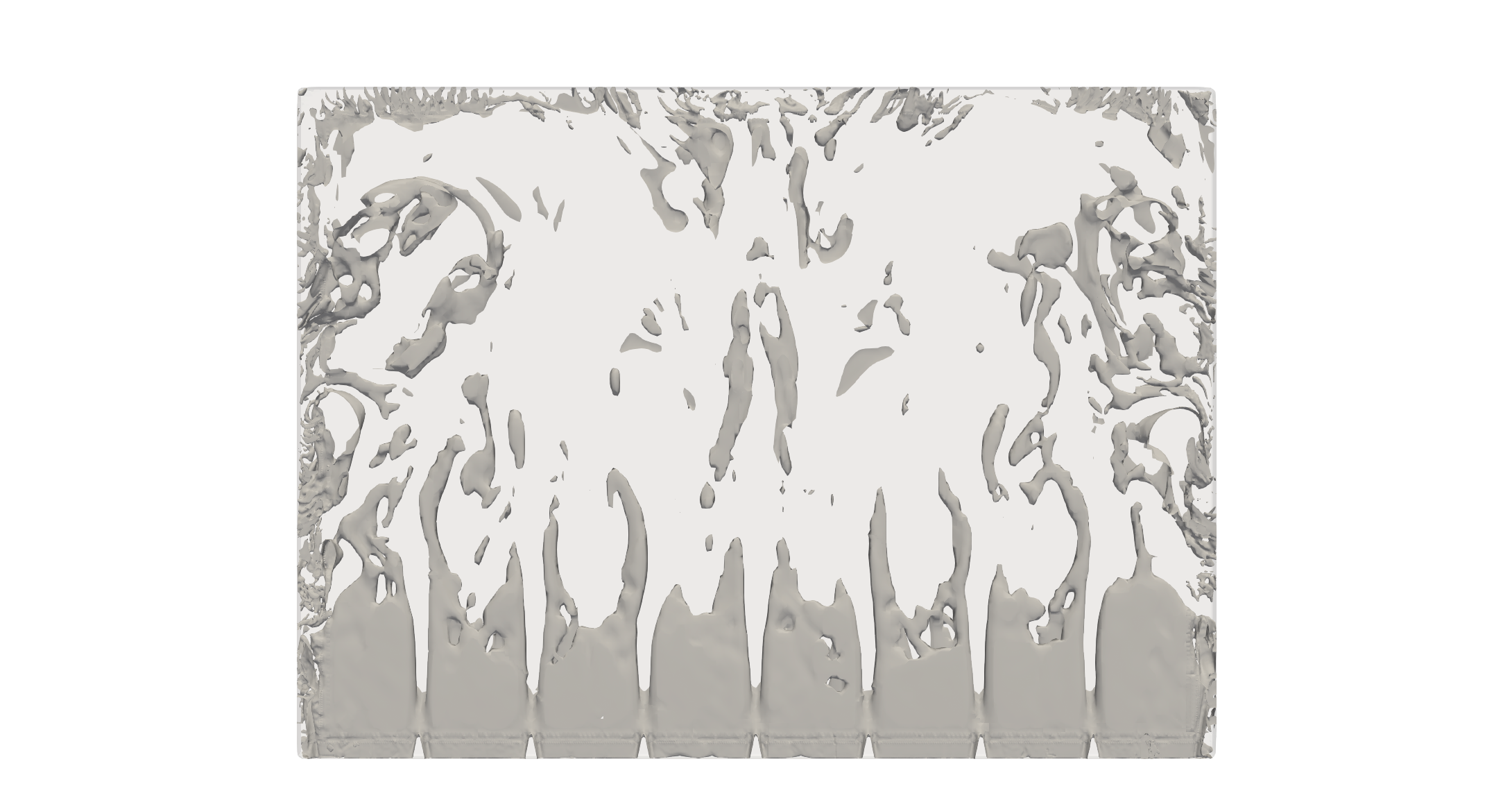}} \hfil
\subfloat[\centering]{\includegraphics[width=.28\linewidth,trim={15cm 2cm 15cm 4cm},clip]{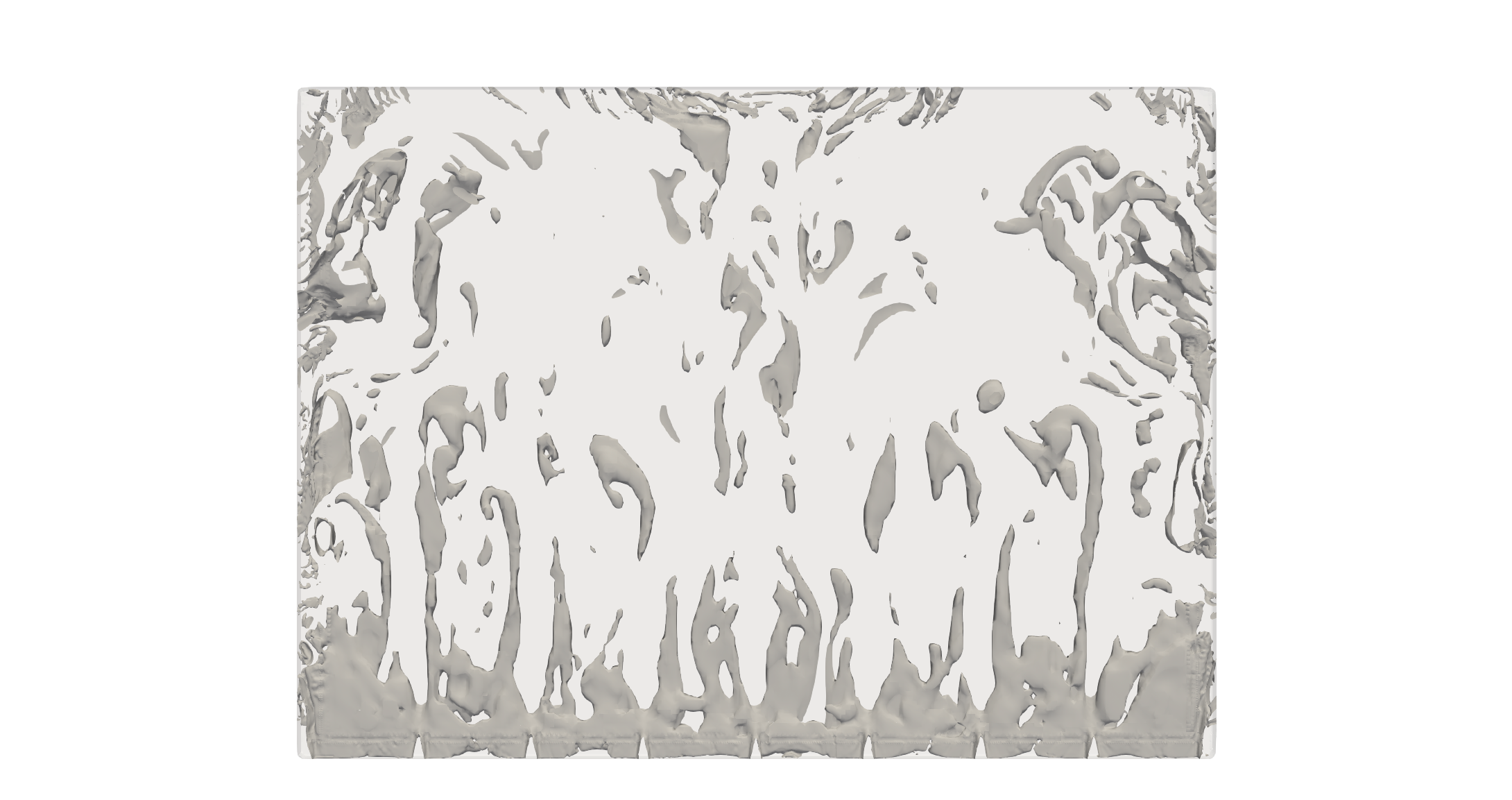}} \hfil
\subfloat[\centering]{\includegraphics[width=.28\linewidth,trim={15cm 2cm 15cm 4cm},clip]{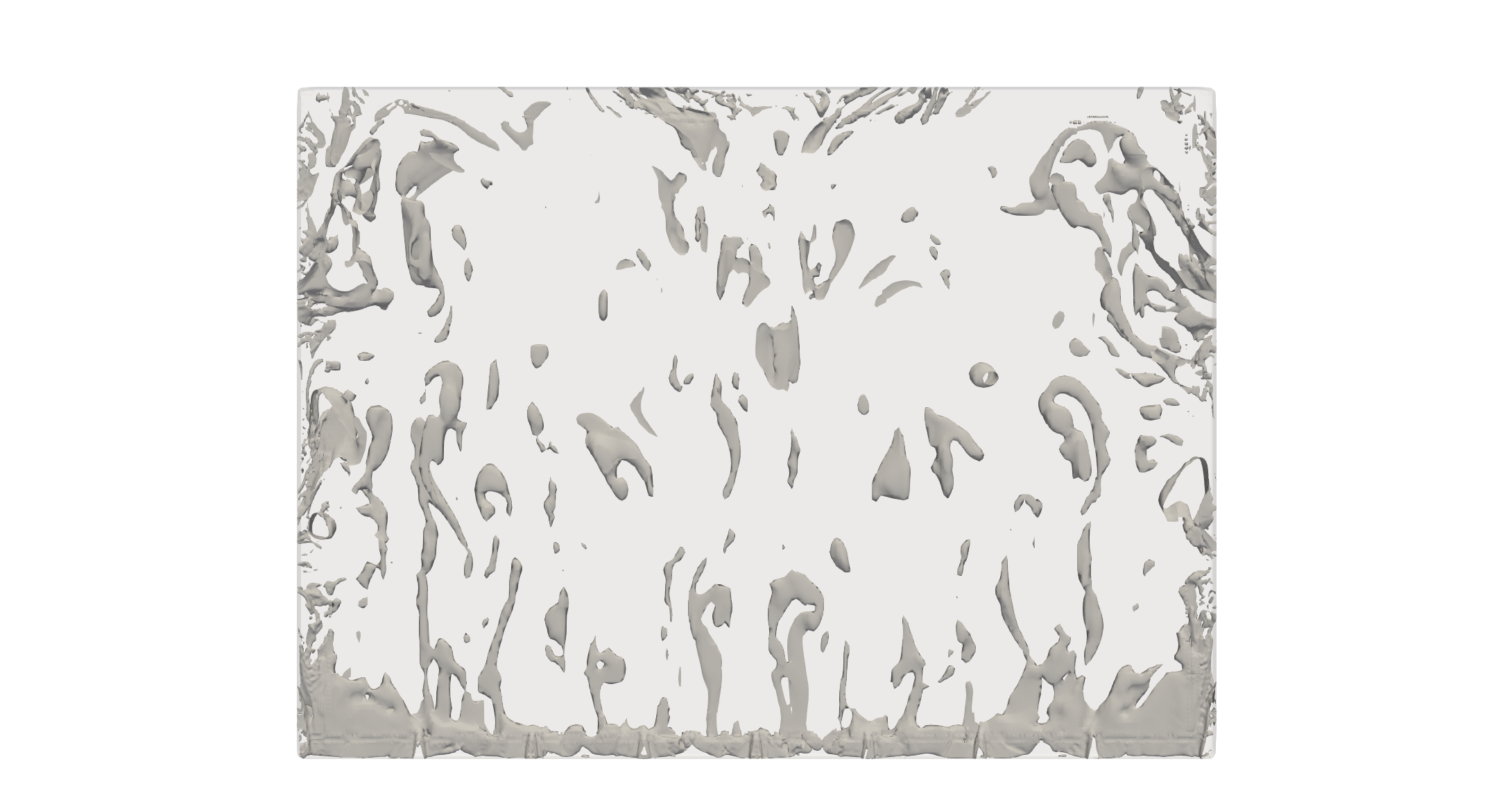}} \\
\subfloat[\centering]{\includegraphics[width=.28\linewidth,trim={15cm 2cm 15cm 4cm},clip]{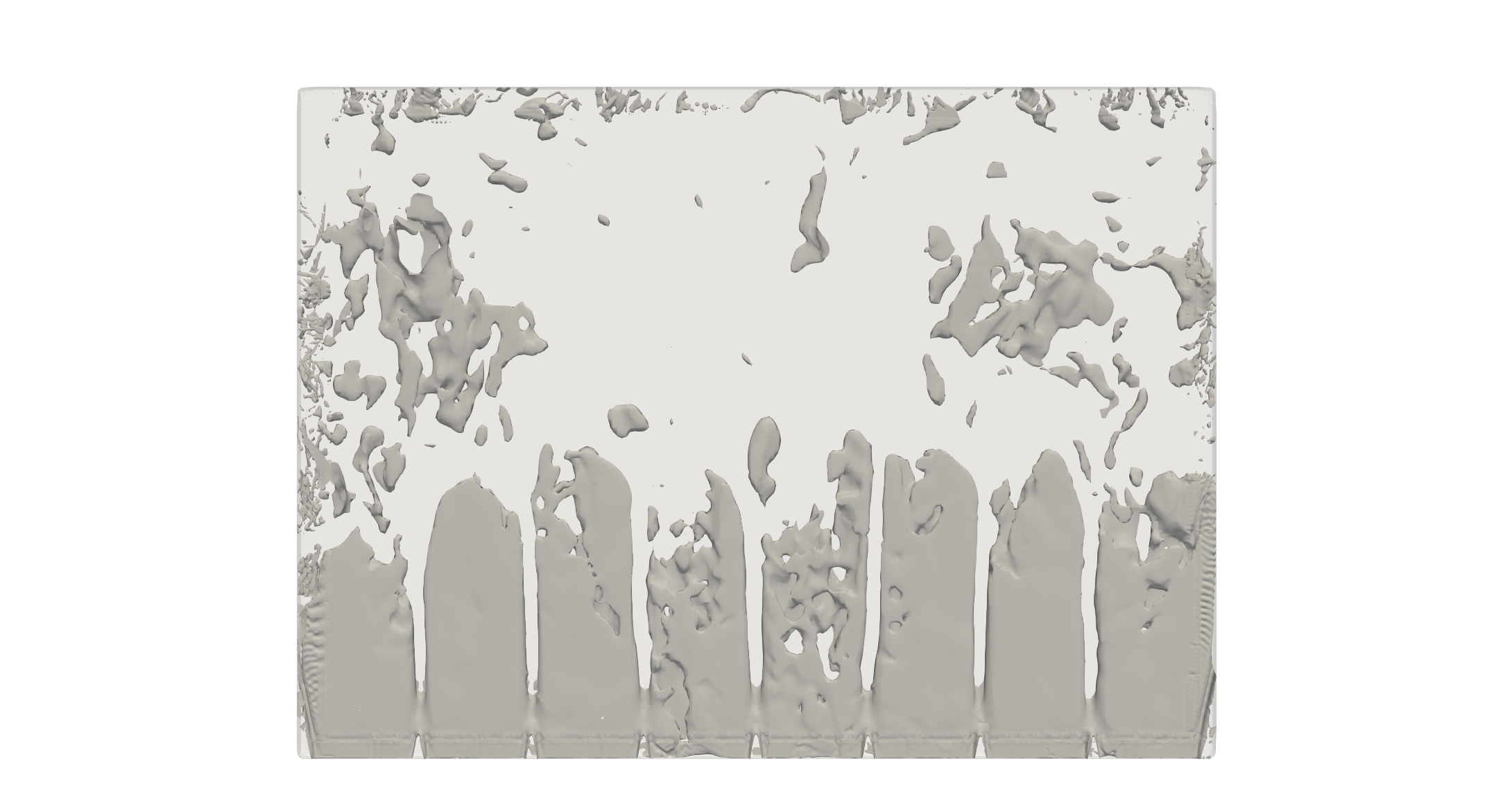}} \hfil
\subfloat[\centering]{\includegraphics[width=.28\linewidth,trim={15cm 2cm 15cm 4cm},clip]{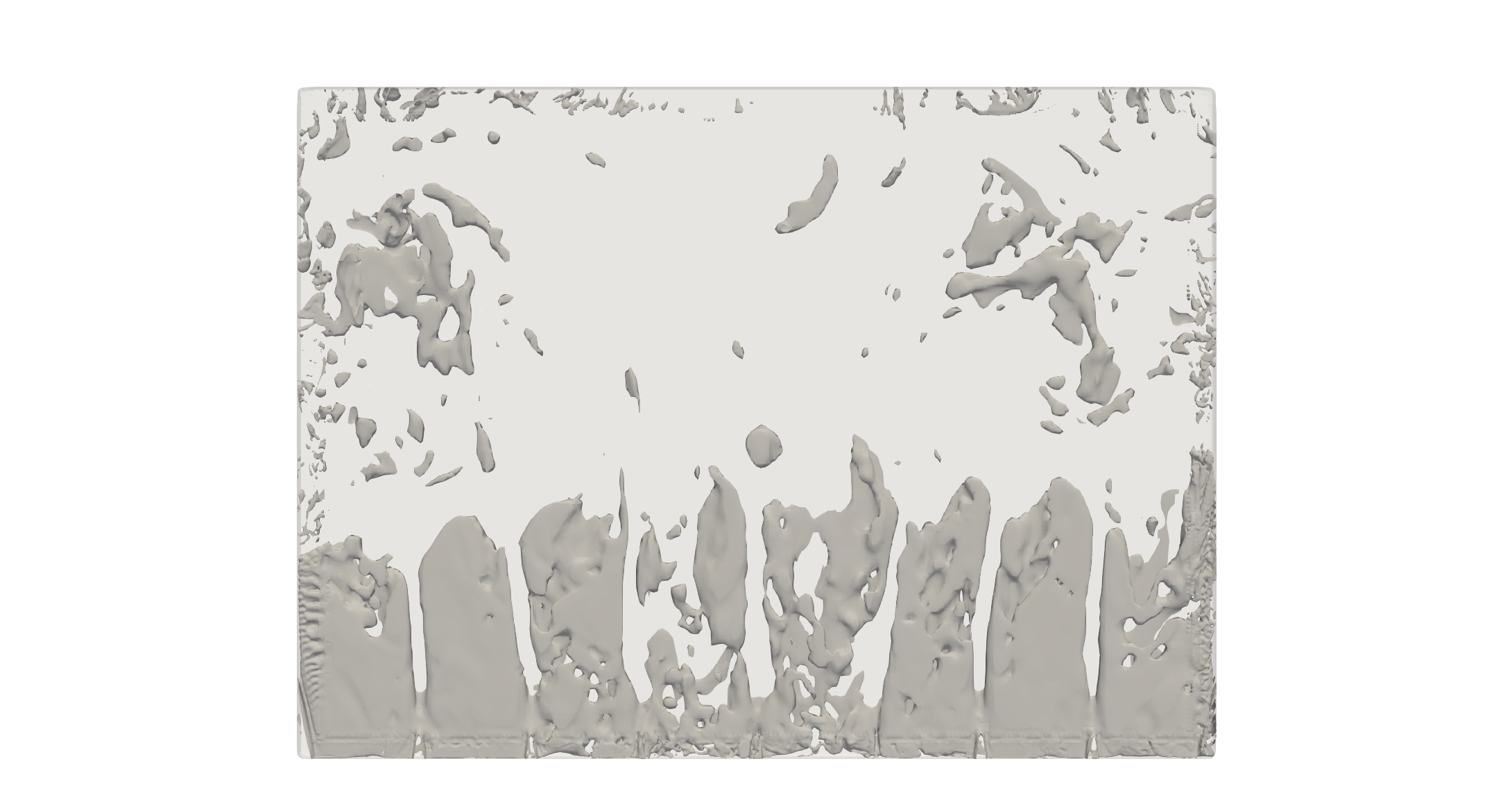}} \hfil
\subfloat[\centering]{\includegraphics[width=.28\linewidth,trim={15cm 2cm 15cm 4cm},clip]{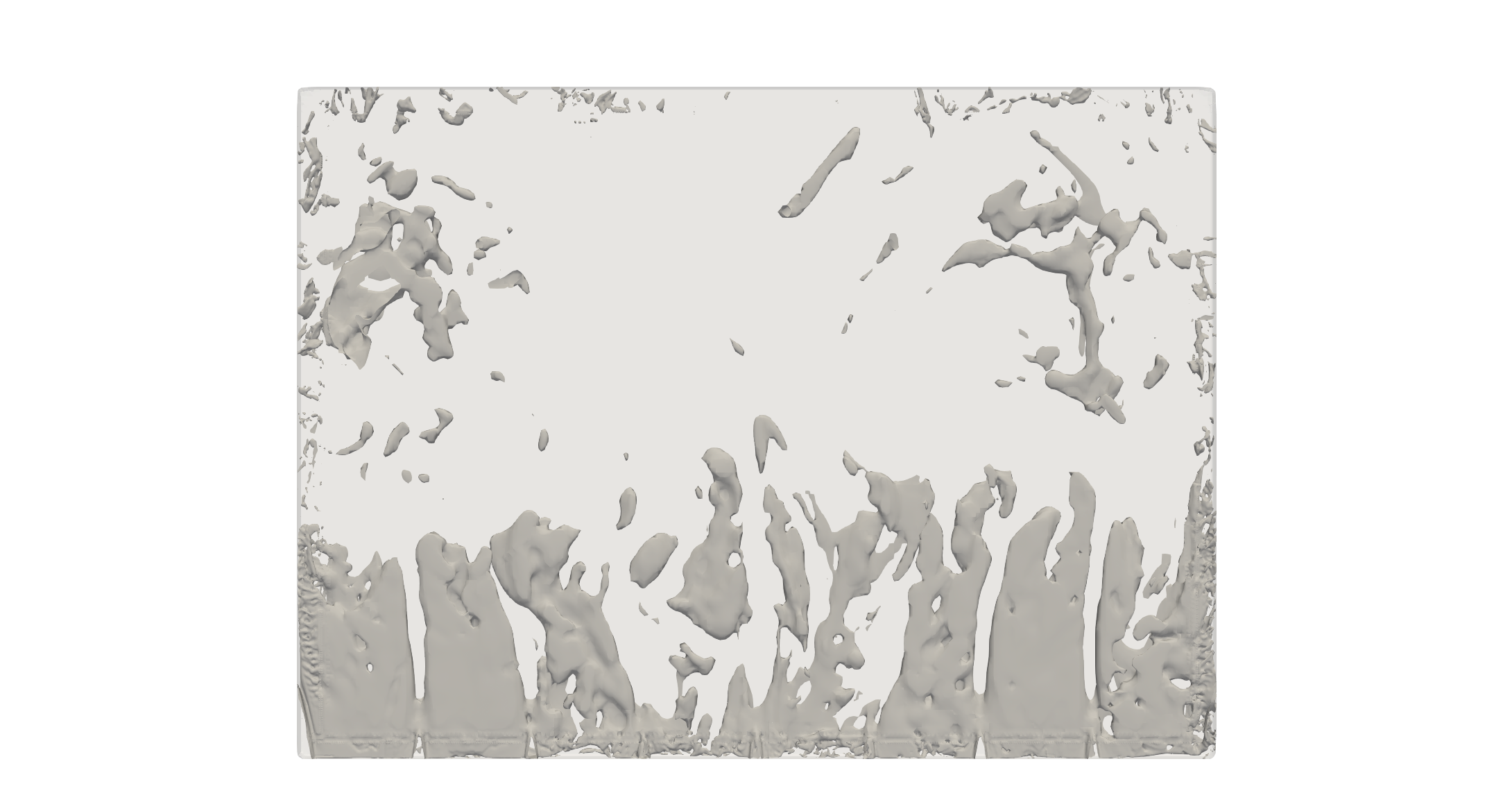}} \hfil
\caption{Intermediate filling stages of the \SI{3}{\milli\metre} plate showing the free surface at $t=\SI{17}{\milli\second}$, $t=\SI{19}{\milli\second}$, and $t=\SI{21}{\milli\second}$ without turbulence modeling (a–c) and with $k$-$\varepsilon$ turbulence modeling (d–f).}
\label{fig:figureI}
\end{figure}

\begin{figure}[bp]
\centering 
\subfloat[\centering]{\includegraphics[width=.28\linewidth,trim={15cm 2cm 15cm 4cm},clip]{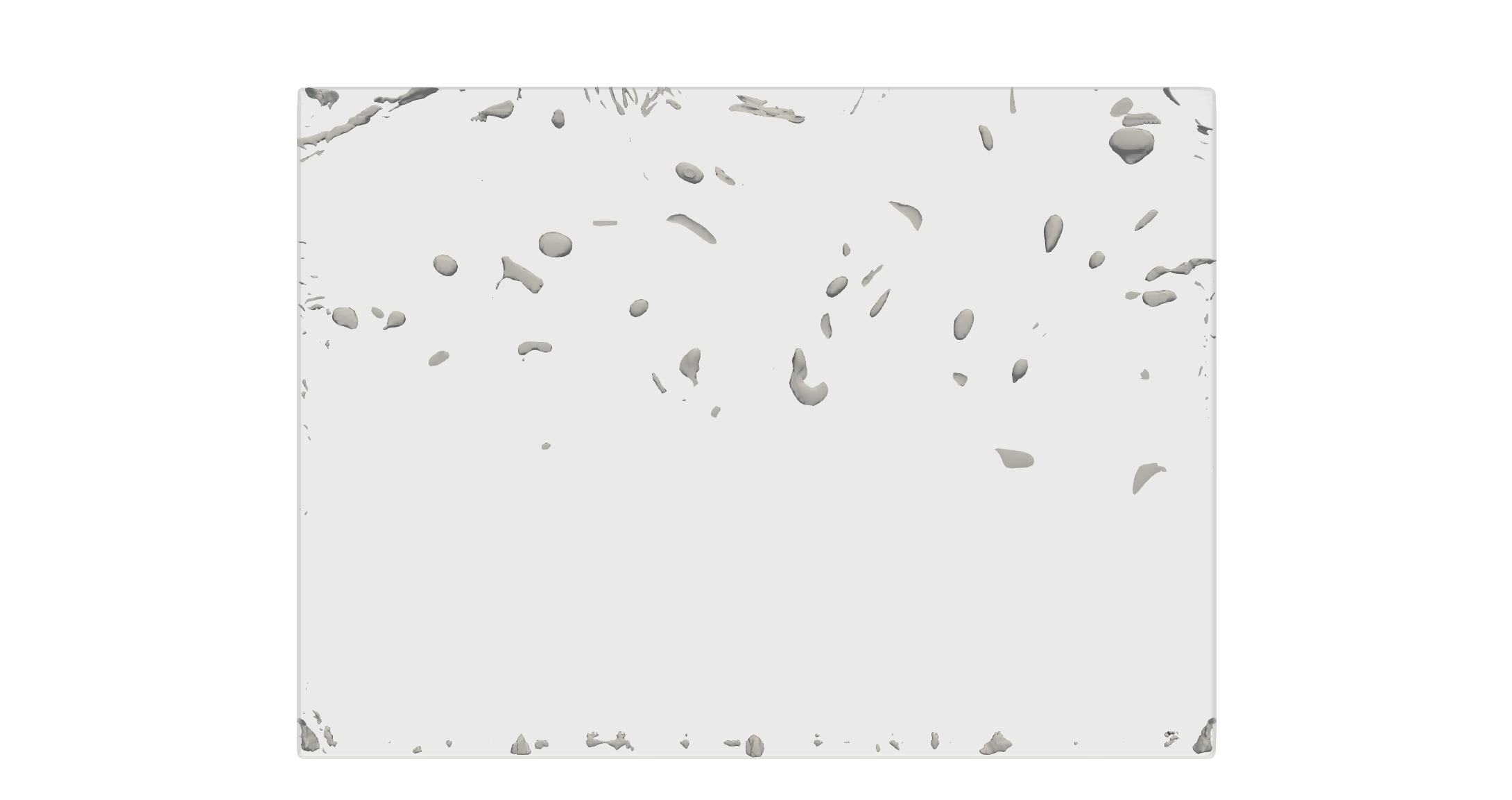}} \hfil
\subfloat[\centering]{\includegraphics[width=.28\linewidth,trim={15cm 2cm 15cm 4cm},clip]{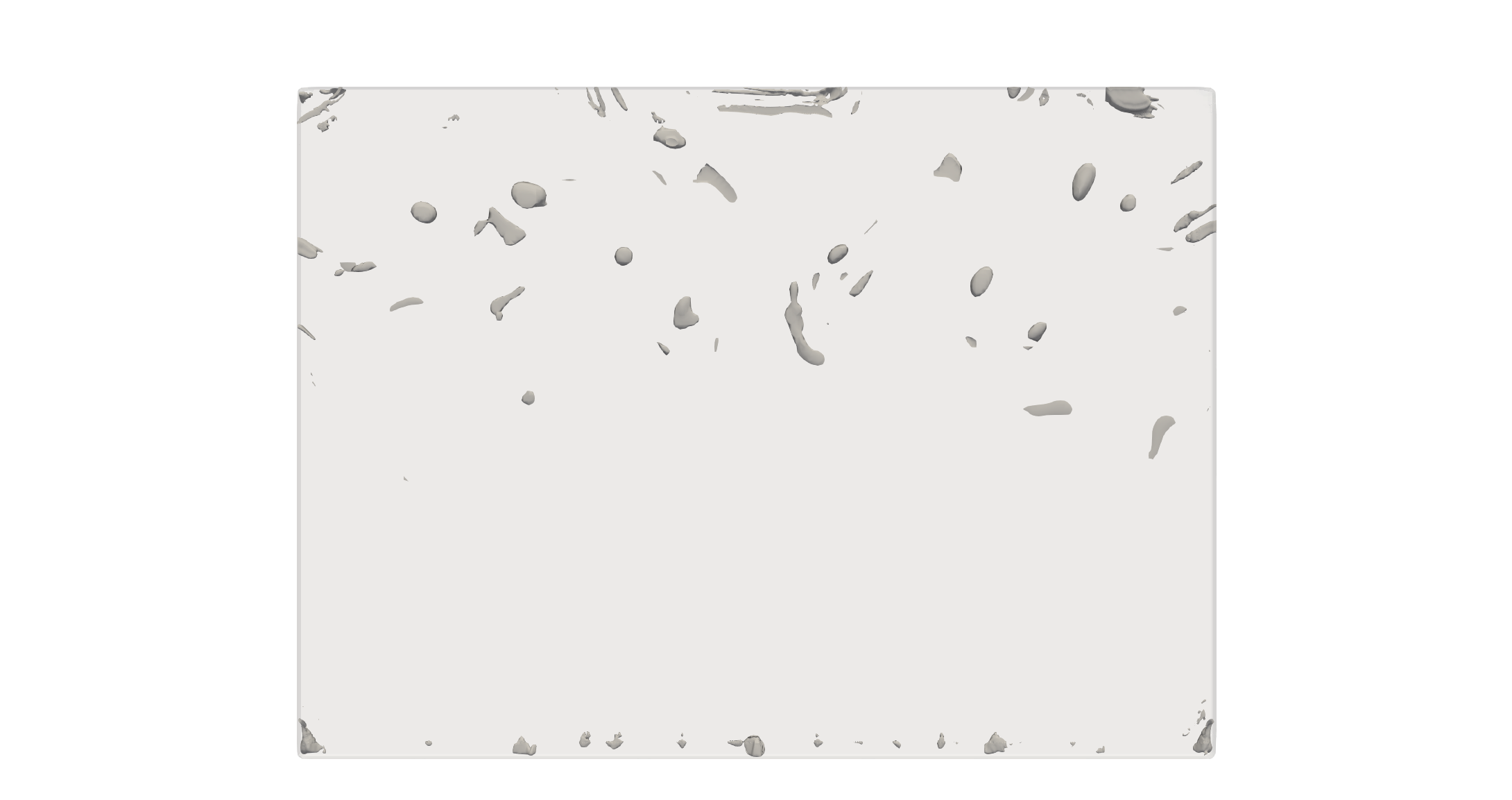}} \hfil
\subfloat[\centering]{\includegraphics[width=.28\linewidth,trim={15cm 2cm 15cm 4cm},clip]{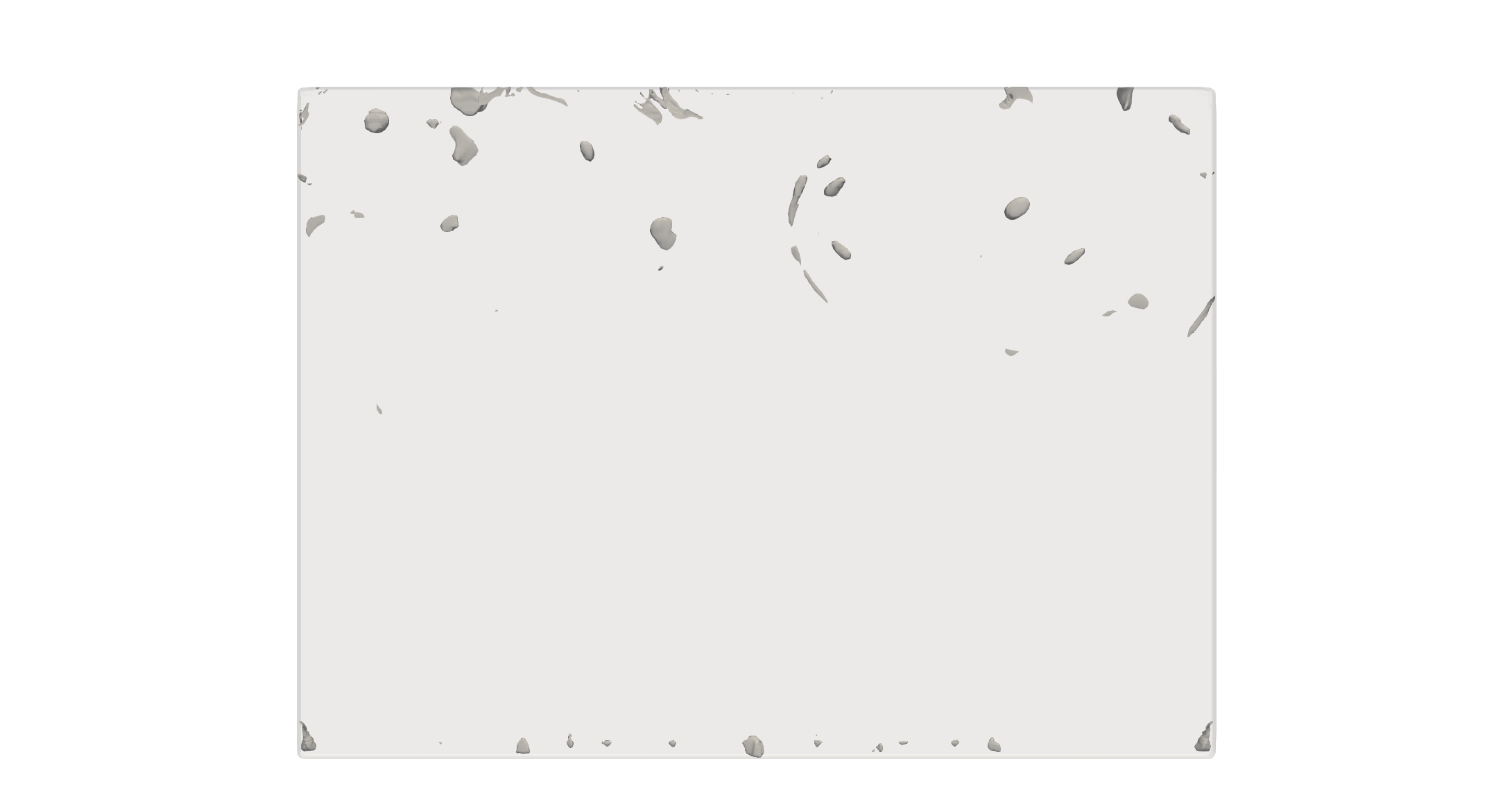}} \\
\subfloat[\centering]{\includegraphics[width=.28\linewidth,trim={15cm 2cm 15cm 4cm},clip]{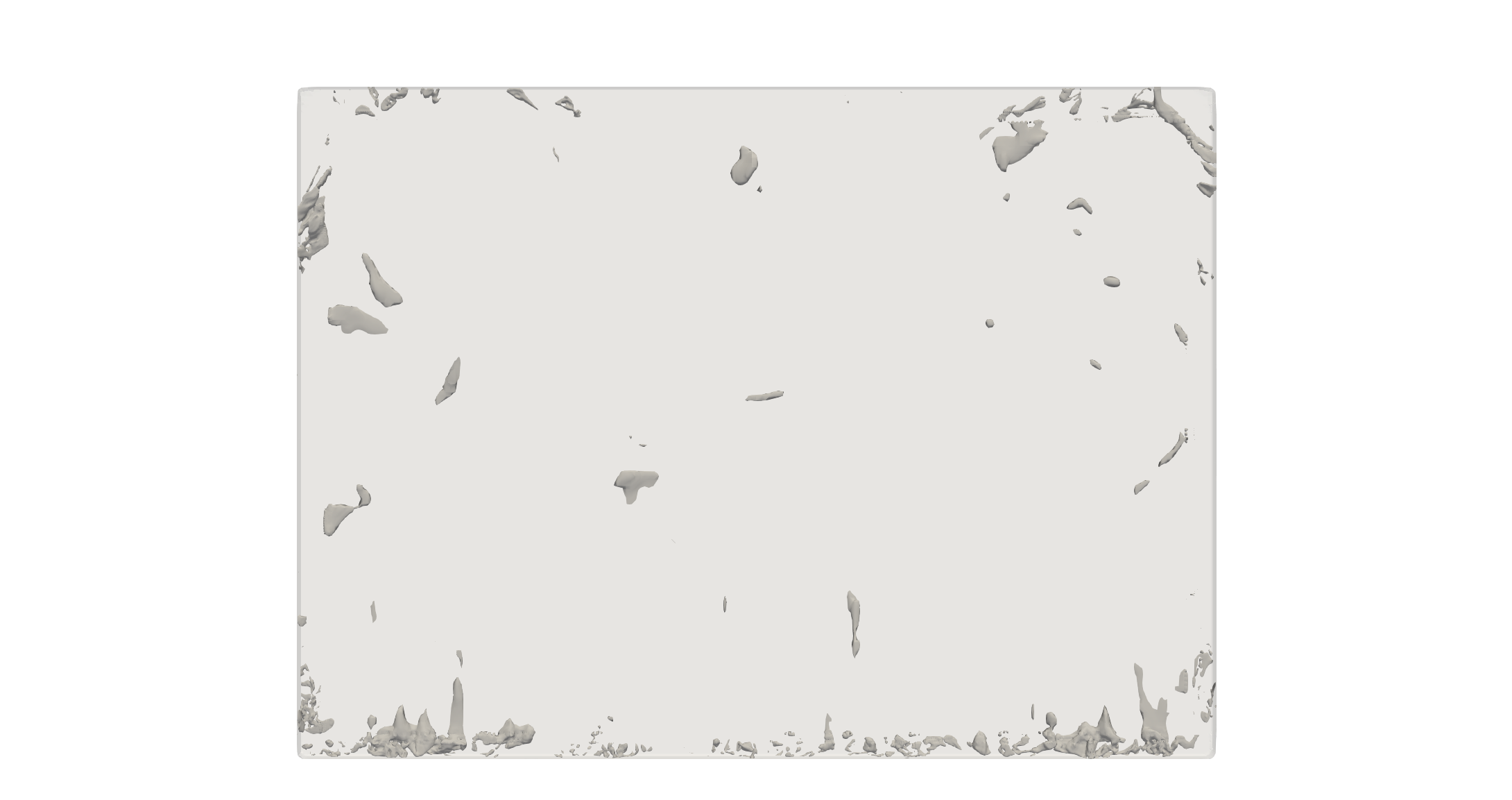}} \hfil
\subfloat[\centering]{\includegraphics[width=.28\linewidth,trim={15cm 2cm 15cm 4cm},clip]{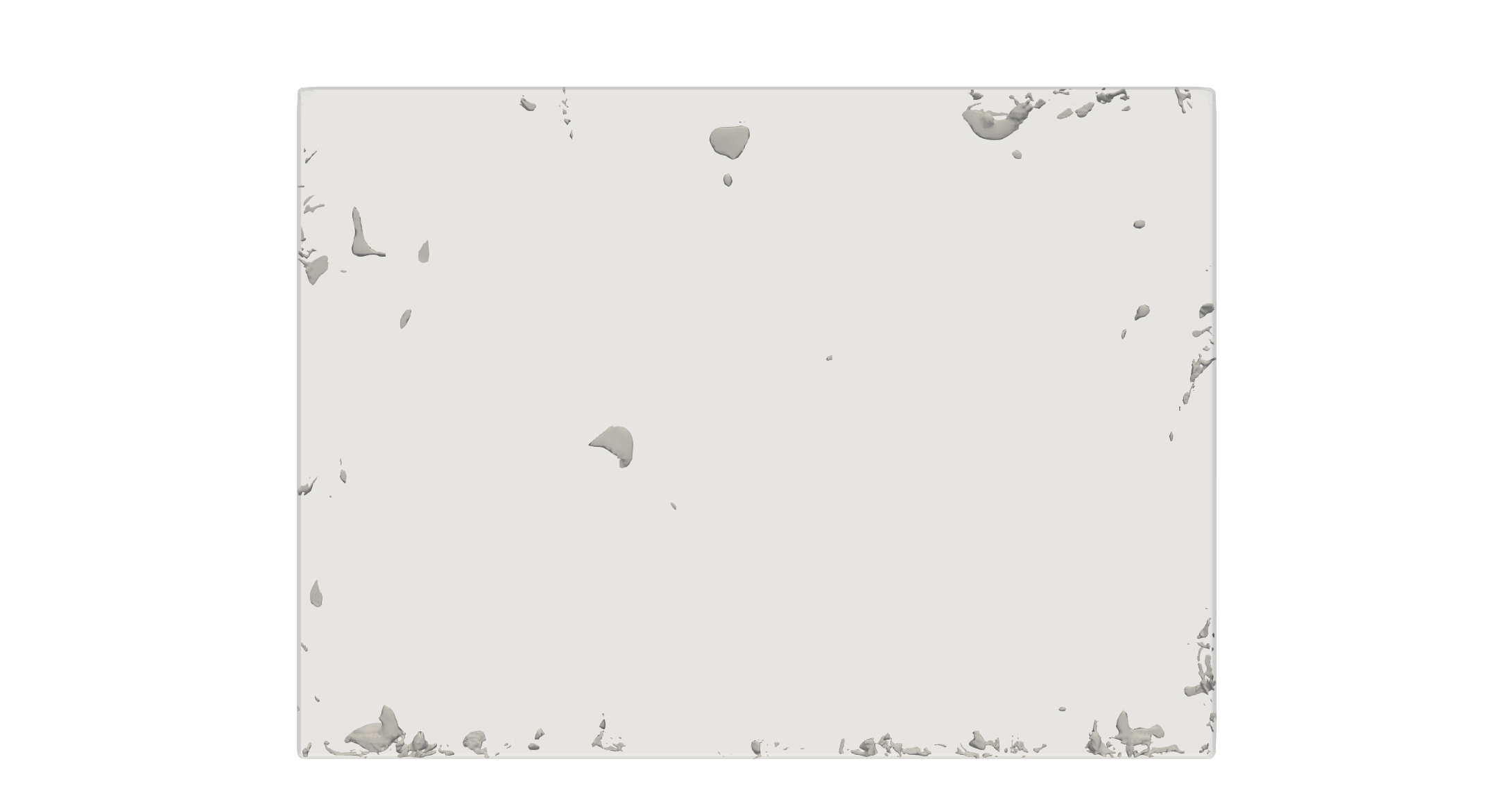}} \hfil
\subfloat[\centering]{\includegraphics[width=.28\linewidth,trim={15cm 2cm 15cm 4cm},clip]{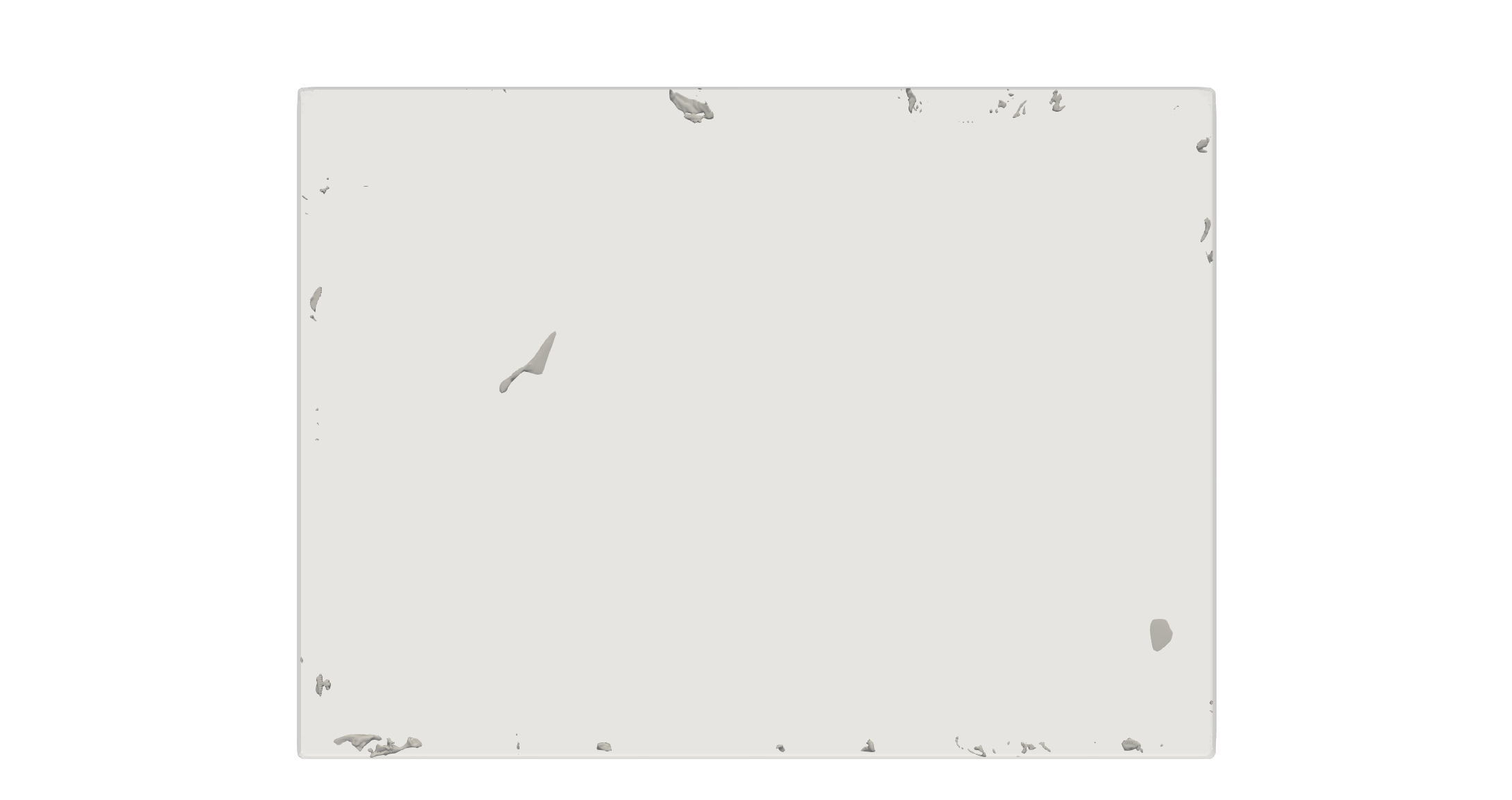}} \hfil
\caption{Final filling stages of the \SI{3}{\milli\metre} plate showing the free surface at $t=\SI{28}{\milli\second}$, $t=\SI{32}{\milli\second}$, and $t=\SI{35}{\milli\second}$ without turbulence modeling (a-c) and with $k$-$\varepsilon$ turbulence modeling (d-f).}
\label{fig:figureJ}
\end{figure}

\begin{figure}[h]
\small
\centering
\newcommand{\MyMarkerSize}{-2.5pt}
\newcommand{\MyLineWidth}{1pt}
\newcommand{\MyWidth}{.9\linewidth}
\newcommand{\MyHeight}{5cm}

\begin{tikzpicture}
  \begin{axis}[
    width=\MyWidth,
    height=\MyHeight,
    xlabel={Time / \SI{}{\milli\second}},
    ylabel={$V_{Air} / \SI{}{\cubic\meter}$},
    axis y line*=left,
    xmin = 26,  
    xmax = 43,
    ymin = 0,
    ymax = 10e-9,
    yticklabel style={
    /pgf/number format/fixed,
    /pgf/number format/precision=1,
    /pgf/number format/fixed zerofill,
    },
    every mark/.append style={mark size=\MyMarkerSize},  
    ]

    \addplot[
    color=col1,
    dashed,
    line width=\MyLineWidth, 
    mark=*, 
    mark options = {solid, draw=black, line width=0.5pt},
    sharp plot, mark size=\MyMarkerSize,
    ]
    table [x=Time, y=Top, col sep=semicolon] {poreVolumeTopMid.dat};
    \label{p1}

    \addplot[
    color=col2,
    line width=\MyLineWidth, 
    mark=square*, 
    dashed,
    mark options = {solid, draw=black, line width=0.5pt},
    sharp plot, mark size=\MyMarkerSize,
    ]
    table [x=Time, y=Mid, col sep=semicolon] {poreVolumeTopMid.dat};
    \label{p2}
  \end{axis}
  
  \begin{axis}[
    width=\MyWidth,
    height=\MyHeight,
    xlabel={Time / \SI{}{\milli\second}},
    ylabel={$\overline{p}$ / \SI{}{\pascal}},
    axis y line*=right,
    axis x line=none,
    xmin = 26,
    xmax = 43,
    ymin = 0,
    ymax = 4e7,
    grid=major,
    xmajorgrids=true,
    xminorgrids=true,
    yticklabel style={
    /pgf/number format/fixed,
    /pgf/number format/precision=0,
    /pgf/number format/fixed zerofill,
    },
    every mark/.append style={mark size=\MyMarkerSize},  
    legend style={
        draw=black, 
        fill=white, 
        font=\small,
        anchor=north,
        legend columns=2,
    },
    legend pos=north east,
    legend cell align=left,
    ]
    
    \addlegendimage{/pgfplots/refstyle=p1}\addlegendentry{$V_{Air}^{TR}$}
    \addlegendimage{/pgfplots/refstyle=p2}\addlegendentry{$V_{Air}^{MR}$}
    
    \addplot[
    line width=\MyLineWidth, 
    color=col3,
    mark=triangle*, 
    dashed,
    mark options = {solid, draw=black, line width=0.5pt},
    sharp plot, mark size=\MyMarkerSize,
    ]
    table [x=Time, y=Average_p, col sep=comma] {top_average_p_by_time_44.dat};
    \addlegendentry{\gls{sym:p_av}$^{VR}$}
    
    \addplot[
    line width=\MyLineWidth, 
    color=col4,
    mark=diamond*,
    dashed,
    mark options = {solid, draw=black, line width=0.5pt},
    sharp plot, mark size=\MyMarkerSize,
    ]
    table [x=Time, y=Average_p, col sep=comma] {mid_average_p_by_time_44.dat};
    \addlegendentry{\gls{sym:p_av}$^{MR}$}

  \end{axis}

\end{tikzpicture}
\caption{Average Pressure \gls{sym:p_av} and cumulative entrapped air volume \gls{sym:vair} in \gls{mr} and \gls{vr} of the plate during the final stages of mold filling.}
\label{fig:figureM}
\end{figure}
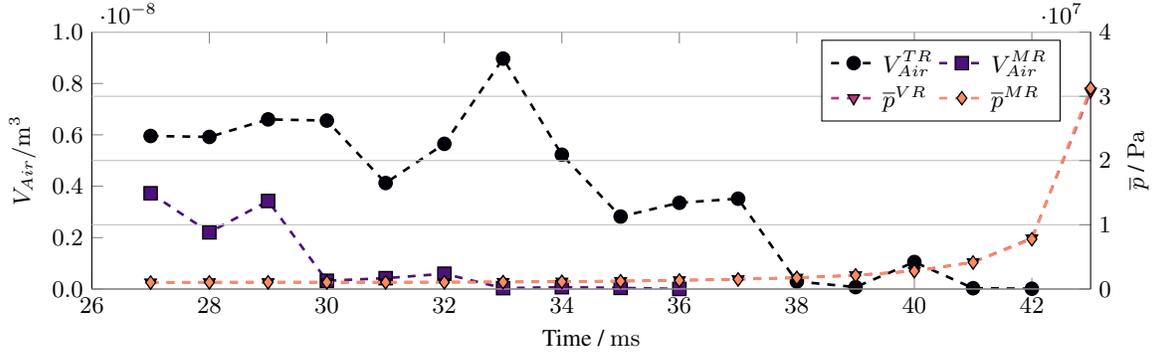

The compressible two-phase framework was applied to the \SI{3}{\milli\metre} plate for the reference case corresponding to the industrial process conditions, i.e.\ an inlet velocity of \gls{sym:vin}$=\SI{2.25}{\metre\per\second}$, in order to examine how the free surface organizes gas transport during filling and how turbulence modeling modifies this evolution (Figs.~\ref{fig:figureH}--\ref{fig:figureJ}). In contrast to many casting simulations that visualize the entire liquid phase, the present analysis isolates and tracks the water-air interface directly. This is critical because it is the interface, rather than the bulk melt, that governs the formation of folds, the entrapment and release of gas, and the establishment of purge pathways toward the headspace. Rendered \gls{sym:alpha} fields (Fig.~\ref{fig:alpha_rendered_grid}) provide a qualitative view of how the interface evolves with different inlet velocities; however, such images cannot quantify the relative severity of surface activity or air entrapment. As filling progresses, apparent interfacial structures diminish in the final stages not because fluid motion ceases, but because distinct interfaces collapse once liquid fronts merge and cavity pressure compresses residual gas. For this reason, the following section introduces quantitative evaluation metrics that condense free surface dynamics into measurable indicators of oxidation risk, filling continuity, and preferential flow paths. The evaluation is based on a compressible multiphase formulation that explicitly resolves the free surface while capturing the evolving pressure field. In practice, the simulations provide both the time resolved morphology of the advancing interface and the corresponding pressure history inside the cavity. Together, these quantities determine whether entrained air pockets survive, shrink, or are vented. The approach therefore treats the free surface as the primary diagnostic for fluid-air interaction: continuous interface connectivity indicates hydraulic communication with the headspace, whereas fragmented or dispersed fronts signal the presence of isolated regions prone to porosity formation. By focusing on interface dynamics rather than on the bulk velocity field alone, the evaluation directly targets the mechanisms most critical for defect prediction in \gls{hpdc}. In this study, the cavity filling process is divided into three stages (early, intermediate, and late filling) in order to resolve how interface morphology and turbulence modeling interact at successive phases of cavity filling. The three filling stages are characterized as follows:
\begin{itemize} 
  \item \textbf{Early filling stage $t=\SIrange[range-phrase=-,range-units=single]{2}{5}{\milli\second}$} (Fig.~\ref{fig:figureH}): Both the laminar and the $k$-$\varepsilon$ simulations predict compact ingate plumes and nearly identical front shapes while inertia dominates. The interface grows symmetrically from the ingates with limited lateral spreading; at these times differences between the two models remain small.
    
  \item \textbf{Intermediate filling stage} $t=\SIrange[range-phrase=-,range-units=single]{17}{21}{\milli\second}$ (Fig.~\ref{fig:figureI}): Systematic deviations appear as the front interacts with the walls and with recirculating flow. In the laminar run, the plumes lose coherence and the interface fragments into irregular fingers; small closed recirculation cells are visible downstream of the jets. In contrast, the $k$-$\varepsilon$ run maintains a more continuous, energetic front. The added eddy viscosity $\mu_t = C_\mu \rho k^2/\varepsilon$ enhances cross stream momentum diffusion, which damps small vortices that would otherwise trap gas, sustains jet coherence over a longer distance, and spreads pressure more uniformly across the sheet.
  
  \item \textbf{Final filling stage} $t=\SIrange[range-phrase=-,range-units=single]{28}{35}{\milli\second}$ (Fig.~\ref{fig:figureJ}): The divergence between models becomes decisive. The laminar interface is highly dispersed in the upper half of the plate, with numerous disconnected liquid islands. Such fragmentation delays hydraulic contact between the fluid and the headspace; pressure therefore rises later and more locally, and air pockets persist in the \gls{mr}. The $k$-$\varepsilon$ case preserves a connected front that couples earlier to the headspace. This promotes faster and stronger pressurization and establishes continuous purge paths: entrained gaps are swept toward the upper boundary or corners where outflow is favored, or are compressed below the grid resolution and no longer detectable. Consistently, by $t=\SIrange[range-phrase=-,range-units=single]{32}{35}{\milli\second}$  the turbulence modeled frames concentrate the remaining voids near the top edge, whereas the laminar frames still show multiple interior pockets.
\end{itemize}
At first sight it might appear counterintuitive that the $k$-$\varepsilon$ simulations exhibit lower visible porosity, since turbulence is often associated with stronger free surface formation. This reduction, however, is confined to the late filling stages ($t=\SIrange[range-phrase=-,range-units=single]{28}{35}{\milli\second}$), when cavity pressurization dominates. A separate evaluation of the integral free surface area (Fig.~\ref{fig:figureK} and Table~\ref{tab:table4}) shows that the turbulence model consistently produces a larger total interfacial area throughout filling. The apparent contradiction is resolved by recognizing that a more connected free surface improves hydraulic communication with the headspace, which in turn accelerates cavity pressurization and promotes the collapse or expulsion of entrained air pockets. Three coupled mechanisms explain this behavior: First, the eddy viscosity transports momentum laterally across the \SI{3}{\milli\metre} section, damping small recirculation cells, maintaining coherent ingate plumes, and distributing pressure more uniformly; low velocity niches that sustain air are therefore reduced in number and persistence. 
Second, a continuous interface establishes earlier hydraulic contact with the headspace, allowing the cavity to pressurize more rapidly and strongly, which compresses or expels entrained gas volumes. Third, in a thin plate the main porosity risk arises from stagnant recirculation rather than bulk three-dimensional breakup; turbulent diffusion here helps sweep air toward boundaries and shortens the residence time of free surface folds in the core. Two factors further affect the computed pore volumes. The evaluation timing is critical, as cavity pressure rises during the final phase, pores contract or dissolve, so velocity and pressure histories were used to select analysis windows ($t=\SIrange[range-phrase=-,range-units=single]{28}{35}{\milli\second}$) before maximum compression. Moreover the numerical resolution imposes a detection threshold: despite the relatively fine mesh, only comparatively large voids can be tracked; in the $k$-$\varepsilon$ case, some pockets are likely compressed below this threshold rather than completely removed. Even with these caveats, the spatial redistribution is unambiguous: turbulence modeling reduces the survival of interior air pockets and relocates the remainder toward purge favorable boundaries by the end of filling.
To quantitatively assess the dynamics of entrapped air during the final stages of mold filling in \gls{hpdc}, the cumulative volume of air pockets was computed separately for the \gls{vr} and \gls{mr} of the \SI{3}{\milli\metre} plate, as defined in the CAD geometry (Fig.~\ref{fig:CADmodel}). In parallel, the temporal evolution of the average pressure within these regions was analyzed. As shown in Figure~\ref{fig:figureM}, the entrapped air volume and mean pressure exhibit consistent trends in both regions. With increasing cavity pressure during filling, air gaps contract, leading to a sharp rise in local pressure. In the \gls{mr}, the air volume initially fluctuates due to transient flow effects but decreases steadily thereafter. Between $t \approx \SI{28}{\milli\second}$ and $t \approx \SI{32}{\milli\second}$, a portion of the entrapped air migrates from the \gls{mr} to the \gls{vr}, which is reflected by corresponding increases in the \gls{vr} volume curve. By $t \approx \SI{36}{\milli\second}$, all air pockets in the \gls{mr} have been eliminated. 
In contrast, the \gls{vr} retains residual air until approximately $t \approx \SI{42}{\milli\second}$. During this period, the mean pressure in the \gls{vr} rises sharply, reaching values above \SI{30}{\mega\pascal} immediately before the final collapse of the remaining air pockets. A critical aspect in this context is the timing of porosity evaluation. If assessment is performed too early, the fluid front may not yet have completed filling, and thus the predicted pore distribution can be misleading. If performed too late, increasing cavity pressure compresses or even dissolves small pores, driving them below the numerical resolution of the grid. Hence, it is essential to monitor both pressure and velocity histories in the critical regions to identify the time window in which porosity evaluation remains physically meaningful. This requirement is further illustrated in Figures~\ref{fig:figureG}, \ref{fig:figureO}, and \ref{fig:figureON}, which show the concurrent evolution of cavity pressure and velocity. These quantities provide the necessary reference to align the evaluation of porosity with the physical mechanisms controlling air entrapment and collapse. This late stage pressure surge is consistent with rapid compression and potential dissolution of trapped gases into the fluid, and it coincides with the onset of backflow effects in the gating system. Moreover, the regional differences identified here, with higher and more persistent porosity in the \gls{vr} compared to the \gls{mr}, are consistent with the free surface structures observed in Figure~\ref{fig:figureJ}, where turbulence modeling revealed more compact interfaces in the midsection but a more fragmented surface in the upper region. Together, these observations highlight the necessity of a coupled evaluation of air volume, pressure, and velocity to correctly identify the critical window for porosity analysis and to link pore evolution with the underlying flow dynamics.

\subsection{Time Integrated Free Surface Area (TIFSA)}
\label{ssec:tifsa}
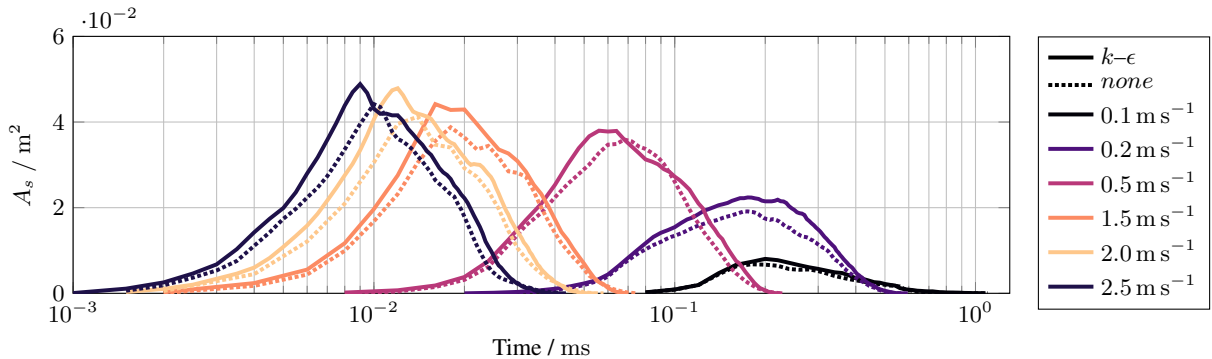
\begin{figure}
\small
\centering
\newcommand{\MyMarkerSize}{1.5pt}
\newcommand{\MyLineWidth}{1.5pt}
\newcommand{\MyWidth}{.85\linewidth}
\newcommand{\MyHeight}{5cm}
\newcommand{\MyLabel}{test}

\tikzset{
  ransstyle/.style={solid, line width=\MyLineWidth},
  lstyle/.style={densely dotted, line width=\MyLineWidth}
}

\begin{tikzpicture}
\centering
\begin{axis}[
    width=\MyWidth,
    height=\MyHeight,
    xmode = log,
    xlabel={Time / \SI{}{\milli\second}},
    ylabel={$A_s~/~\SI{}{\metre\squared}$},
    grid=major,
    xmajorgrids=true,
    xminorgrids=true,
    xmin = 0.001,
    xmax = 1.3,
    ymin = 0,
    ymax = 0.06,
    every mark/.append style={mark size=5pt},
    legend style={draw=black, fill=white, font=\footnotesize, anchor=north east, legend columns=1},
    legend cell align={left},
    legend pos=outer north east,
    yticklabel style={
        /pgf/number format/fixed,
        /pgf/number format/precision=0,
        /pgf/number format/fixed zerofill
    },
]


\addlegendimage{black, ransstyle, sharp plot};
\addlegendentry{$k$--$\epsilon$}
\addlegendimage{black, lstyle, sharp plot};
\addlegendentry{$none$}

\addplot[col1, ransstyle, sharp plot] table[x=Time010RANS, y=Area010RANS, col sep=semicolon]{surface_area.dat};
\addlegendentry{$\SI{0.1}{\metre\per\second}$}
\addplot[col1, lstyle, sharp plot, forget plot] table[x=Time010L, y=Area010L, col sep=semicolon]{surface_area.dat};

\addplot[col2, ransstyle, sharp plot] table[x=Time020RANS, y=Area020RANS, col sep=semicolon]{surface_area.dat};
\addlegendentry{$\SI{0.2}{\metre\per\second}$}
\addplot[col2, lstyle, sharp plot, forget plot] table[x=Time020L, y=Area020L, col sep=semicolon]{surface_area.dat};

\addplot[col3, ransstyle, sharp plot] table[x=Time050RANS, y=Area050RANS, col sep=semicolon]{surface_area.dat};
\addlegendentry{$\SI{0.5}{\metre\per\second}$}
\addplot[col3, lstyle, sharp plot, forget plot] table[x=Time050L, y=Area050L, col sep=semicolon]{surface_area.dat};

\addplot[col4, ransstyle, sharp plot] table[x=Time150RANS, y=Area150RANS, col sep=semicolon]{surface_area.dat};
\addlegendentry{$\SI{1.5}{\metre\per\second}$}
\addplot[col4, lstyle, sharp plot, forget plot] table[x=Time150L, y=Area150L, col sep=semicolon]{surface_area.dat};

\addplot[col5, ransstyle, sharp plot] table[x=Time200RANS, y=Area200RANS, col sep=semicolon]{surface_area.dat};
\addlegendentry{$\SI{2.0}{\metre\per\second}$}
\addplot[col5, lstyle, sharp plot, forget plot] table[x=Time200L, y=Area200L, col sep=semicolon]{surface_area.dat};

\addplot[col6, ransstyle, sharp plot] table[x=Time250RANS, y=Area250RANS, col sep=semicolon]{surface_area.dat};
\addlegendentry{$\SI{2.5}{\metre\per\second}$}
\addplot[col6, lstyle, sharp plot, forget plot] table[x=Time250L, y=Area250L, col sep=semicolon]{surface_area.dat};

\end{axis}
\end{tikzpicture}
\caption{Time evolution of the free surface area at various filling velocities; comparison between simulations with and without $k$-$\varepsilon$ turbulence modeling.}
\label{fig:figureK}
\end{figure}

\begin{figure}[h]
\small
\begin{minipage}[b]{.45\textwidth} 
\subfloat[\centering]{

\newcommand{\MyMarkerSize}{3pt}
\newcommand{\MyLineWidth}{1pt}
\newcommand{\MyWidth}{\linewidth}
\newcommand{\MyHeight}{5cm}

\begin{tikzpicture}
\centering
\begin{axis}[
    width=\MyWidth,
    height=\MyHeight,
    xlabel={$We_g~/~-$},
    ylabel={$A_{s,max}~/~\SI{}{\metre\squared}$},
    xmajorgrids=true,
    xminorgrids=true,
    xmin = 10,
    xmax = 1e5,
    xmode=log,
    ymin=0,
    ymax=0.055,
    log basis x=10,
    legend style={anchor=south west, font=\small, fill=none, draw=none},
    tick align=inside,
    tick style={black},
    font=\small,
    clip=false,
    yticklabel style={
    /pgf/number format/fixed,
    /pgf/number format/precision=2,
    /pgf/number format/fixed zerofill,
    every mark/.append style={mark size=\MyMarkerSize},  
    scaled y ticks=false,
    legend style={draw=black, fill=white, font=\small},
    legend pos=north west,
    legend cell align=left,
    },
]

\addplot[
    color=col1,
    mark=*,
    thick,
    dashed,
    mark options = {solid, draw=black, line width=0.5pt},
    line width=\MyLineWidth, 
    sharp plot, mark size=\MyMarkerSize,
    ] table [x=we, y=Amax_kep, col sep=space]{Amax_vs_We.dat};
\addlegendentry{k-$\varepsilon$}

\addplot[
    color=col2,
    mark=square*,
    thick,
    dashed,
    mark options = {solid, draw=black, line width=0.5pt},
    line width=\MyLineWidth, 
    sharp plot, mark size=\MyMarkerSize,
    ] table [x=we, y=Amax_noTurb, col sep=space]{Amax_vs_We.dat};
\addlegendentry{none}
\end{axis}

\begin{axis}[
    width=\MyWidth,
    height=\MyHeight,
    hide y axis,
    axis x line=top,
    axis y line=none,
    ymin=0,
    ymax=0.055,
    xmin=0.492999087, 
    xmax=49.29990872,
    xmode=log,
    log basis x=10,
    xlabel={$v_{g}~/~\SI{}{\meter\per\second}$},
    x label style={anchor=south, font=\small},
    tick align=outside,
    tick style={black},
    font=\small,
    every outer x axis line/.append style={black},
    every tick label/.append style={font=\small},
    axis line style={-}, 
    xtick={0.5,1,2,5,10,20,40},
    xticklabels={0.5,1.0,2.0,5.0,10.0,20.0,40.0},
]
\end{axis}
\end{tikzpicture}\label{fig:WebervsSurface}}
\end{minipage}
\hfil
\begin{minipage}[b]{.45\textwidth} 
\subfloat[\centering]{

\newcommand{\MyMarkerSize}{3pt}
\newcommand{\MyLineWidth}{1pt}
\newcommand{\MyWidth}{\linewidth}
\newcommand{\MyHeight}{5cm}

\begin{tikzpicture}
\centering

\begin{axis}[
    width=\MyWidth,
    height=\MyHeight,
    xlabel={$v_{in}~/~\SI{}{\meter\per\second}$},
    ylabel={$TIFSA~/~\SI{}{\metre\squared\second}$},
    ymajorgrids=true,
    xmajorgrids=true,
    xmin = 0,
    xmax = 2.6,
    ymin=0,
    ymax=0.007,
    log basis x=10,
    legend style={anchor=south west, font=\small, fill=none, draw=none},
    tick align=inside,
    tick style={black},
    font=\small,
    clip=false,
    ytick={0.000,0.001,0.002,0.003,0.004,0.005,0.006,0.007},
    yticklabel style={
    /pgf/number format/fixed,
    /pgf/number format/precision=1,
    /pgf/number format/fixed zerofill,
    every mark/.append style={mark size=\MyMarkerSize},  
    legend style={draw=black, fill=white, font=\small},
    legend pos=north east,
    legend cell align=left,
    },
]

\addplot[
    color=col1,
    mark=*,
    thick,
    dashed,
    mark options = {solid, draw=black, line width=0.5pt},
    line width=\MyLineWidth, 
    sharp plot, mark size=\MyMarkerSize,
    ] table [x=vin, y=RANS, col sep=space]{tifsa_comb.dat};
\addlegendentry{k-$\varepsilon$}

\addplot[
    color=col2,
    mark=square*,
    thick,
    dashed,
    mark options = {solid, draw=black, line width=0.5pt},
    line width=\MyLineWidth, 
    sharp plot, mark size=\MyMarkerSize,
    ] table [x=vin, y=L, col sep=space]{tifsa_comb.dat};
\addlegendentry{none}
\end{axis}

\begin{axis}[
    width=\MyWidth,
    height=\MyHeight,
    hide y axis,
    axis x line=top,
    axis y line=none,
    ymin=0,
    xmin=0, 
    xmax=34.666666666666666666,
    ymax=0.007,
    log basis x=10,
    xlabel={$v_{g}~/~\SI{}{\meter\per\second}$},
    x label style={anchor=south, font=\small},
    tick align=outside,
    tick style={black},
    font=\small,
    every outer x axis line/.append style={black},
    every tick label/.append style={font=\small},
    axis line style={-}, 
]

\end{axis}

\end{tikzpicture}\label{fig:figureL}}
\end{minipage}
\caption{Influence of filling velocity on (a) the maximum free surface area \gls{sym:as}$_{max}$ and (b) the \gls{tifsa} criterion}
\label{fig:figure12}
\end{figure}
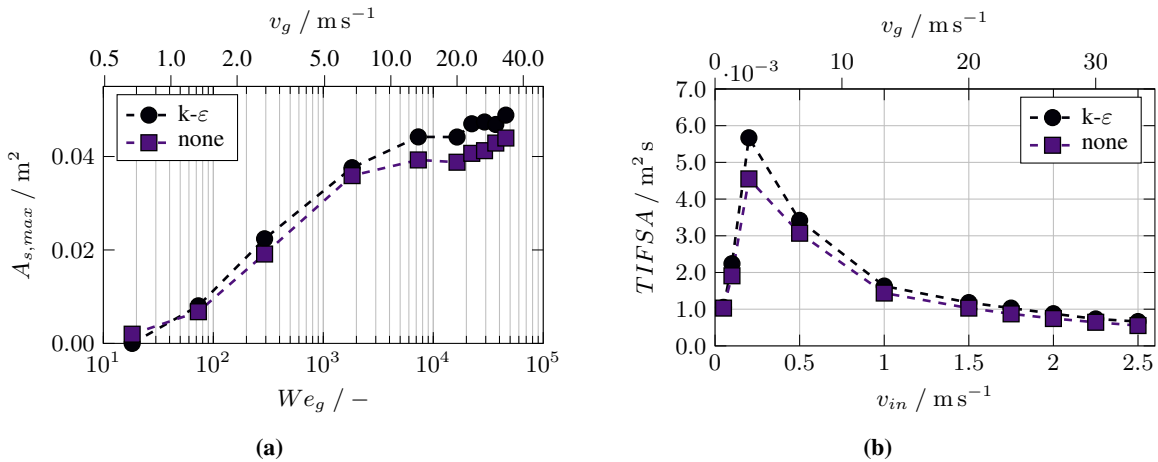

The temporal evolution of the free surface area for different inlet velocities is shown in Figure~\ref{fig:figureK} for the \SI{3}{\milli\metre} plate geometry, consistent with the evaluation domain used in Figs.~\ref{fig:figureH}-\ref{fig:figureJ}. 
The curves illustrate how free surface formation develops during filling rather than only at isolated instants. At low velocities, the free surface grows gradually and reaches relatively small peak values, followed by a long decay phase. With increasing velocity, the free surface rises more steeply, reaches higher maxima, and decays more abruptly as the cavity fills faster. The overall shape of the curves therefore reflects the interplay between interface generation and the progressively shorter filling times at higher velocities. The comparison of simulations with and without the $k$-$\varepsilon$ turbulence model confirms this trend but reveals systematic differences. In all cases, turbulence modeling predicts larger peak areas, on average about \SI{13}{\percent} higher than without turbulence modeling, see Table~\ref{tab:table4}. 
\begin{table}
\footnotesize
\centering
\caption{Differences in integral free surface area between simulations with and without turbulence modeling for various inlet velocities.}
\label{tab:table4}
\begin{tabularx}{\textwidth}{lcXXXXXXXXXX}
\toprule
\textbf{Inlet velocity} \gls{sym:vin} 
& \SI{}{\meter\per\second} & 0.05 & 0.10 & 0.20 & 0.50 & 1.00 & 1.50 & 1.75 & 2.00 & 2.25 & 2.50 \\ 
\textbf{Abs. integral difference $\mathbf{10^{-4}}$}
& \SI{}{\meter\squared\second} & 0.20 & 3.30 & 11.0 & 3.50 & 1.90 & 1.50 & 1.60 & 1.30 & 0.91 & 1.10 \\
\textbf{Rel. integral difference}
& \SI{}{\percent} & 1.90 & 14.7 & 19.8 & 10.2 & 11.7 & 12.7 & 15.1 & 15.0 & 12.4 & 16.5 \\ 
\bottomrule
\end{tabularx}
\end{table}
Moreover, the timing and sharpness of the peaks differ: some cases reach their maximum earlier, while others display broader or more irregular profiles. These differences underline that free surface dynamics cannot be reduced to a linear scaling with velocity. Instead, they emerge from a combination of turbulence intensity, vortex formation, and local flow separation, which evolve in a strongly nonlinear manner as the filling conditions change. The dependence of the maximum free surface area on the ingate Weber number \gls{sym:weg} is shown in Figure~\ref{fig:WebervsSurface}. At low velocities \gls{sym:weg} is small (\gls{sym:weg}~$ \approx 18$), and surface tension still suppresses interface deformation, resulting in relatively smooth filling with only limited folding. As velocity increases, \gls{sym:weg} rises quadratically, strengthening inertial forces and driving interface fragmentation. This produces a sharp growth in free surface area up to intermediate values (\gls{sym:weg}~$ \approx 1.8\times10^{3}$), where turbulent breakup becomes dominant and extensive interfacial corrugation develops. Once \gls{sym:weg} exceeds this range, however, the maximum free surface area no longer increases proportionally. At very high velocities (\gls{sym:weg}~$ \approx 4.6\times10^{4}$) inertia completely overwhelms surface tension, but the extremely short filling times limit the opportunity for further surface formation. The resulting behavior is a degressive increase that approaches saturation. This interpretation is consistent with breakup studies of oxidizing droplets, where mode transitions occur around \gls{sym:we}~$ \sim 15$–$35$~\cite{hopfes2021a,hopfes2021b}. In the present geometry, the lowest case lies in this transition regime, while the intermediate and high cases fall well into the turbulence controlled domain typical of \gls{hpdc} gating conditions. Importantly, in the high-\gls{sym:weg} regime oxide films are expected to be entrained in a highly dispersed form, which is considered less detrimental to mechanical performance~\cite{campbell2011,tiryakioglu2020}. Numerical effects must also be acknowledged: higher velocities increase the Courant number, requiring smaller timesteps to resolve the interface. In this study, stability criteria were satisfied (Fig.~\ref{fig:figureB}), ensuring that the observed saturation is physical. The relatively simple plate geometry further limits turbulence development, which reduces differences between laminar and turbulence modeled cases but does not alter the overall Weber number scaling. The \gls{tifsa} criterion provides a global measure of total fluid-air interface exposure during mold filling (Fig.~\ref{fig:figureL}). Unlike the maximum values, which increase monotonically with \gls{sym:weg} until saturation, the \gls{tifsa} exhibits a non-monotonic trend: it rises sharply at intermediate velocities before declining again at higher values. This behavior reflects the competition between free surface generation and filling duration. At moderate \gls{sym:we}, turbulence creates large interfacial areas while filling times remain sufficiently long for these surfaces to persist, yielding the peak in exposure. At very high \gls{sym:we}, turbulence is stronger but the cavity is filled so rapidly that the overall exposure is reduced. From this trend, two process windows can be identified:  
\begin{enumerate}
    \item \textbf{Low-velocity regime (\gls{sym:we}~$ \sim 10$):} Filling is smooth and tranquil, but long filling times promote oxidation and premature solidification. Such conditions are unfavourable for \gls{hpdc} but are typical of slower processes such as sand casting.  
    \item \textbf{High-velocity regime (\gls{sym:we}~$ \gg 10^{3}$):} Turbulence and breakup are intense, yet the short filling times limit cumulative exposure. This regime is characteristic of \gls{hpdc} and underpins the practical preference for rapid filling.  
\end{enumerate}
Between these regimes lies a transitional range corresponding to the \gls{tifsa} peak, which combines the disadvantages of both: substantial turbulence induced surface creation together with prolonged atmospheric exposure. Beyond this point, further increases in velocity reduce the integrated exposure despite higher instantaneous free surface areas.

\subsection{X-ray computed tomography and Temporal Mean Volume Fraction (TMVF)}
\label{ssec:xray}
\begin{figure}[hbt]
\scriptsize
\centering 
\begin{minipage}[b]{\textwidth} 
\subfloat[\centering]{\def\svgwidth{210pt}\input{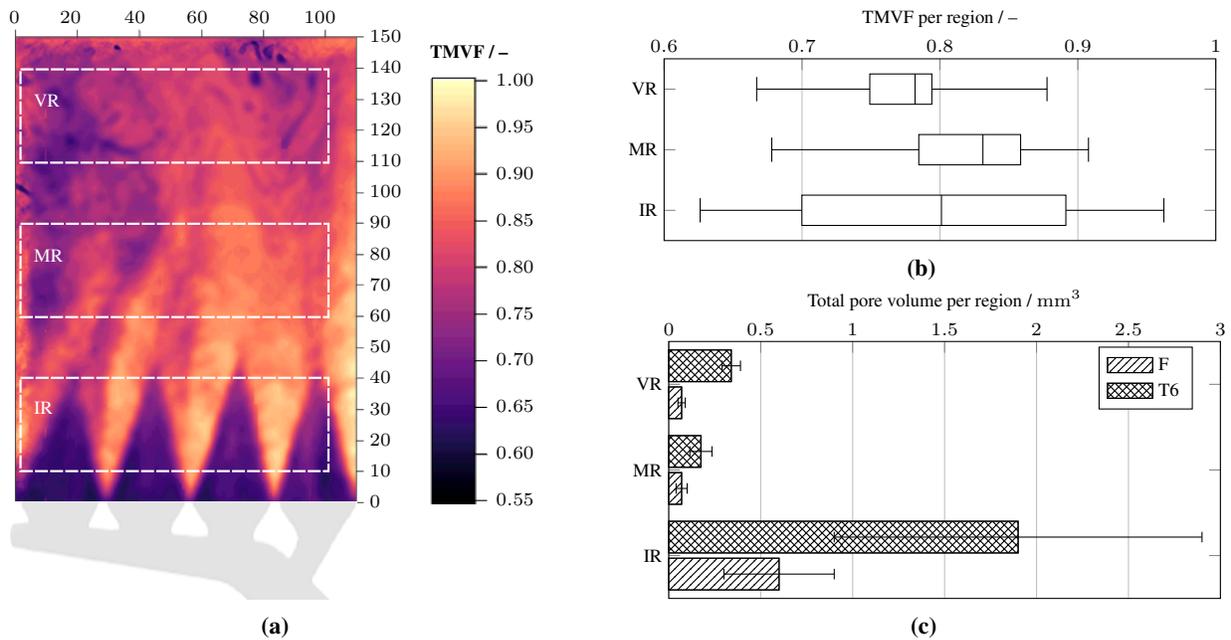}\label{fig:tmvf}} \hfil
\begin{minipage}[b]{0.45\textwidth} 
\subfloat[\centering]{
\begin{tikzpicture}
\begin{axis}[
    xmajorgrids,
    width=1.2\linewidth,
    height=4cm,
    xmin=0.6,
    xmax=1,
    ytick={1,2,3},
    yticklabels={\gls{ir}, \gls{mr}, \gls{vr}},
    xtick pos=top,
    xlabel={\gls{tmvf} per region / --},
    ytick align=inside,
]
    \addplot+ [draw=black, boxplot prepared={
        lower whisker=0.6261, 
        lower quartile=0.69991,
        median=0.80107, 
        upper quartile=0.8913,
        upper whisker=0.96238,
        box extend=.5},
        draw=black,
        fill=white,
    ] coordinates {};

    \addplot+ [boxplot prepared={
        lower whisker=0.67804, 
        lower quartile=0.78467,
        median=0.83115, 
        upper quartile=0.85849,
        upper whisker=0.9077,
        box extend=.5},
        draw=black,
        fill=white,
    ] coordinates {};

    \addplot+ [boxplot prepared={
        lower whisker=0.66711, 
        lower quartile=0.74913,
        median=0.78194, 
        upper quartile=0.79413,
        upper whisker=0.87763,
        box extend=.5},
        draw=black,
        fill=white,
    ] coordinates {};

\end{axis}
\end{tikzpicture}\label{fig:boxplotDistTMVF}} \vspace{-1pt}
\subfloat[\centering]{\definecolor{orange}{rgb}{1,0.5,0}

\begin{tikzpicture}
  \begin{axis}[
    xbar,
    area legend,
    bar width=12pt,
    width=1.2\linewidth,
    height=5cm,
    symbolic y coords={\gls{ir}, \gls{mr}, \gls{vr}},
    ytick=data,
    xlabel={Total pore volume per region / \SI{}{\cubic\milli\meter}},
    enlarge y limits=0.25,
    legend style={at={(0.95,0.98)}, anchor=north east, draw=black},
    legend cell align={left},
    xmajorgrids,
    xtick pos=top,
    xmin=0,
    xmax=3,
    ytick align=inside,
    legend style={/pgfplots/single xbar legend},
  ]
    \addplot+[
      xbar,
      fill=white,
      postaction={pattern=north east lines},
      draw=black,
      error bars/.cd,
      x dir=both, x explicit,
      error bar style={black},
    ] coordinates {
        (0.6,\gls{ir}) +- (0.3,0)
        (0.07,\gls{mr}) +- (0.03,0)
        (0.07,\gls{vr}) +- (0.02,0)
    };
    \addlegendentry{F}
    
    \addplot+[
      xbar,
      fill=white,
      postaction={pattern=crosshatch},
      draw=black,
      error bars/.cd,
      x dir=both, x explicit,
      error bar style={black},
    ] coordinates {
        (1.9,\gls{ir}) +- (1,0)
        (0.175,\gls{mr}) +- (0.06,0)
        (0.34,\gls{vr}) +- (0.05,0)
    };
    \addlegendentry{T6}
    
  \end{axis}
\end{tikzpicture}

\label{fig:porosity_bar}}
\end{minipage}
\end{minipage}
\caption{Comparison of \gls{tmvf} and \gls{ct} pore volume with (a) \gls{tmvf} field for the simulation conducted with $v_{inlet} = \SI{2.25}{\meter\per\second}$ and turbulence modeling, (b) boxplots of the \gls{tmvf} distribution in the evaluation regions and (c) the total pore volume in said regions, taken from the \gls{ct} analysis performed on as cast (F) and heat treated (T6) specimens.}
\end{figure}

\gls{ct} was performed on thirty specimens, comprising five heat treated (T6) and five as-cast (F) samples for each of the three defined regions: \gls{ir}, \gls{mr}, and \gls{vr} (Fig.~\ref{fig:platte}). As shown in Figure~\ref{fig:porosity_bar}, the total pore volume is highest in the \gls{ir}, followed by the \gls{vr}, while the \gls{mr} exhibits the lowest values. This ordering is consistent across both as-cast and T6 conditions, although absolute pore volumes increase after heat treatment due to the thermal expansion and recombination of pre-existing gas inclusions, particularly in the \gls{vr}. For the simulations, the cumulative mass of entrained air was tracked throughout the filling phase. The resulting evolution is presented in Figure~\ref{fig:figureM}, which compares entrapped air volume and mean pressure histories in the \gls{mr} and \gls{vr}. The \gls{cfd} results reproduce the experimental observations, predicting higher air content in \gls{vr} than in \gls{mr}. A slope based regression analysis confirms that porosity in \gls{vr} is reduced less effectively during pressurization. In absolute terms, the simulated entrapped volume in \gls{vr} is nearly twice that of \gls{mr}, closely resembling the factor-of-two difference observed in \gls{ct}. The \gls{tmvf} distribution (Fig.~\ref{fig:tmvf}) provides further mechanistic insight into filling behavior. Unlike surface-only diagnostics such as \gls{tivf}, this three-dimensional field captures the cumulative liquid residence across the plate thickness, highlighting differences in wetting continuity, recirculation, and intermittent exposure. In the \gls{ir}, \gls{tmvf} variance is highest (Fig.~\ref{fig:boxplotDistTMVF}), with elevated values in zones where overlapping jets ensure continuous wetting and low values in adjacent areas near the ingates that remain intermittently gas filled. This bimodal distribution identifies the \gls{ir} as a continuous mixing zone with sharp interfaces and strong surface renewal. Free surface renderings (Fig.~\ref{fig:figureI}) support this interpretation, revealing high interfacial activity that favors porosity despite strong local flow. In the \gls{mr}, \gls{tmvf} values are comparatively homogeneous and plateau at intermediate-to-high levels, reflecting stable filling and limited recirculation. This corresponds to the lowest porosity detected by both \gls{ct} and simulations (Fig.~\ref{fig:figureJ},~\ref{fig:figureM}), confirming the \gls{mr} as the least defect prone section. By contrast, the \gls{vr} exhibits mottled low-to-intermediate \gls{tmvf} values and eddy like patterns. These signatures, consistent with the fragmented interface structures in Figure~\ref{fig:figureJ}, indicate repeated surface re-exposure and late filling, which promote persistent pore survival. The statistical evaluation reinforces these observations. As shown in Figure~\ref{fig:boxplotDistTMVF}, the lower percentiles of the \gls{tmvf} distribution inversely correspond to the pore volumes measured by \gls{ct}. Regions with extended low-\gls{tmvf} tails (\gls{ir} and \gls{vr}) exhibit higher porosity, while the \gls{mr}, characterized by uniformly high \gls{tmvf}, remains largely defect free.  

Although \gls{tmvf} is stationary in nature and cannot resolve the transient migration of entrained air or the advection of the filling front, it provides a robust local footprint of filling history. Its strength lies in delineating regions characterized by (i) strong mixing and eddies, (ii) sharp, repeatedly renewed interfaces, and (iii) extended flow paths leading to intermittent wetting. In combination with \gls{ct}, it offers a consistent mechanistic explanation of porosity formation across the three regions and supports its use as a diagnostic indicator of casting quality in thin-walled \gls{hpdc} components. Nonetheless, further validation on more complex geometries is recommended to generalize this approach.

\subsection{Experimental validation of flow patterns}
\begin{figure}[h]
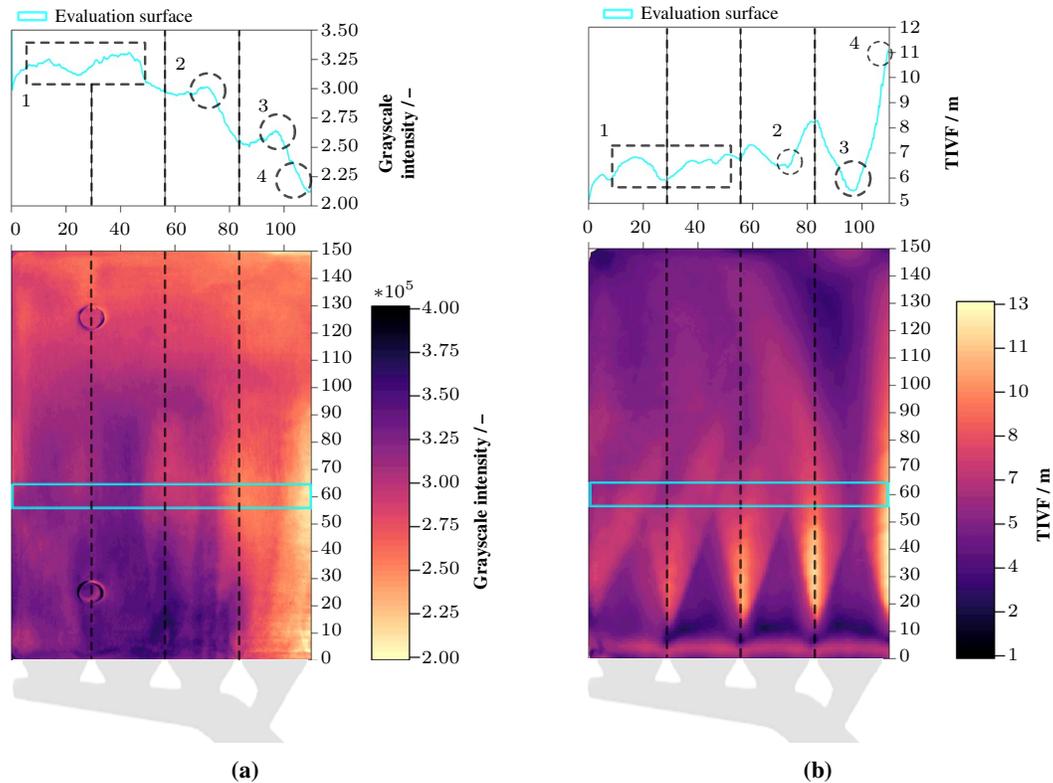

\scriptsize
\centering 
\subfloat[\centering]{\def\svgwidth{180pt}\input{figure14_a}\label{sfig:exp_flow_a}} \hfil
\subfloat[\centering]{\def\svgwidth{180pt}\input{figure14_b}\label{sfig:exp_flow_b}} 
\caption{Experimental-numerical comparison of surface modification patterns in \gls{hpdc} as (a) Normalized grayscale intensity map and (b) Simulated \gls{tivf}}
\label{fig:exp_flow}
\end{figure}

Figure~\ref{fig:exp_flow} compares normalized grayscale intensity maps with the simulated \gls{tivf} distribution, linking experimentally observed surface modification to numerically predicted cumulative melt flux. The grayscale intensity (Fig.~\ref{sfig:exp_flow_a}) highlights discoloration and blank exposure caused by oxide transport, lubricant removal, and repeated impingement. 

The scale has been inverted so that bright regions correspond to low grayscale values, indicating strongly modified, oxide free surfaces, whereas darker tones mark less affected areas. Averaging across twenty mirrored plates suppresses sample-to-sample variability and emphasizes robust patterns. The inclusion of the gating system at the bottom of the figure contextualizes the jet entry points and subsequent flow development. The simulated \gls{tivf} field (Fig.~\ref{sfig:exp_flow_b}) quantifies the cumulative volumetric flux across surface adjacent cells and thus highlights zones repeatedly swept by melt. Triangular high value lobes appear above the ingates and broaden upward as jets expand and overlap. This morphology agrees with the instantaneous jet structures in Fig.~\ref{fig:figureI}d–f, where the $k$-$\varepsilon$ model resolves coherent upward plumes. Whereas Fig.~\ref{fig:figureI} captures single filling stages, \gls{tivf} integrates their effect over time, revealing the cumulative imprint of jet impingement on cavity surfaces. The annotated zones (1–4) illustrate specific correspondences between experiment and simulation:  

\begin{itemize}
  \item \textbf{Zone 1:} A broad, diffuse dark region in grayscale coincides with a low-\gls{tivf} plateau, reflecting weak surface renewal where jets diverge and flux is limited.  
  \item \textbf{Zone 2:} A localized dark band corresponds to a distinct \gls{tivf} minimum, suggesting that laterally spreading jets, visible in Fig.~\ref{fig:figureI}, bypass this region.  
  \item \textbf{Zone 3:} Similar to Zone~2, both grayscale and \gls{tivf} show local minima, representing an inter-jet gap with reduced impingement and weaker modification.  
  \item \textbf{Zone 4:} A bright experimental zone (very low grayscale) aligns with a pronounced \gls{tivf} maximum, where jets converge and overlap, producing blank, oxide free surfaces.  
\end{itemize}  
The agreement across these regions shows that \gls{tivf} resolves not only primary jet footprints but also inter-jet gaps and overlap zones, capturing both strong and weak surface modification. Quantitatively, the line plots (top panels) confirm this correspondence: local minima in grayscale intensity (Zones~2–3) coincide with local minima in \gls{tivf}, while maxima in \gls{tivf} (Zone~4) align with bright experimental zones. Some discrepancies remain. Simulated \gls{tivf} peaks are sharper and more contrasted, reflecting the numerical resolution of coherent jets, while the grayscale field is smoother due to averaging, optical effects, and variations in release agent coverage. Global slopes also differ: \gls{tivf} increases slightly toward the right hand ingates, consistent with the mild imbalance in ingate velocity histories (Fig.~\ref{fig:figureO}), whereas grayscale intensity decreases, likely influenced by local surface chemistry and lubricant distribution. Further refinement, such as coupling \gls{tivf} with local pressure or shear rate information, may enhance predictive accuracy.

\section{Discussion}
\label{sec:disc}
The numerical framework developed in this study provides a consistent basis for simulating free surface dynamics, cavity pressurization, and defect related flow phenomena in \gls{hpdc}. Verification through mesh and residual analyses confirmed that accurate prediction of flow separation and interface stability requires sufficient spatial resolution. In particular, the third mesh case, containing approximately $1.4\times10^5$ cells, achieved convergence in free surface predictions at less than half the computational cost of the finest mesh. This highlights that overly coarse meshes, still common in industrial practice, can severely underestimate air entrapment and overpredict casting quality.

The compressible multiphase formulation proved essential to capture the cavity pressurization and the interaction between the advancing fluid and entrained air. The analysis of ten inlet velocities showed that filling speed strongly controls pressure evolution: lower velocities delayed and weakened pressurization, whereas higher velocities produced earlier, sharper pressure peaks. These findings reflect the central trade-off in \gls{hpdc}: avoiding premature solidification at low velocities while minimizing turbulence driven air entrapment at high velocities. Porosity evaluation was shown to be time sensitive, since pore volume contracts as cavity pressure rises. Meaningful assessment therefore requires velocity and pressure histories at critical locations to select physically consistent evaluation windows.

The turbulence study indicated that laminar and $k$-$\varepsilon$ closures predict comparable global filling patterns but diverge in local surface dynamics. Without turbulence modeling, free surface creation was underestimated and interface evolution appeared overly smooth, while the $k$-$\varepsilon$ model preserved surface coherence and exposed localized accumulations of entrapped air. For engineering purposes, this suggests that laminar simulations may approximate bulk filling but risk overlooking small scale surface structures that seed porosity. At the same time, even with the applied resolution, only relatively large pores could be tracked numerically; finer scale porosity or bifilm defects remain below grid resolution. Advanced closures such as \gls{les} or \gls{des} may offer higher fidelity for complex gating systems, though at significantly increased computational cost.

Validation against experimental data further confirmed the predictive capability of the solver. \gls{ct} based porosity mapping and simulation derived \gls{tmvf} fields consistently identified the top region of the plate as prone to pore accumulation, while the mid region remained relatively free of defects. Beyond absolute pore counts, the \gls{cfd} reproduced the relative \gls{vr}-\gls{mr} difference observed in \gls{ct}, demonstrating that regional porosity trends can be reliably captured. Together with complementary evaluation metrics such as \gls{tifsa} and \gls{tivf}, the approach provides a practical diagnostic basis for defect prediction, supporting virtual prototyping and early stage process evaluation.

Despite these results, several limitations remain: water was used as a surrogate fluid, ensuring Reynolds number similarity but omitting the thermal-rheological coupling of molten aluminum. Solidification shrinkage, gas solubility, and oxide film effects were neglected, and the simplified plate geometry does not include gating, runners, or vents. These simplifications mean that while qualitative trends are transferable, quantitative defect predictions in industrial parts require additional modeling complexity. Addressing these aspects through coupled solidification models, higher fidelity turbulence closures, and extension to real die geometries defines the next steps toward robust virtual design of \gls{hpdc} processes.

\section{Conclusion}
This work established and validated a \gls{cfd} based framework for analyzing compressible multiphase flow, free surface dynamics, and porosity formation in \gls{hpdc}. The main findings can be summarized as follows:

\begin{itemize}
  \item Accounting for compressibility was essential to capture cavity pressurization and pore contraction. Inlet velocity directly controlled both the magnitude and timing of pressure peaks, and with rising pressure pores were compressed or dissolved. This highlights the need to define evaluation windows carefully, based on velocity and pressure histories, rather than on a fixed filling fraction.
  
  \item Turbulence modeling influenced free surface morphology and porosity prediction. While laminar and $k$-$\varepsilon$ closures produced similar global filling behavior, the turbulence model preserved front coherence, accelerated pressurization, and reduced the persistence of interior air pockets. This demonstrates that turbulence representation is decisive for reliable porosity assessment.
  
  \item Comparison with \gls{ct} based porosity mapping confirmed the predictive value of the free surface criteria and \gls{tmvf} metric. Both consistently identified the \gls{vr} as most defect prone, while the \gls{mr} remained relatively defect free. Beyond absolute pore counts, the simulations reproduced the relative \gls{vr}-\gls{mr} difference observed experimentally, underscoring the ability of the framework to capture regional defect trends relevant for design.
  
  \item \gls{tifsa}, \gls{tmvf}, and \gls{tivf} provide a complementary set of indicators: \gls{tifsa} condenses surface exposure and oxidation risk into a single measure; \gls{tmvf} captures wetting continuity and mixing zones; and \gls{tivf} highlights preferential flow paths and wall loading. Together, they form a practical toolkit for early stage process assessment without exhaustive filling investigations.
  
  \item The simulations revealed the inherent trade-off between filling speed, turbulence, and defect risk: higher inlet velocities suppress misruns but intensify turbulence and pressure peaks, whereas lower velocities reduce turbulence but risk incomplete filling. The framework allows this balance to be quantified and explored systematically.
\end{itemize}

\noindent
Overall, the study demonstrates that compressible multiphase \gls{cfd}, combined with targeted evaluation metrics, can reproduce experimentally observed porosity distributions and provide physics based guidance for process optimization in \gls{hpdc}. Despite the relatively fine mesh applied, only larger pores could be resolved, while smaller bifilms or dissolved gases remain below grid scale, underscoring the resolution dependence of porosity prediction. Likewise, the simplified plate geometry does not reflect the complexity of industrial castings. These limitations open clear avenues for future work: extending the framework to full industrial geometries, integrating solidification and gas solubility models, and assessing higher fidelity turbulence closures such as LES to better capture transient vortex dynamics.








\FloatBarrier

\bibliography{09_References.bib}

\end{document}